\documentclass[authoryear,1p]{elsarticle}

\pdfoutput=1

\usepackage{amsmath}
\usepackage{amssymb}
\usepackage{graphicx}
\usepackage[colorlinks=false,citecolor=red,linkcolor=black]{hyperref}
\usepackage{lineno}
\usepackage{lscape}
\usepackage{xcolor}

\newcommand{\pdx}[1]{\frac{\partial {#1}}{\partial x}}
\newcommand{\pdt}[1]{\frac{\partial {#1}}{\partial t}}
\newcommand{\bb}[1]{\mathrm{\textbf{#1}}}

\journal{Advances in Water Resources}

\begin{document}
	
	\begin{frontmatter}
		
		
		\title{Analytical implementation of Roe solver for two-layer shallow water equations with accurate treatment for loss of hyperbolicity}

		\author[gradri]{Nino Krvavica\corref{mycorrespondingauthor}}
		\cortext[mycorrespondingauthor]{Corresponding author}
		\ead{nino.krvavica@uniri.hr}
		\author[gradri]{Miran Tuhtan}
		\author[gradri]{Gordan Jeleni\'c}

		\address[gradri]{University of Rijeka, Faculty of Civil Engineering, Radmile Matejcic 3, 51000 Rijeka, Croatia}

		\fntext[fn1]{Accepted manuscript, DOI:10.1016/j.advwatres.2018.10.017}
		
		\begin{abstract}
			
			A new implementation of the Roe scheme for solving two-layer shallow-water equations is presented in this paper. The proposed A-Roe scheme is based on the analytical solution to the characteristic quartic of the flux matrix, which is an efficient alternative to a numerical eigensolver. Additionally, an accurate method for maintaining the hyperbolic character of the governing system is proposed. The efficiency of the quartic closed-form solver is examined and compared to numerical eigensolvers. Furthermore, the accuracy and computational speed of the A-Roe scheme is compared to the Roe, Lax-Friedrichs, GFORCE, PVM, and IFCP schemes. Finally, numerical tests are presented to evaluate the efficiency of the iterative treatment for the hyperbolicity loss. The proposed A-Roe scheme is as accurate as the Roe scheme, but much faster, with computational speeds closer to the GFORCE and IFCP scheme.
			
		\end{abstract}
		
		
	\end{frontmatter}
	
%	\linenumbers

\section{Introduction}

Shallow-water equations (SWE) are widely used to simulate geophysical flows with dominantly horizontal processes. These equations can be extended to a two-layer system that describes the flow of two superimposed and immiscible layers of fluid with different densities or even different phases. For example, a two-layer configuration is found in sea straits \citep{castro2001q,castro2004numerical}, highly stratified estuaries \citep{krvavica2017numerical,krvavica2016salt}, gravity currents \citep{la2012two,adduce2011gravity}, mudflows \citep{canestrelli2012mass}, debris flows \citep{pelanti2008roe,majd2014lhllc}, submarine avalanches \citep{fernandez2008new,luca2009two}, and pyroclastic flows \citep{doyle2011two}. Although such processes can be described more accurately by 3D Navier-Stokes equations, two-layer models make a popular alternative because of their simplicity and a significantly lower computational cost.

Two-layer SWE are defined as a coupled system of conservation laws with source terms, or so-called balance laws \citep{castro2001q}. These equations are challenging to solve numerically because of the layer coupling and non-conservative source terms accounting for the variable geometry or friction. In recent years, numerical methods for solving two-layer equations have received great attention and have been an object of intense research \citep{castro2001q,castro2004numerical,kurganov2009central,castro2010some,bouchut2010robust,fernandez2011intermediate,canestrelli2012restoration}. A number of authors have presented different numerical schemes for non-conservative hyperbolic systems based on the finite-difference method \citep{fjordholm2012energy,liu2015influence}, finite-element method \citep{ljubenkov2015hydrodynamic} or, more often, finite-volume method (FVM) \citep{castro2001q,kurganov2009central,bouchut2010robust,canestrelli2012restoration}. 

Among the most popular and robust FVM schemes are Roe schemes, which belong to a family of approximate Riemann solvers \citep{bermudez1994upwind,castro2001q,pares2004well}. Roe schemes have good well-balanced properties and in comparison to incomplete Riemann solvers, such as Lax-Friedrichs, HLL or FORCE/GFORCE methods, are less diffusive and provide better resolution of discontinuities \citep{castro2010some,kesserwani2008riemann}. However, Roe schemes require computation of the full eigenstructure of the flux matrix at each time step \citep{castro2010some}. When analytical expressions for the eigenstructure are unavailable, a spectral decomposition of the flux matrix is needed, making Roe schemes computationally expensive and, therefore, less attractive for practical applications, such as simulating complex geophysical flows in sea straits, stratified estuaries, submarine avalanches, etc.

In this research field, there do not exist explicit formulations for eigenvalues of coupled two-layer SWEs which are directly expressed in terms of the conserved variables \citep{castro2004numerical}.
Because of the coupling and the corresponding $4 \times 4$ flux matrix, some authors suggest that it is not possible to derive the explicit form of eigenvalues, \textit{e.g.}, \textit{"...simple explicit expressions of the system's eigenvalues cannot be derived..."} \citep{pelanti2008roe}, \textit{"...the explicit expression for the eigenvalues cannot be found."} \citep{kim2008two}, \textit{"The coupling between the layers... does not provide explicit access to the system eigenstructure"} \citep{abgrall2009two}, whereas others are aware of the existence of the analytical solutions to quartic equations but considered them to be too complicated or less efficient, \textit{e.g.}, \textit{"...there is not an easy explicit expression of the eigenvalues..."} \citep{fernandez2011intermediate}, \textit{"...a direct calculation of its eigenvalues can be hard..."} \citep{fjordholm2012energy}, \textit{"...a closed form of the eigenvalues is non-trivial..."} \citep{sarno2017some}, \textit{etc}.
On the other hand, Cardano-Vieta formula for cubic equations has been used as a more efficient approach in comparison to numerical solvers when computing eigenstructure of Saint Venant-Exner models, defined by a cubic characteristic equation (see \cite{castro2009two} and \cite{carraro2018efficient}).

Considering the computational cost of spectral decomposition and the prevailing opinion that explicit eigenvalues are "unavailable", \cite{fernandez2011intermediate} and \cite{castro2012class} have recently proposed new Riemann solvers based on the polynomial approximation of the viscosity matrix, which should represent a good compromise between the computational speed and accuracy. 

Taking all these specific concerns into account, the main goal of this paper is to present a more efficient implementation of the Roe scheme for a coupled two-layer SWE system, which is based on a compact analytical solution to the eigenstructure. New analytical formulae are proposed, which may be used instead of numerical tools and algorithms when computing eigenvalues and eigenvectors at each time step. Additionally, a numerical treatment for the hyperbolicity loss is presented that always leads to a state that is close to the boundary of the hyperbolicity region but inside its interior, which avoids the appearance of both complex and double eigenvalues.

This paper is organized as follows: first, the governing system of a coupled two-layer SWE system is defined; next, a path-conserving numerical scheme is presented with an analytical solution to the eigenstructure; several results are also presented, namely, the computational cost and accuracy analysis of the closed-form quartic solver, as well as several performance tests of the proposed scheme; and finally, the results are discussed and conclusions are drawn.

\section{Two-layer shallow-water flow: theory, Roe scheme and analytical eigenvalue resolution}

\subsection{Governing system of equations}

Let us consider the following PDE system derived for a one-dimensional (1D) two-layer shallow-water flow in prismatic channels with rectangular cross-sections of constant width, written in a general vector form \citep{castro2001q}:
\begin{equation}
\pdt{\bb{w}} + \pdx{\bb{f}(\bb{w})} = 
\bb{B}(\bb{w})\pdx{\bb{w}} 
+ \bb{g}(\bb{w}),
\label{eq:consys1}
\end{equation}
where $x$ refers to the axis of the channel and $t$ is time. The vector of conserved quantities $\bb{w}$, the flux vector $\bb{f(w)}$ and the bathymetry source term $\bb{g(w)}$ are respectively defined as follows \citep{castro2001q}:
\begin{equation}
\bb{w} = 
\begin{Bmatrix}
h_{1} \\  q_{1} \\ h_{2} \\ q_{2}
\end{Bmatrix},
\quad 
\bb{f}(\bb{w}) = 
\begin{Bmatrix}
q_{1} \\
\tfrac{q^{2}_{1}}{h_{1}} + \tfrac{g}{2}h^{2}_{1}  \\
q_{2} \\
\tfrac{q^{2}_{2}}{h_{2}} + \tfrac{g}{2}h^{2}_{2}
\end{Bmatrix},
\quad
\bb{g}(\bb{w}) = 
\begin{Bmatrix}
0 \\
-gh_1\frac{\textrm{d}b}{\textrm{d}x}\\
0 \\
-gh_2\frac{\textrm{d}b}{\textrm{d}x} \\
\end{Bmatrix},
\end{equation}
where $h_j$ is the layer thickness (or depth), $q_j=h_j u_j$ is the layer flow rate per unit width, $u_j$ is the layer-averaged horizontal velocity, $g$ is acceleration of gravity, $b$ is the bed elevation, and index $j=1,2$ denotes the respective upper and lower layer. Matrix $\bb{B(w)}$ is a result of coupling the two-layer system, defined as \citep{castro2001q}:
\begin{equation}
\bb{B}(\bb{w}) = 
\begin{bmatrix}
0 & 0 & 0 & 0 \\
0 & 0 & -gh_1 & 0 \\
0 & 0 & 0 & 0 \\
-grh_2 & 0 & 0 & 0
\end{bmatrix},
\end{equation}
where $r=\rho_1/\rho_2 < 1$ is the ratio between the upper layer density $\rho_1$ and the lower layer density $\rho_2$.

The right-hand side of Eq.~(\ref{eq:consys1}) contains the terms describing the momentum exchange between two layers, and source terms for channel bathymetry. The system can be rewritten in the following quasi-linear form \citep{castro2001q}:
\begin{equation}
\frac{\partial \mathbf{w}}{\partial t} + \boldsymbol{\mathcal{A}}(\mathbf{w})\frac{\partial \mathbf{w}}{\partial x} = 
\mathbf{g}(\mathbf{w}),
\label{eq:consys2}
\end{equation}
where 
\begin{equation}
\boldsymbol{\mathcal{A}(\mathbf{w}}) = 
\frac{\partial \mathbf{f(w)}}{\partial \mathbf{w}} - \mathbf{B}(\mathbf{w}) = 
\mathbf{J}(\mathbf{w}) - \mathbf{B}(\mathbf{w})
\end{equation} is the pseudo-Jacobian matrix that contains the flux gradient terms as well as the coupling terms:
\begin{equation}
\boldsymbol{\mathcal{A}(\mathbf{w}}) =
\begin{bmatrix}
0 				& 1 	& 0 			& 0 \\
c_1^2 - u_1^2 	& 2u_1 	& c_1^2 		& 0 \\
0 				& 0 	& 0 			& 1 \\
rc_2^2 			& 0 	& c_2^2 - u_2^2	& 2u_2.
\end{bmatrix}
\label{eq:A}
\end{equation}
where $c_j^2=gh_j$, is propagation celerity of internal and external perturbations (waves), for $j=1,2$. 

The characteristic polynomial of $\boldsymbol{\mathcal{A}(\mathbf{w})}$ is defined as $p(\lambda) = \textrm{det} \left( \boldsymbol{\mathcal{A}} - \lambda \bb{Id} \right)$, where $\lambda$ is the eigenvalue of $\boldsymbol{\mathcal{A}(\mathbf{w})}$ and $\bb{Id}$ is a $4\times4$ identity matrix. The coefficients of the 4th order characteristic polynomial
\begin{equation}
p(\lambda) = \lambda^4 + a\lambda^3 + b\lambda^2 + c\lambda + d
\label{eq:polynomial}
\end{equation} 
are derived from Eq.~(\ref{eq:A}):
\begin{equation}
a = -2 \left( u_1 + u_2 \right), \\
\label{eq:a}
\end{equation}
\begin{equation}
b = u_1^2  - c_1^2 + 4u_1u_2 +  u_2^2 - c_2^2,
\label{eq:b}
\end{equation}
\begin{equation}
c = - 2u_2 \left(u_1^2 - c_1^2 \right) - 2u_1 \left( u_2^2 - c_2^2 \right) ,
\label{eq:c}
\end{equation}
\begin{equation}
d = \left( u_1^2 - c_1^2 \right) \left( u_2^2 - c_2^2 \right) - rc_1^2c_2^2 .
\label{eq:d}
\end{equation}
Substituting coefficients $a$, $b$, $c$, and $d$, Eq.~(\ref{eq:polynomial}) can be written in the form
\begin{equation}
p(\lambda) = \left( \lambda^2 - 2u_1 \lambda + u_1^2 - c_1^2 \right) \left( \lambda^2 - 2u_2 \lambda + u_2^2 - c_2^2 \right) - r c_1^2c_2^2,
\label{eq:charpoly}
\end{equation}
where four roots $\lambda_{k}$, $k=1,..,4$, of $p(\lambda)$ are the eigenvalues of matrix $\boldsymbol{\mathcal{A}(\mathbf{w})}$.

The eigenvalues define the propagation speeds of barotropic (external) and baroclinic (internal) perturbations.
External eigenvalues $\lambda_{ext}^{\pm}$ are always real \citep{castro2001q}; however, at sufficiently large relative velocities $\Delta u = \lvert u_1 - u_2 \rvert$, the internal eigenvalues $\lambda_{int}^{\pm}$ may become complex and the governing system may lose its hyperbolic character \citep{castro2011numerical}. 

Since explicit eigenvalues of a two-layer system are considered too complicated and unavailable \citep{pelanti2008roe,kim2008two,abgrall2009two,fernandez2011intermediate,fjordholm2012energy,sarno2017some}, the following approximations derived under the assumption of $r\approx 1$ and $u_1\approx u_2$ are usually used for internal and external eigenvalues \citep{schijf1953theoretical}:
\begin{equation}
\lambda_{ext}^{\pm} = U_1 \pm \sqrt{g(h_1+h_2)}
\label{eq:eig_ext}
\end{equation}
\begin{equation}
\lambda_{int}^{\pm} = U_2 \pm \sqrt{g(1-r) \frac{h_1 h_2}{h_1 + h_2} \left[ 1 - \frac{(u_1 - u_2)^2}{g(1-r)(h_1 + h_2)}\right]},
\label{eq:eig_int}
\end{equation}
with
\begin{equation}
U_1 = \frac{h_1 u_1 + h_2 u_2}{h_1 + h_2} \quad \textrm{and} \quad
U_2 = \frac{h_1 u_2 + h_2 u_1}{h_1 + h_2}.
\end{equation}

From Eq.~(\ref{eq:eig_int}) it follows that internal eigenvalues become complex for
\begin{equation}
\frac{(u_1 - u_2)^2}{g(1-r)(h_1 + h_2)} > 1.
\label{eq:condition}
\end{equation}
Note that Eq.~(\ref{eq:condition}) is valid only when dealing with two layers of similar densities ($r = \rho_1/\rho_2 \approx 1$) and when velocities in both layers are comparable ($u_1\approx u_2$). These conditions are found in some stratified flows in nature, such as exchange flows through sea straits \citep{castro2004numerical,chakir2009roe} or some cases of highly stratified estuaries \cite{krvavica2016salt}. For a general application, however, this condition may not be necessary to ensure hyperbolicity, as demonstrated by \cite{sarno2017some}.

\subsection{Numerical scheme}

A family of Roe schemes is considered here, which represent a particular case of path-conservative numerical schemes based on the finite volume method. Path-conservative schemes are used to approximate general conservation systems with non-conservative terms \citep{pares2006numerical}. A first order accurate path-conservative scheme for Eq.~(\ref{eq:consys2}) without the bathimetry source term is written as follows \citep{pares2006numerical}:
\begin{equation}
\bb{w}_i^{n+1} = \bb{w}_i^n - \frac{\Delta t}{\Delta x} \left(\bb{D}_{i-1/2}^{+} + \bb{D}_{i+1/2}^{-} \right)
\label{eq:pathcons}
\end{equation}
where $\Delta x$ and $\Delta t$ are the respective spatial and time increment (considered constant here for simplicity), $\bb{w}_i^n$ denotes the approximate cell-averaged values of the exact solution obtained by the numerical scheme at cell $I_i = [x_{i-1/2}, x_{i+1/2}]$ in time $t^n = n\Delta t$, and matrices $\bb{D}_{i+1/2}^{\pm}$ are intermediate functions defined at the cell interface $x_{i+1/2}$:
\begin{equation}
\bb{D}_{i+1/2}^{\pm} = \boldsymbol{\mathcal{A}}_{i+1/2}^{\pm}(\bb{w}_{i}^n,\bb{w}_{i+1}^n) \cdot (\bb{w}_{i+1}^n - \bb{w}_{i}^n)
\label{eq:path_cons}
\end{equation}
with $\boldsymbol{\mathcal{A}}_{i+1/2}^{\pm}$ defined by a decomposition of the Roe linearisation of the form \citep{pares2006numerical}:
\begin{equation}
\boldsymbol{\mathcal{A}}_{i+1/2}^{\pm} = \frac{1}{2} \left( 
\boldsymbol{\mathcal{A}}_{i+1/2} \pm \boldsymbol{Q}_{i+1/2} \right)
\label{eq:roelin}
\end{equation}
where
\begin{equation}
	\boldsymbol{\mathcal{A}}_{i+1/2} = \boldsymbol{\mathcal{A}}_{i+1/2}^{+} + \boldsymbol{\mathcal{A}}_{i+1/2}^{-}
\end{equation}
and $\boldsymbol{Q}_{i+1/2}$ represents a numerical viscosity matrix, whose choice depends on a particular numerical scheme. 

For a two-layer system defined by Eq.~(\ref{eq:consys1}), Roe linearisation is performed at the cell interfaces $x_{i+1/2}$ between cells $I_i$ and $I_{i+1}$ as follows \citep{castro2001q}:
\begin{equation}
\bb{w}_{i+1/2} = 
\begin{Bmatrix}
h_{1,i+1/2} & q_{1,i+1/2} & h_{2,i+1/2} & q_{2,i+1/2}
\end{Bmatrix}^\text{T},
\label{eq:wi+1/2}
\end{equation}
where
\begin{equation}
h_{j,i+1/2} = \frac{h_{j,i} + h_{j,i+1}}{2}, j=1..2,
\end{equation}
\begin{equation}
u_{j,i+1/2} = \frac{u_{j,i}\sqrt{h_{j,i}} + u_{j,i+1}\sqrt{h_{j,i+1}}}{\sqrt{h_{j,i}} + \sqrt{h_{j,i+1}}}, j=1..2,
\end{equation}
\begin{equation}
q_{j,i+1/2} = h_{j,i+1/2}u_{j,i+1/2}, j=1,2
\end{equation}
and also
\begin{equation}
\boldsymbol{\mathcal{A}}_{i+1/2} = \bb{J}_{i+1/2} - \bb{B}_{i+1/2},
\label{eq:Ai+1/2}
\end{equation}
where matrices $\bb{J}_{i+1/2}$ and $\bb{B}_{i+1/2}$ correspond to $\bb{J}(\bb{w}_{i+1/2})$ and $\bb{B}(\bb{w}_{i+1/2})$, respectively. The viscosity matrix in Roe methods corresponds to \citep{castro2001q}:
\begin{equation}
\boldsymbol{Q}_{i+1/2} = \lvert \boldsymbol{\mathcal{A}}_{i+1/2} \rvert
\label{eq:Q_Roe}
\end{equation}
with
\begin{equation}
\lvert \boldsymbol{\mathcal{A}}_{i+1/2} \rvert = 
\bb{K}_{i+1/2}
\lvert \boldsymbol{\Lambda}_{i+1/2} \rvert
\bb{K}_{i+1/2}^{-1}.
\label{eq:A_abs}
\end{equation}
where $\lvert \boldsymbol{\Lambda}_{i+1/2} \rvert$ is a $N\times N$ diagonal matrix whose coefficient are the absolute eigenvalues $\lvert \lambda_k \rvert, k=1,..,N$, $\bb{K}_{i+1/2}$ is the same-size matrix whose columns are right eigenvectors corresponding to those eigenvalues and $\bb{K}^{-1}_{i+1/2}$ is the inverse of $\bb{K}_{i+1/2}$.
To achieve good well-balanced properties, the source terms are upwinded using projection matrices \citep{castro2001q}:
	\begin{equation}
	\bb{P}_{i+1/2}^{\pm} = \frac{1}{2} \bb{K}_{i+1/2}
	\left( \bb{Id} \pm \textrm{sign}( \boldsymbol{\Lambda}_{i+1/2}) \right)
	\bb{K}_{i+1/2}^{-1}.
	\end{equation}
	where $\textrm{sign}( \boldsymbol{\Lambda}_{i+1/2})$ is a $N\times N$ diagonal matrix whose coefficient are $\textrm{sign}( \lambda_k ), k=1,..,N$.

To finally solve a coupled two-layer system, the Roe scheme is written in the following form \citep{castro2001q}:
\begin{equation}
\begin{aligned}
	\bb{w}_i^{n+1} = \bb{w}_i^n &  -\frac{\Delta t}{\Delta x} \left( \bb{f}_{i-1/2} - \bb{f}_{i+1/2} \right) \\
					& + \frac{\Delta t}{2 \Delta x} \left[ \bb{B}_{i-1/2} 
						\left( \bb{w}_i^n - \bb{w}_{i-1}^n \right) 
						+ \bb{B}_{i+1/2} 
						\left( \bb{w}_{i+1}^n - \bb{w}_{i}^n \right) \right]			 \\
					& + \frac{\Delta t}{\Delta x} \left( \bb{P}^{+}_{i-1/2} \bb{g}_{i-1/2}
						+  \bb{P}^{-}_{i+1/2} \bb{g}_{i+1/2}\right)	,					
\end{aligned}
\label{eq:Q_scheme}
\end{equation}
with the numerical flux
\begin{equation}
	\bb{f}_{i+1/2} = \frac{1}{2} \left( \bb{f}_{i}^n + \bb{f}_{i+1}^n \right) - \frac{1}{2} \left| \boldsymbol{\mathcal{A}}_{i+1/2} \right| \left( \bb{w}_{i+1}^n - \bb{w}_i^n \right).
\label{eq:F_num}
\end{equation}

To prevent the numerical viscosity of the Roe scheme from vanishing when any of the eigenvalues of the matrix $\lvert \boldsymbol{\mathcal{A}}_{i+1/2} \rvert$ are zero, the Harten regularization (entropy fix) is applied \citep{castro2001q}. 
Numerical difficulties may also appear in Roe scheme when one of the layers vanish and when wet-dry fronts develop at the interface. The former issue is resolved by setting a wet-dry parameter ($\varepsilon$), so that when the depth of one of the layers in a cell is lower than $\varepsilon$, the cell is considered as a one-layer system and a corresponding two-equation PDE system \citep{bermudez1994upwind} is solved instead of Eq.~(\ref{eq:consys1}). The well-balanced property of the numerical scheme in the presence of wet-dry fronts is achieved by a source term modification for the two-layer system introduced by \cite{castro2005numerical}.

Note that Eq.~(\ref{eq:roelin}) can also be applied to other numerical schemes from the family of path-conserving schemes, such as Lax-Friedrichs (LF) \citep{toro2013riemann}, where
\begin{equation}
	\boldsymbol{Q}_{i+1/2} = \frac{\Delta x}{\Delta t} \bb{Id},
\end{equation}
or FORCE and GFORCE schemes \citep{toro2013riemann}, where 
\begin{equation}
	\boldsymbol{Q}_{i+1/2} = (1- \omega)\frac{\Delta x}{\Delta t} \bb{Id} + \omega \frac{\Delta t}{\Delta x}  \boldsymbol{\mathcal{A}}_{i+1/2}^2,
\end{equation}
with $\omega = 0.5$ and $\omega = 1/(1+CFL)$, respectively. The $CFL$ number is defined as \citep{castro2010some}:
\begin{equation}
CFL = \frac{\Delta t}{\Delta x} \max (\lambda_k), \quad k=1,..,N.
\end{equation}
where CFL stands for Courant-Friedrichs-Lewy number.

As stated earlier, in comparison to incomplete Riemman solvers, Roe schemes are less diffusive and have good well-balanced properties \citep{castro2010some}. However, Roe schemes require the numerical computation of the whole eigenstructure of matrix $\boldsymbol{\mathcal{A}}_{i+1/2}$, which can be computationally very expensive. A possible alternative to the spectral decomposition required in the Roe scheme is the redefinition of the viscosity matrix $\boldsymbol{Q}_{i+1/2}$ by the Polynomial Viscosity Matrix (PVM), which can be written as \citep{castro2012class}:
\begin{equation}
\boldsymbol{Q}_{i+1/2} = \lvert \boldsymbol{\mathcal{A}}_{i+1/2} \rvert = \sum_{k=0}^{3}\alpha_k \mathcal{A}^{k}_{i+1/2}
\end{equation}
where $\alpha_k$ are the solutions of the following linear system:
\begin{equation}
\begin{bmatrix}
1 & \lambda_1 & \lambda_1^2 & \lambda_1^3 \\
1 & \lambda_2 & \lambda_2^2 & \lambda_2^3 \\
1 & \lambda_3 & \lambda_3^2 & \lambda_3^3 \\
1 & \lambda_4 & \lambda_4^2 & \lambda_3^3 
\end{bmatrix}
\begin{Bmatrix}
\alpha_0 \\ \alpha_1 \\ \alpha_2 \\ \alpha_3
\end{Bmatrix}
=
\begin{Bmatrix}
\lvert \lambda_1 \rvert \\
\lvert \lambda_2 \rvert \\
\lvert \lambda_3 \rvert \\
\lvert \lambda_4 \rvert \\
\end{Bmatrix}
\label{eq:alpha4}
\end{equation}
The eigenvalues are computed by approximate expressions given by Eq.~(\ref{eq:eig_ext}) and (\ref{eq:eig_int}). This scheme will be denoted here as the PVM-Roe scheme.

Since the CPU time needed to compute Eq.~(\ref{eq:alpha4}) is similar to the one required to obtain Eq.~(\ref{eq:A_abs}), a simpler and faster Intermediate Field Capturing Parabola (IFCP) scheme was derived from the family of PVM schemes, given by \cite{fernandez2011intermediate}:
\begin{equation}
\boldsymbol{Q}_{i+1/2} = \alpha_0 \bb{Id} + \alpha_1 \boldsymbol{\mathcal{A}}_{i+1/2} + \alpha_2 \boldsymbol{\mathcal{A}}_{i+1/2}^2,
\end{equation}
where $\alpha_k$ are defined as:
\begin{equation}
\begin{bmatrix}
1 & \lambda_1 & \lambda_1^2  \\
1 & \lambda_2 & \lambda_2^2  \\
1 & \chi_{int} & \chi_{int}^2 \\
\end{bmatrix}
\begin{Bmatrix}
\alpha_0 \\ \alpha_1 \\ \alpha_2 
\end{Bmatrix}
=
\begin{Bmatrix}
\lvert \lambda_1 \rvert \\
\lvert \lambda_2 \rvert \\
\lvert \chi_{int} \rvert
\end{Bmatrix},
\label{eq:alpha3}
\end{equation}
with
\begin{equation}
\chi_{int} = \mathcal{S}_{ext} \max \left( \lvert \lambda_3 \rvert, \lvert \lambda_4 \rvert \right),
\end{equation}
and
\begin{equation}
\mathcal{S}_{ext} =
\begin{cases}
\mathrm{sign} (\lambda_3 + \lambda_4), \quad \mathrm{if} (\lambda_3 + \lambda_4) \neq 0 \\
1, \quad \mathrm{otherwise}
\end{cases}.
\end{equation}
As in the original PVM scheme, the approximate expressions given by Eq.~(\ref{eq:eig_ext}) and (\ref{eq:eig_int}) are used to compute the eigenvalues.  However, in this case, the coefficients $\alpha_k$ can be explicitly defined (see \cite{fernandez2011intermediate}).

\subsection{Definition of the A-Roe numerical scheme}

We propose a new implementation of the Roe scheme named A-Roe. The A-Roe scheme is defined by Eqs.~(\ref{eq:Q_scheme}) and (\ref{eq:F_num}), where the viscosity matrix is given by Eq.~(\ref{eq:A_abs}), but instead of using a numerical solver (denoted here as N-Roe) or approximating the viscosity matrix, the eigenstructure is solved analytically -- eigenvalues are computed by a closed-form solution to the roots of the characteristic quartic polynomial given by Eq.~(\ref{eq:charpoly}), and then the corresponding eigenvectors are easily obtained. The proposed scheme shares the same properties as the Q-scheme of Roe regarding the well-balanced properties and the capability to deal with wet-dry fronts (the same numerical techniques and modifications designed for Roe methods are directly applicable to the A-Roe method proposed here). 

\subsubsection{Eigenvalues and a closed-form quartic solver}

An analytical solution for quartic equations has been derived by Ferrari in the 16th century \citep{abramowitz1972}. This closed-form solution is obtained by the method of radicals and it depends on the solution of a residual cubic equation, which can be solved by the Cardano's method \citep{abramowitz1972}. Although this classical method is the fastest \citep{strobach2015}, it is considered problematic due to cancellation errors for certain combinations of polynomial coefficients \citep{strobach2010,strobach2015,flocke2015}. 

No theoretical analysis of the cancellation error for the closed-form quartic solver has been made so far, but several studies found that the analytical solution produces inferior results for small roots in case of a large root spread, \textit{i.e.}, when one of the roots is several orders of magnitude larger than the others \citep{strobach2010,strobach2015,flocke2015}. For example, \cite{strobach2010} demonstrated that a closed-form quartic solver produced an average error between $10^{-14}$ and $10^{-15}$ for root spreads in range 1 to 1000, but for some individual cases with extreme root spreads in the range of $10^{18}$, the quartic solver produced completely corrupted results. For this reason, Ferrari's analytical solution is considered unreliable and is usually avoided in computational use. 

Although the closed-form quartic solver is unsuitable for general use, its accuracy should be re-evaluated in the context of this study to assess if it could still be considered reliable for computing the eigenstructure of the pseudo-Jacobian matrix of the governing SWE system given by Eq.~(\ref{eq:A}). First of all, high accuracy (error $<10^{-14}$) of the quartic solver is not imperative because: (\textit{i}) there are many viable alternatives to complete Riemann solvers that only approximate the viscosity matrix \citep{castro2012class}, and (\textit{ii}) the traditional approach in developing these models is based on a matrix eigensolver, such as the LAPACK subroutine \textit{dgeev.f} \citep{lapack}, which also shows a similar average error as the closed-form quartic solver (although, it is more reliable for extreme root spreads) \citep{strobach2010}. More importantly, the eigenvalues of the pseudo-Jacobian matrix have a physical meaning - they represent the propagation speeds of the internal and external gravity waves. Considering that the propagation speeds of these waves depend mainly on the flow velocity and water depth \citep{schijf1953theoretical}, extreme eigenvalue spreads should not be expected since they are not physically possible in real geophysical flows.

Ferrari's method for solving quartic equations \citep{abramowitz1972} is given by a series of simple algebraic equations involving one root of a cubic equation (see Appendix \ref{sec:A1}). Although it is possible to combine these equations into a single explicit expression, it is too extensive to be presented in a journal format, and certainly not optimized to be implemented in a computational algorithm. To our knowledge, such formulation is available only on \cite{wikipedia}. 
Therefore, in this study, we present a simple closed-form approach for finding real roots of the quartic equation (\ref{eq:polynomial}) consisting of eight simple algebraic evaluations. A detailed derivation of these equations is given in \ref{sec:A1}. 

Given the coefficients $a, b, c$ and $d$ of the characteristic 4th order polynomial, defined by Eqs.~(\ref{eq:a}) - (\ref{eq:d}), the real eigenvalues are computed by the following expressions:
	\begin{equation}
	\lambda_{1,2} = \lambda_{ext}^{\pm} = \frac{ - \frac{a}{2} \pm \sqrt{Z} - \sqrt{- A - Z \mp \frac{B}{\sqrt{Z}} } }{2} ,
	\label{eq:solution1}
	\end{equation}
	\begin{equation}
	\lambda_{3,4} = \lambda_{int}^{\pm} = \frac{- \frac{a}{2} \pm \sqrt{Z} + \sqrt{- A - Z \mp \frac{B}{\sqrt{Z}} }}{2} .
	\label{eq:solution2}
	\end{equation}
where
\begin{equation}
	Z =  \frac{1}{3} \left( 2 \sqrt{\Delta_0} \cos \frac{\phi}{3}  - A \right),
	\label{eq:Zcoeff}
\end{equation}
\begin{equation}
	\phi = \arccos \left( \frac{\Delta_1}{2 \sqrt{\Delta_0^3}}\right),
	\label{eq:S_cubic}
\end{equation}
with
\begin{equation}
A = 2b - \frac{3a^2}{4},
\label{eq:Acoeff}
\end{equation}
\begin{equation}
B = 2c - ab + \frac{a^3}{4} .
\label{eq:Bcoeff}
\end{equation}
and
\begin{equation}
\Delta_0 = b^2 + 12d - 3ac, 
\label{eq:D0}
\end{equation}
\begin{equation}
\Delta_1 = 27a^2d - 9abc + 2b^3 - 72bd + 27c^2.
\label{eq:D1}
\end{equation}

\subsubsection{Eigenvectors}

The $4 \times 4$ matrix $\bb{K}$  whose columns are right eigenvectors $\bb{k}_k$ corresponding to eigenvalues $\lambda_k, k=1,..,4$ are found by solving the following equation:
\begin{equation}
\left( \boldsymbol{\mathcal{A}} - \lambda \bb{Id} \right) \bb{K} = 0
\label{eq:eigenvector}
\end{equation}
Since $\boldsymbol{\mathcal{A}} - \lambda \bb{Id}$ is singular there are infinite solutions to Eq.~(\ref{eq:eigenvector}), \textit{i.e.}, for an assumed value for one component of the eigenvector, the remaining components are easily computed. For example, if we assume $\bb{k}_{k,[1]} = 1$, the remaining eigenvector components are obtained from Eq.~(\ref{eq:eigenvector}) as:
\begin{equation}
\bb{k}_k = 
\begin{Bmatrix}
1 &  \lambda_k & \mu_k & \lambda_k\mu_k
\end{Bmatrix}^{T},
\end{equation}
where
\begin{equation}
%\mu_k = - \frac{ \lambda_k^2 - 2u_1 \lambda_k - (c_1^2 - u_1^2)}{c_1^2}
\mu_k = 1 - \frac{ \left(\lambda_k - u_1 \right)^2}{c_1^2}	
\end{equation}
and
\begin{equation}
\bb{K} =
\begin{bmatrix}
	\bb{k}_1 & \bb{k}_2 & \bb{k}_3 & \bb{k}_4
\end{bmatrix}.
\label{eq:Keigenvectors}
\end{equation}

Note that the associated eigenvectors can alternatively be derived as proposed by \cite{rosatti2008generalized} or \cite{murillo2010exner} for a cubic characteristic polynomial.

\subsubsection{The numerical viscosity matrix}

Once the eigenstructure has been computed, the viscosity matrix $\vert \boldsymbol{\mathcal{A}} \vert$ can be obtained from Eq.~(\ref{eq:A_abs}) as:
\begin{equation}
	\vert \boldsymbol{\mathcal{A}} \vert =  \bb{K} \vert \boldsymbol{\Lambda} \vert  \bb{K}^{-1}
\label{eq:Aabs}
\end{equation}
where 
\begin{equation}
 \vert \boldsymbol{\Lambda} \vert = 
 \begin{bmatrix}
 \lvert \lambda_1 \rvert & & 0 \\
  & \ddots  &  \\
  0 & &   \lvert \lambda_4 \rvert \\
 \end{bmatrix}
\end{equation}

To avoid computationally expensive numerical calculation of the inverse matrix, $\bb{K}^{-1}$ can be obtained from:
\begin{equation}
\bb{K}^{-1} = \frac{1}{\textrm{det}(\bb{K})}\textrm{adj}(\bb{K}).
\end{equation}
Full explicit expressions for $\bb{K}^{-1}$ are given in \ref{sec:A2}.
However, we found that it is computationally less demanding to rewrite Eq.~(\ref{eq:Aabs}) as
\begin{equation}
\bb{K}^T \vert \boldsymbol{\mathcal{A}} \vert^T  = (\bb{K} \vert \boldsymbol{\Lambda} \vert)^T ,
\label{eq:Aabs_T}
\end{equation} 
which corresponds to a general matrix equation $\bb{Ax}=\bb{B}$, solve it numerically for $\bb{x}$ (for example, by a LAPACK routine \textit{gesv} \citep{lapack}), and then transpose it.

\subsection{Numerical treatment for the loss of hyperbolicity}

Since the proposed A-Roe scheme is valid only for real eigenvalues, an appropriate numerical treatment is required in the case of hyperbolicity loss when complex eigenvalues appear.
In the past, the problem of the hyperbolicity loss has been bypassed by applying a real Jordan decomposition of the pseudo-Jacobian matrix; however, such numerical workaround may still cause un-physical oscillations or unrealistic results \citep{castro2011numerical}. Introducing the third intermediate layer seemed promising and physically justified, however, it proved to be only partially successful \citep{castro2012hyperbolicity}. 

Recently, several more physically realistic treatments have been proposed. \cite{castro2011numerical} have introduced a simple numerical algorithm, which adds an extra amount of friction at every cell where complex values are detected. The amount of friction is computed at each cell to satisfy the approximate hyperbolic condition given by Eq.~(\ref{eq:condition}). This approach is physically justified because the friction term may be seen as an approximation of an additional momentum flux which appears locally due to turbulent mixing processes. In real flows, loss of hyperbolicity corresponds to strong shear stress and the development of interfacial instabilities, such as Kelvin-Helmholtz waves \citep{castro2011numerical,sarno2017some}. Once the instabilities appear, turbulent mixing initiates vertical mass and momentum transfer, and an intermediate layer of a finite thickness develops. \cite{krvavica2018relevance} also showed that adding physically realistic friction and entrainment terms may prevent the loss of hyperbolicity in some situations.

\cite{sarno2017some} improved this idea by computing the discriminant $\mathcal{D}$ of the characteristic polynomial given by Eq.~(\ref{eq:charpoly}). When $\mathcal{D}>0$, roots of the characteristic polynomial, \textit{i.e.}, eigenvalues, are either all real or all complex. Since two (external) eigenvalues are always real, the remaining two (internal) eigenvalues can only be real if $\mathcal{D}>0$.
However, \cite{sarno2017some} computed $\mathcal{D}$ from a formula for a discriminant of a general polynomial $p(x)$ of a degree $n$, as a function of its coefficients $a_n$, given by:
\begin{equation}
\mathcal{D}(p) = (-1)^{n(n-1)/2} \frac{1}{a_n}\textrm{det}(\bb{R}(p,p'))
\label{eq:discriminant}
\end{equation}
where $p'$ is derivative of polynomial $p$, and $\bb{R}(p,p')$ is the Sylvester matrix of $p$ and $p'$ \citep{sarno2017some}. For a quartic equation, this formula yields a rather long expression (for details see \citep{sarno2017some}).

In this work, a similar approach to \cite{sarno2017some} is proposed; however, the choice of the discriminant and the implementation of the hyperbolicity correction differs.
First, the hyperbolicity condition is based on the discriminant of the resolvent cubic equation $\mathcal{D}_{cubic}$ (see \ref{sec:A1}) given by
\begin{equation}
	\Delta = \frac{27}{64} \mathcal{D}_{cubic} = 4\Delta_0^3 - \Delta_1^2  > 0 
	\label{eq:the_condition}
\end{equation}
It is easy to verify that $\Delta = \frac{27}{64} \mathcal{D}_{cubic} = 27 \mathcal{D}_{quartic}$; however, $\Delta$ is more compact and therefore less computationally demanding than $\mathcal{D}_{quartic}$ given by Eq.~(\ref{eq:discriminant}).

Furthermore, to take advantage of the fact that A-Roe method solves $\Delta_0$ and $\Delta_1$ when computing the linearised values at every intercell, the hyperbolicity verification and correction is performed directly at this stage. The optimal correction is then only added as an extra friction source term when computing the values at the next time step. This implementation requires almost no extra computational time for verifying the hyperbolicity. Additional computation is required only when correcting the momentum term if hyperbolicity loss is detected at a specific intercell at some time step. 

The proposed implementation is described as follows:

\begin{enumerate}
	
	\item Once the solutions $\bb{w}_i^n$ are known at each cell $I_i$ at time $t^n$, the first part of the Roe linearisation is computed by Eqs.~(\ref{eq:wi+1/2}) - (\ref{eq:Ai+1/2}) to get conserved values $\bb{w}_{i+1/2}$ at cell interfaces $I_{i+1/2}$ and compute linearised pseudo-Jacobian matrix $\boldsymbol{\mathcal{A}}_{i+1/2}$
	
	\item Coefficients of the characteristic polynomial are then computed for conserved values $\bb{w}_{i+1/2}$ at cell interfaces by Eqs.~(\ref{eq:a})-(\ref{eq:d})
	
	\item At every cell interface, the first step of the explicit quartic solver is computed by Eqs.~(\ref{eq:D0}) and (\ref{eq:D1}) to get $\Delta_0$ and $\Delta_1$
	
	\item The discriminant of the resolvent cubic equation $\Delta$ is computed using Eq.~(\ref{eq:the_condition}) and the hyperbolicity condition is verified at each cell interface:
	\begin{itemize}
		\item If $\Delta > 0$, the quartic solver continues computing Eq.~(\ref{eq:solution1}) - (\ref{eq:solution2}) to obtained the eigenvalues. The eigenvector matrix is constructed using Eq.~(\ref{eq:Keigenvectors}), and finally the viscosity matrix is computed by Eq.~(\ref{eq:Aabs}) (fully analytical) or Eq.~(\ref{eq:Aabs_T}) (semi-analytical, but faster)
		
		\item If $\Delta \leq 0$, the linearised velocities at those interfaces (computed at step 1) are corrected by an optimal friction term:
		\begin{equation}
		\begin{aligned}
		(u_{1,i+1/2}^{n})^{corr} = u_{1,i+1/2}^{n} 
		+ \Delta t F_{corr} \frac{\textrm{sign} \left( u_{2,i+1/2}^{n} - u_{1,i+1/2}^{n} \right)}{h_{1,i+1/2}^{n}}  \\
		(u_{2,i+1/2}^{n})^{corr} = u_{2,i+1/2}^{n} 
		- \Delta t r F_{corr} \frac{\textrm{sign} \left( u_{2,i+1/2}^{n} - u_{1,i+1/2}^{n} \right)}{h_{2,i+1/2}^{n}}  \\
		\label{eq:u_corr}
		\end{aligned}
		\end{equation}
		where $F_{corr}$ is a minimum value that satisfies the condition given by Eq.~(\ref{eq:the_condition}). \cite{sarno2017some} examined several iterative methods and found that the fastest algorithm for this kind of problems is the Illinois method \citep{dowell1971modified}, which is implemented here as follows. First, an interval is chosen so that $F_{corr} \in \left[a_0, b_0\right]$, where $a_0 = 0$ (no correction) and
		\begin{equation}
		b_0 = \frac{\vert u_{2,i+1/2} - u_{1, i+1/2}\vert}{\Delta t \left(\frac{1}{h_{1,i+1/2}} + \frac{r}{h_{2,i+1/2}}\right)}
		\end{equation}
		which yields a hyperbolic state with $u_{1,i+1/2} - u_{2,i+1/2} = 0$. The next guess for $F_{corr,p}$ in the $p$-th iteration is calculated through
		\begin{equation}
		F_{corr,p} = b_p - \frac{f(b_p)(b_p - a_p)}{f(b_p) - f(a_p)},
		\end{equation}
		where $f(b_p)=\Delta(b_p)$ and $f(a_p)=\Delta(a_p)$ are the discriminants corresponding to velocities $(u_{1, i+1/2})^{corr}$ and $(u_{2, i+1/2})^{corr}$, respectively, corrected by $F_{corr,p}$ through Eq.~(\ref{eq:u_corr}). At the next iteration step, the interval pairs are chosen as follows:
		\begin{equation}
		\begin{aligned}
		\left(F_{corr,p}, \Delta(F_{corr,p})\right), \left(b_p, \Delta(b_p)\right) &\quad
		\textrm{if } \Delta(b_p)\Delta(F_{corr,p})<0 \\
		\left(a_p, \Delta(a_p)/2\right), \left(F_{corr,p}, \Delta(F_{corr,p})\right) &\quad
		\textrm{else}.
		\end{aligned}
		\end{equation}
		The algorithm iterates until the condition $\vert a_p - b_p \vert \leq \epsilon$ is satisfied (where $\epsilon$ is a convergence threshold), and the final solution is given by:
		\begin{equation}
		F_{corr,p} = \max\left(a_p, b_p\right).
		\label{eq:iter_solution}
		\end{equation}
		Since it always holds that $\Delta(a_p)\Delta(b_p)<0$, Eq. (\ref{eq:iter_solution}) and appropriate $\epsilon$ guarantee that the discriminant is always positive and larger than zero $\Delta(F_{corr,p}) > 0$, which prevents possible problems with singular eigenvector matrix due to double roots when $\Delta = 0$.
		
		After the correction is performed, the analytic solver continues to compute the eigenstructure for the viscosity matrix through Eqs.~(\ref{eq:solution1}), (\ref{eq:solution2}), (\ref{eq:Keigenvectors}), and (\ref{eq:Aabs}) or (\ref{eq:Aabs_T}).
		
	\end{itemize}		
	
	\item Finally, the conserved values are computed for the next time step using, for example, the Q-scheme of Roe, where the friction source term $F_{corr}$ is added as an extra source term describing the vertical momentum transfer between the layers:
	\begin{equation}
	\begin{aligned}
	\bb{w}_i^{n+1} = \bb{w}_i^n &  -\frac{\Delta t}{\Delta x} \left( \bb{f}_{i-1/2} - \bb{f}_{i+1/2} \right) \\
	& + \frac{\Delta t}{2 \Delta x} \left[ \bb{B}_{i-1/2} 
	\left( \bb{w}_i^n - \bb{w}_{i-1}^n \right) 
	+ \bb{B}_{i+1/2} 
	\left( \bb{w}_{i+1}^n - \bb{w}_{i}^n \right) \right]			 \\
	& + \frac{\Delta t}{\Delta x} \left( \bb{P}^{+}_{i-1/2} \bb{g}_{i-1/2}
	+  \bb{P}^{-}_{i+1/2} \bb{g}_{i+1/2}\right)		\\
	& + {\Delta t} \left( \bb{P}^{+}_{i-1/2} \bb{s}_{f,i-1/2}
	+  \bb{P}^{-}_{i+1/2} \bb{s}_{f,i+1/2} \right)				
	\end{aligned}
	\label{eq:num_scheme_fric}
	\end{equation}
	where $\bb{s}_{f,i+1/2}$ is the friction source term, defined as:
	\begin{equation}
	\bb{s}_{f,i+1/2} = 
	\begin{Bmatrix}
	0 \\
	F_{corr} {\textrm{sign} \left( u_{2,i+1/2}^{n} - u_{1,i+1/2}^{n} \right)}\\
	0 \\
	-rF_{corr} {\textrm{sign} \left( u_{2,i+1/2}^{n} - u_{1,i+1/2}^{n} \right)} \\
	\end{Bmatrix}.
	\end{equation}
	The friction source term is introduced to account for the momentum exchange occurring as a result of the hyperbolicity loss (turbulent mixing in real flows). Practically, it decreases the velocity difference between the layers at the cell adjacent to the interface where hyperbolicity loss was detected, and hence prevents a  transfer of the hyperbolicity loss conditions to the next time step.
	
\end{enumerate}

\section{Results}

To evaluate the performance of the proposed A-Roe scheme several numerical tests are presented. First, the accuracy and computational speed of the closed-form quartic solver are analysed. Next, several numerical results are given to analyse the performance of the implemented algorithm in computing a two-layer flow, with a special focus on the computational speed and accuracy of the hyperbolicity correction algorithm.

All numerical algorithms have been implemented in Python 3.6, using the Numpy package. The tests have been performed on 64-bit Windows 10 machine with Intel Core i7-3770 3.4 GHz processor.

\subsection{Computational accuracy and speed of the closed-form quartic solver}

This subsection examines the performance and reliability of the analytical approach to eigenstructure of the governing system. The accuracy and computational speed of the proposed closed-form quartic solver are analysed for one million root combinations. 

\begin{figure}[htbp]
	\center
	\includegraphics[width=6.5cm]{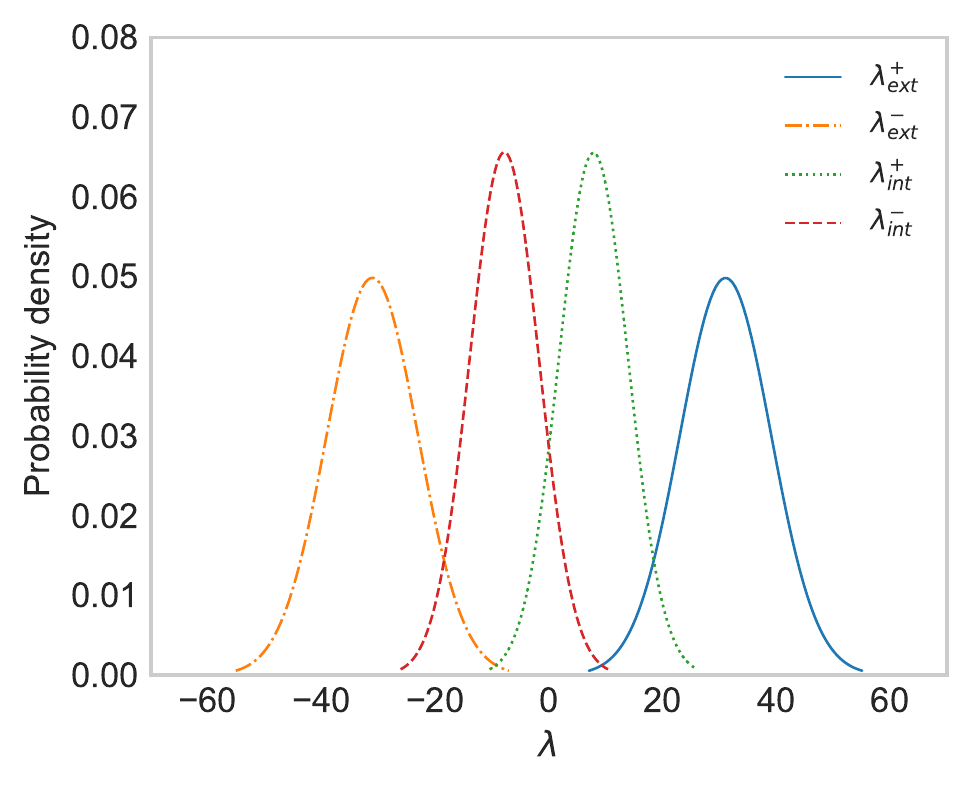}
	\caption{Probability distributions of four real roots representing the eigenvalues of the two-layer SWE system}
	\label{fig:distributions}
\end{figure}

Since the main idea is to apply this quartic solver to the pseudo-Jacobian matrix of the two-layer SWE system, physically realistic roots are examined. Therefore, a large set of flow parameters, namely layer depths $0 < h_{1,2} < 100$ m and velocities $-20 < u_{1,2} < 20$ m s$^{-1}$, as well as density ratios $0.1< r <1$, have been randomly generated from a uniform distribution. Based on these parameters, approximate roots have been calculated by Eq.~(\ref{eq:eig_ext}) and (\ref{eq:eig_int}). Only the solutions with all real roots are then selected and statistically analysed to obtain a corresponding probability distribution for each eigenvalue (Fig.~\ref{fig:distributions}).

Next, one million set of test roots $\lambda_{1,2,3,4}$ are randomly generated as statistically independent samples of each probability distribution presented in Fig.~\ref{fig:distributions}. The coefficients of the characteristic quartic Eq.~(\ref{eq:polynomial}) are then computed according to the following expressions \cite{strobach2010}:
\begin{align}
a &= - (\lambda_1 + \lambda_2 + \lambda_3 + \lambda_4) \\
b &= \lambda_1 \lambda_2 + (\lambda_1 + \lambda_2)(\lambda_3 + \lambda_4) + \lambda_3 \lambda_ 4 \\
c &= - \lambda_1 \lambda_2 (\lambda_3 + \lambda_4) - \lambda_3 \lambda_4 (\lambda_1 + \lambda_2)\\
d &= \lambda_1 \lambda_2 \lambda_3 \lambda_4.
\label{eq:char_coeff}
\end{align}
The closed-form quartic solver (AnalyticQS) given by Eqs.~(\ref{eq:solution1}) and (\ref{eq:solution2}) is then applied to re-compute the roots of the quartic equation defined by coefficients $a, b, c$ and $d$.

For a comparison, the roots of this quartic are also computed by a numerical eigenstructure solver (NumericQS). In this case, the \textit{eig} function from the \textit{numpy.linalg} package has been applied to a companion matrix derived from the same coefficients. Note that the \textit{eig} function directly calls the LAPACK subroutine \textit{dgeev.f} written in Fortran 90 \citep{lapack}.

The errors in both computations are estimated using an absolute error measure:
\begin{equation}
E_k = \lvert \lambda^{ref}_k - \lambda_k \rvert, \quad \textrm{for} \quad k = 1,..,4,
\label{eq:errors}
\end{equation}
where $\lambda^{ref}_k$ is the test root and $\lambda_k$ is the root computed by a specific algorithm. 

\begin{figure}[htbp]
	\center
	\includegraphics[width=6.5cm]{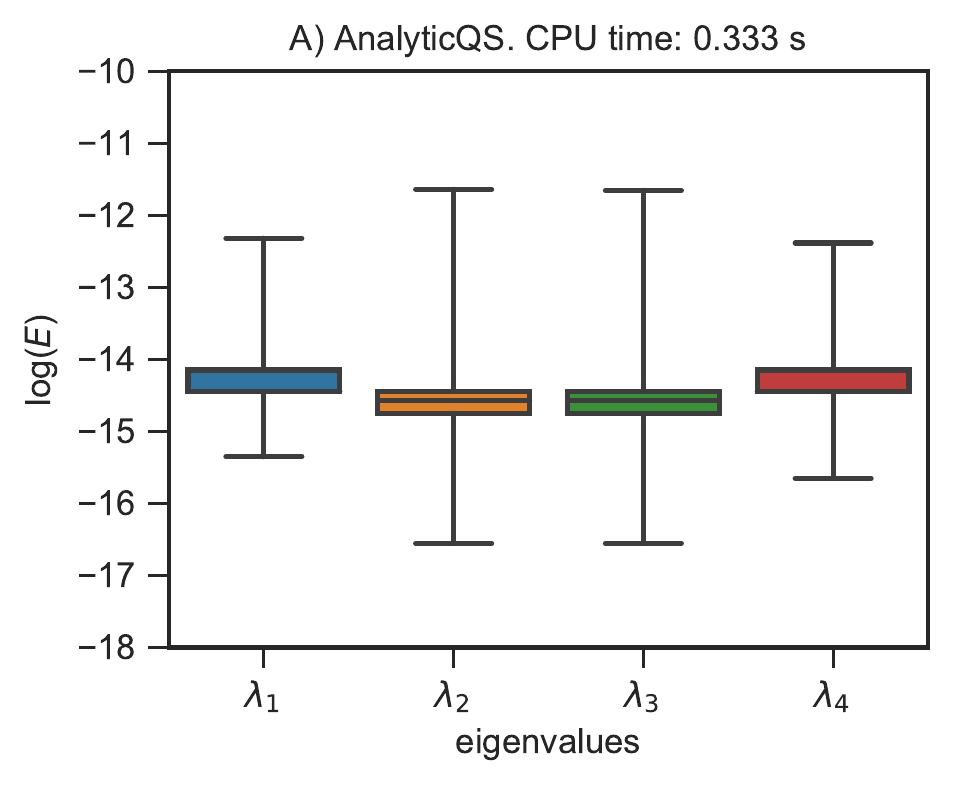}
	\hfill	
	\includegraphics[width=6.5cm]{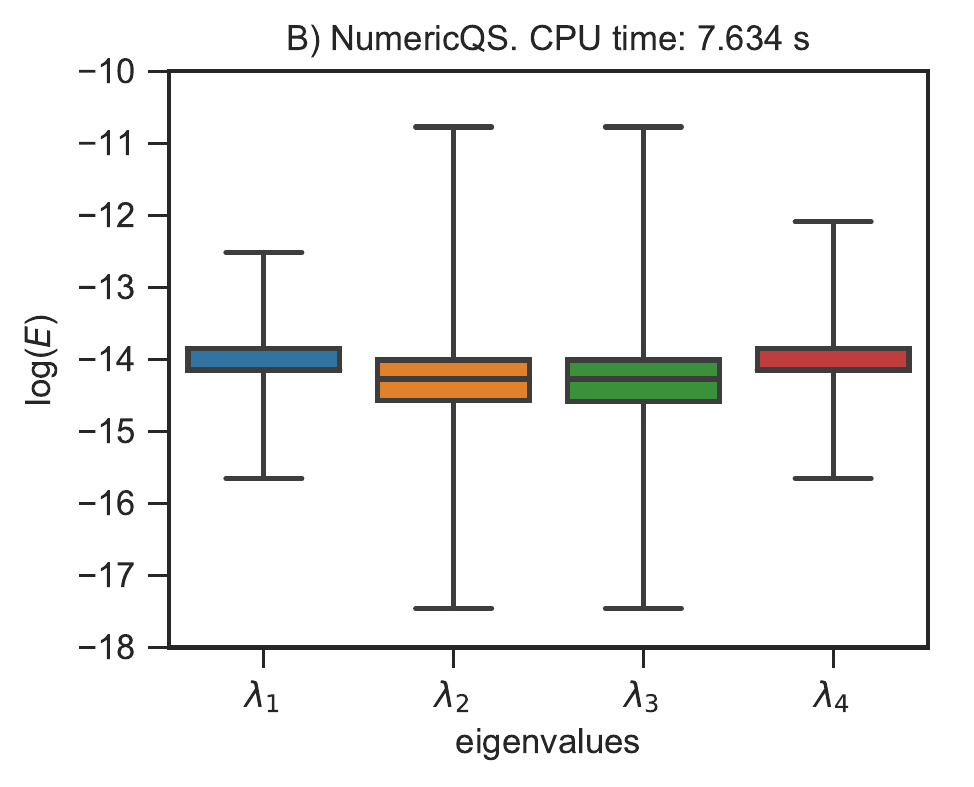}
	\caption{Boxplot of the errors in computing the roots by: A) the proposed closed-form quartic solver (AnalyticQS) and B) numerical eigenstructure solver (NumericQS). Boxes denote the interquartile range and median value, while whiskers denote min and max values.}
	\label{fig:errors}
\end{figure}

Figure \ref{fig:errors} illustrates the statistical representation of the absolute errors computed by Eq.~(\ref{eq:errors}) for $N=10^6$ independent root samples obtained by AnalyticQS and by NumericQS.
The root spread is computed as the ratio of the largest to the smallest root:
\begin{equation}
RS_j = \frac{\max (\lvert \lambda_j \rvert)}{\min (\lvert \lambda_j \rvert)}, \quad \textrm{for} \quad j = 1,..,N.
\end{equation}
where $N$ is the number of samples in the set of independent roots (one million). The spread of computed roots ranges from 1 to $10^7$.

From Fig.~\ref{fig:errors} we observe that the average error for AnalyticQS lies between $10^{-14} < E_k < 10^{-15}$, and that the maximum errors are always below $10^{-11}$. Both the maximum and the average errors are lower in the proposed analytic method (AnalyticQS) than in the NumericQS. Furthermore, AnalyticQS has produced 19.1\% of perfect results ($E_k=0$) over one million trials, while NumericQs has produced 6\% of such results (these were excluded from the set presented by a boxplot in Fig.~\ref{fig:errors}).

More importantly, not only is AnalyticQS more accurate than NumericQS, but it is significantly faster. Best of five runs revealed that AnalyticQS takes 0.333 s and NumericQS 7.634 s of computational time to solve one million quartic equations, which represents more than one order of magnitude improvement. \cite{strobach2010} found similar errors and computational speed-ups ($13 \times$) when comparing these two approaches for randomly generated real roots with $RS<10^5$.

\subsection{Test I: The internal dam-break}

In the following two tests, the efficiency of the proposed A-Roe scheme is evaluated by comparing its accuracy and CPU times against Lax-Friedrichs (LF), GFORCE, PVM-Roe, IFCP, and the N-Roe scheme. Both A-Roe and N-Roe schemes correspond to the generalized Q-scheme of Roe with upwinded source terms and Harten's entropy fix. The only difference between them is the implementation of the eigenstructure solver; N-Roe scheme uses numerical solver (NumericQS), whereas the A-Roe scheme uses the proposed analytical closed-form solver (AnalyticQS).

A two-layer flow through a rectangular channel with flat bottom topography is considered. This test was introduced by \cite{fernandez2011intermediate} to evaluate the accuracy of numerical schemes in simulating an internal dam-break problem over a flat bottom topography $b(x)=0$ m. The spatial domain is set to [0, 10], and the initial condition is given by:
\begin{equation}
h_1(x,0) = 
\begin{cases}
0.2 \textrm{ m}, \quad \textrm{if } x < 5 \textrm{ m} \\
0.8 \textrm{ m}, \quad \textrm{otherwise}
\end{cases}
\quad
h_2(x,0) = 
\begin{cases}
0.8 \textrm{ m}, \quad \textrm{if } x < 5 \textrm{ m} \\
0.2 \textrm{ m}, \quad \textrm{otherwise}
\end{cases}
\end{equation}
\begin{equation}
	u_1(x,0) = u_2(x,0) =0  \textrm{ m s}^{-1}
\end{equation}

Non-reflective conditions are imposed at the boundaries, and the relative density ratio is set to $r=0.98$. 
Several grid densities are considered, namely $\Delta x$ = 1/5, 1/10, 1/20, 1/40, 1/80, and 1/160 m. A fixed time step $\Delta t$ was chosen to allow for a more direct comparison of CPU times between numerical schemes. A constant ratio of $\Delta t = 0.15 \Delta x$ s m$^{-1}$ was used in this test, which gives $CFL \approx 0.6$, depending on the scheme and corresponding maximum eigenvalues. 
The reference solution is computed using the N-Roe scheme and a dense grid of 3200 points.

\begin{figure}[htbp]
	\center
	\includegraphics[width=6.5cm]{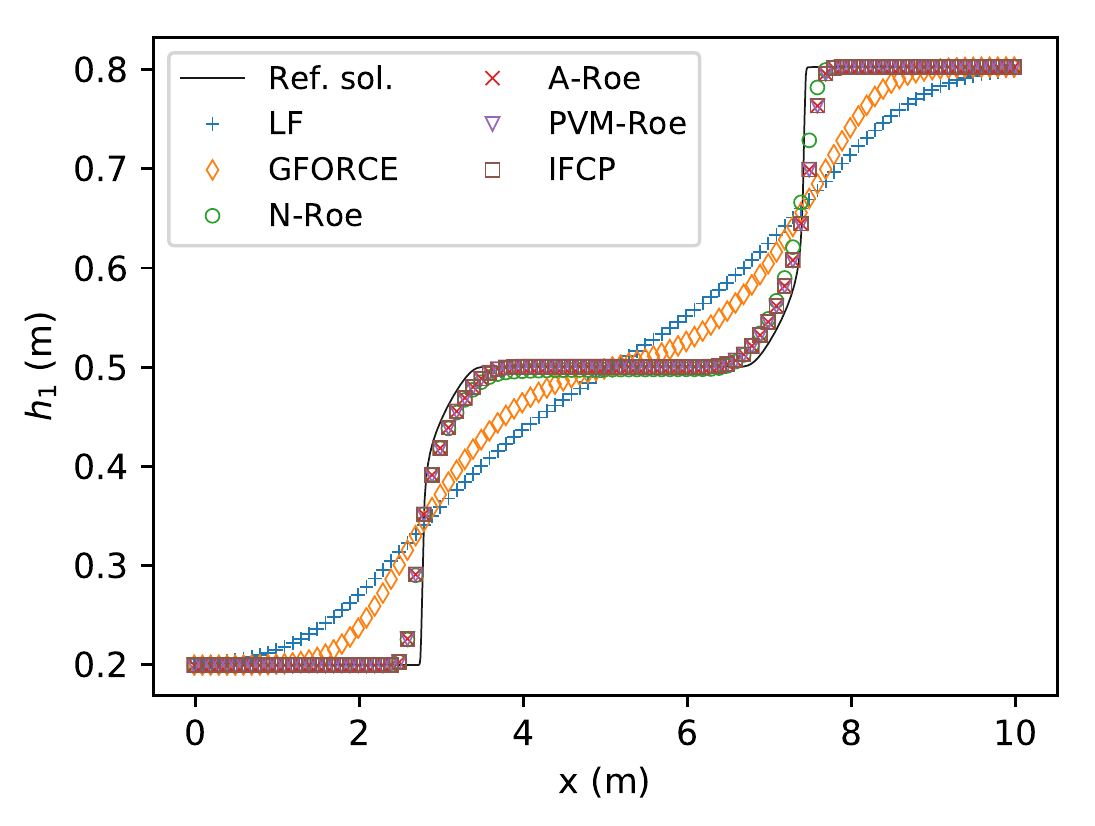}
	\hfill	
	\includegraphics[width=6.5cm]{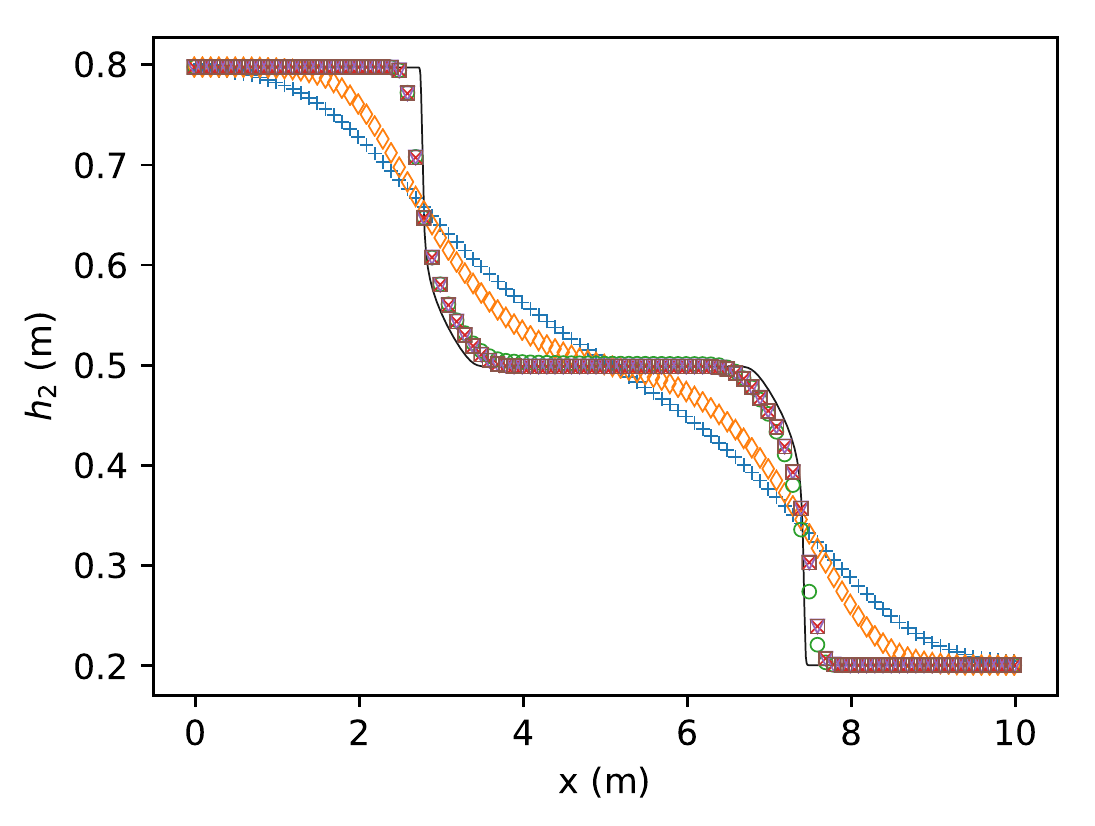}
	\vfill
	\includegraphics[width=6.5cm]{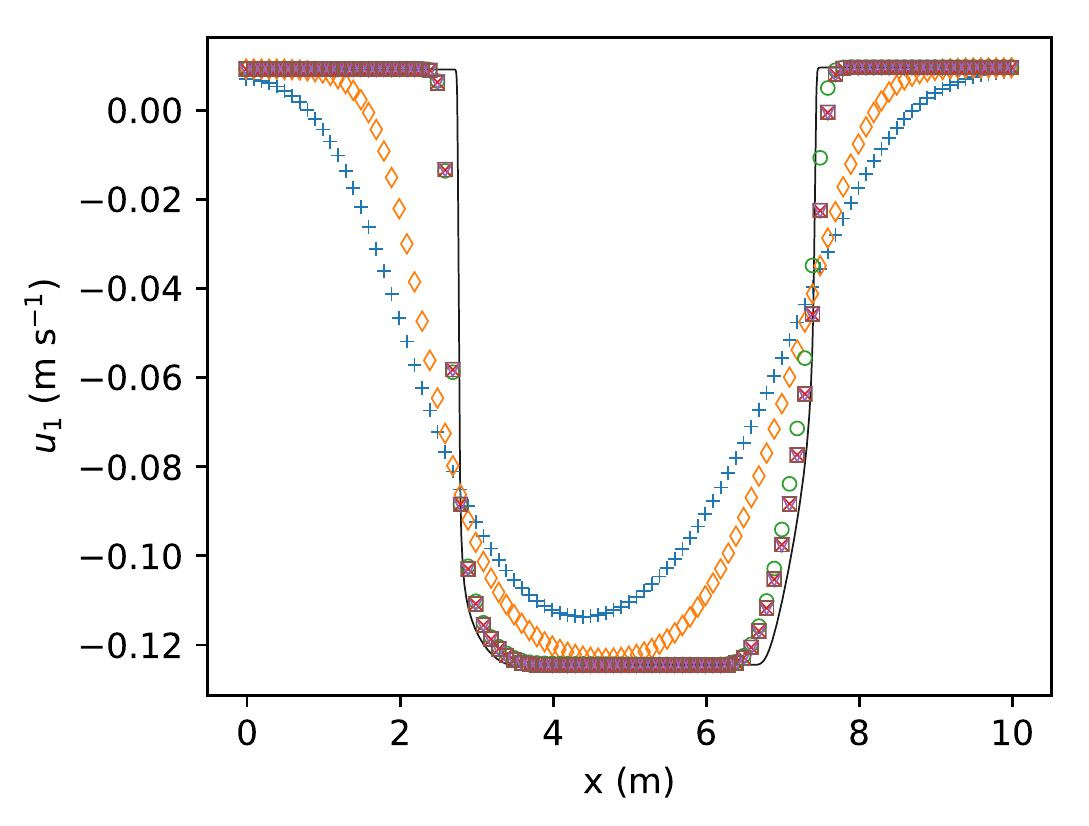}
	\hfill	
	\includegraphics[width=6.5cm]{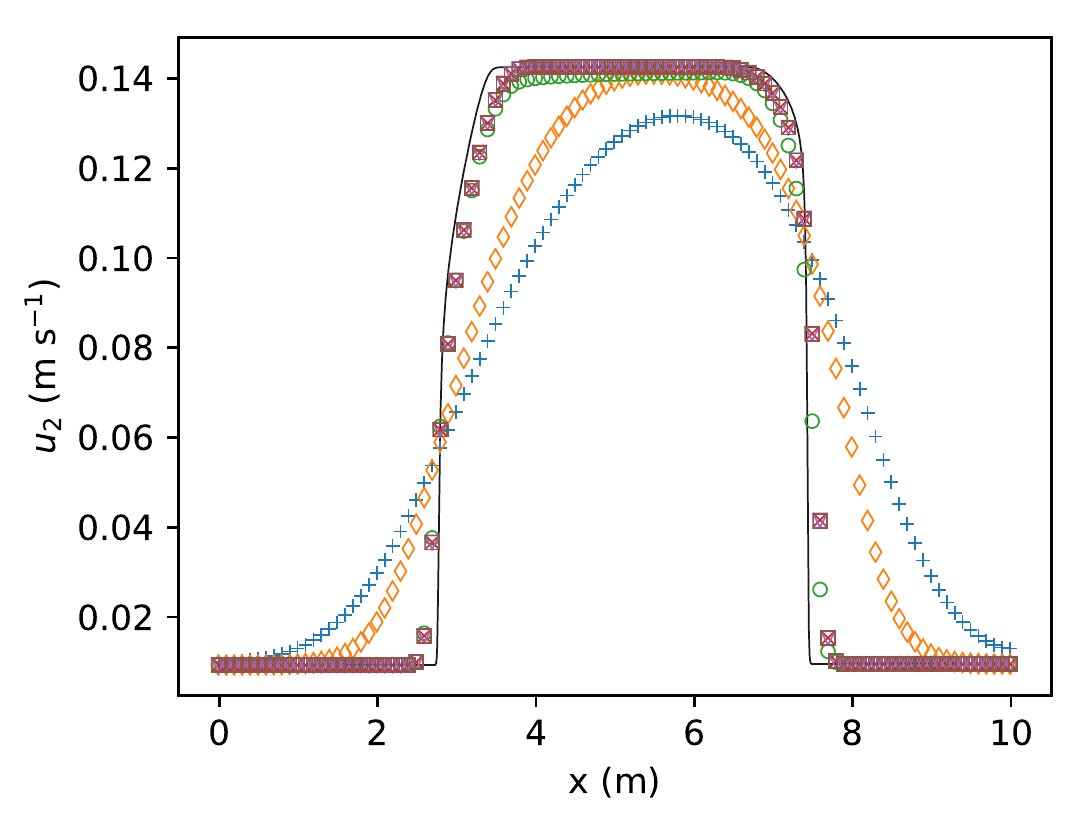}
	\caption{Test I: Results obtained by LF, GFORCE, N-Roe, PVM-Roe, IFCP and A-Roe scheme, compared to the reference solution, at $t=10$ s with $\Delta x = 1/40$ m}
	\label{fig:Results_Test01}
\end{figure}

Figure \ref{fig:Results_Test01} compares LF, GFORCE, N-Roe, PVM-Roe, IFCP, and A-Roe numerical schemes at $t=10$ s with $\Delta x = 1/40$ m against the reference solution. The results clearly show that the A-Roe scheme, similarly as the N-Roe, PVM-Roe, and IFCP schemes, provides more accurate and less diffused interface and velocities in comparison to GFORCE, and especially LF method, for the same grid density. This is in agreement with the results presented by \cite{castro2010some}, who evaluated several first-order numerical schemes. Furthermore, N-Roe and A-Roe scheme produce almost identical results, some differences occur only due to round-off errors when computing eigenstructures, as demonstrated in the previous example.

Table \ref{tab:CPU_Test01} shows CPU times in (s) for different grid densities. As expected, LF and GFORCE have similar CPU times, which are several times lower than the N-Roe and PVM-Roe scheme. However, the A-Roe scheme is significantly faster than the N-Roe (up to 4.1 times) and the PVM-Roe scheme (up to 75\%), with the CPU times comparable to GFORCE and IFCP schemes. As expected, the differences in simulation times between the N-Roe and A-Roe schemes increase with the number of spatial points because of a larger number of eigenvalues that are required at each time step. That is, as the number of spatial points increases, the ratio of the CPU time needed to compute eigenvalues to the total CPU time increases, so the speed-up of the A-Roe method becomes more pronounced.

% Table generated by Excel2LaTeX from sheet 'List2'
\begin{table}[htbp]
	\small
	\centering
	\caption{Test I: CPU times in (s) for different grid sizes obtained by the LF, GFORCE, N-Roe, PVM-Roe, IFCP, and A-Roe schemes (best of 5 runs)}
	\begin{tabular}{rcccccc}
		\hline
		No. of points & LF    & GFORCE & N-Roe   & PVM-Roe & IFCP & A-Roe \\
		\hline
	    50    & 0.24  & 0.27  & 0.49  & 0.36  & 0.30  & 0.32 \\
		100   & 0.53  & 0.63  & 1.45  & 0.95  & 0.71  & 0.73 \\
		200   & 1.38  & 1.62  & 4.89  & 2.81  & 1.93  & 1.89 \\
		400   & 3.65  & 4.76  & 16.81 & 8.69  & 5.47  & 5.46 \\
		800   & 12.10 & 16.42 & 66.46 & 30.42 & 18.21 & 18.08 \\
		1600  & 43.23 & 57.57 & 257.86 & 108.87 & 57.42 & 62.11 \\
		\hline
	\end{tabular}%
	\label{tab:CPU_Test01}%
\end{table}%

\begin{figure}[htbp]
	\center
	\includegraphics[width=6.5cm]{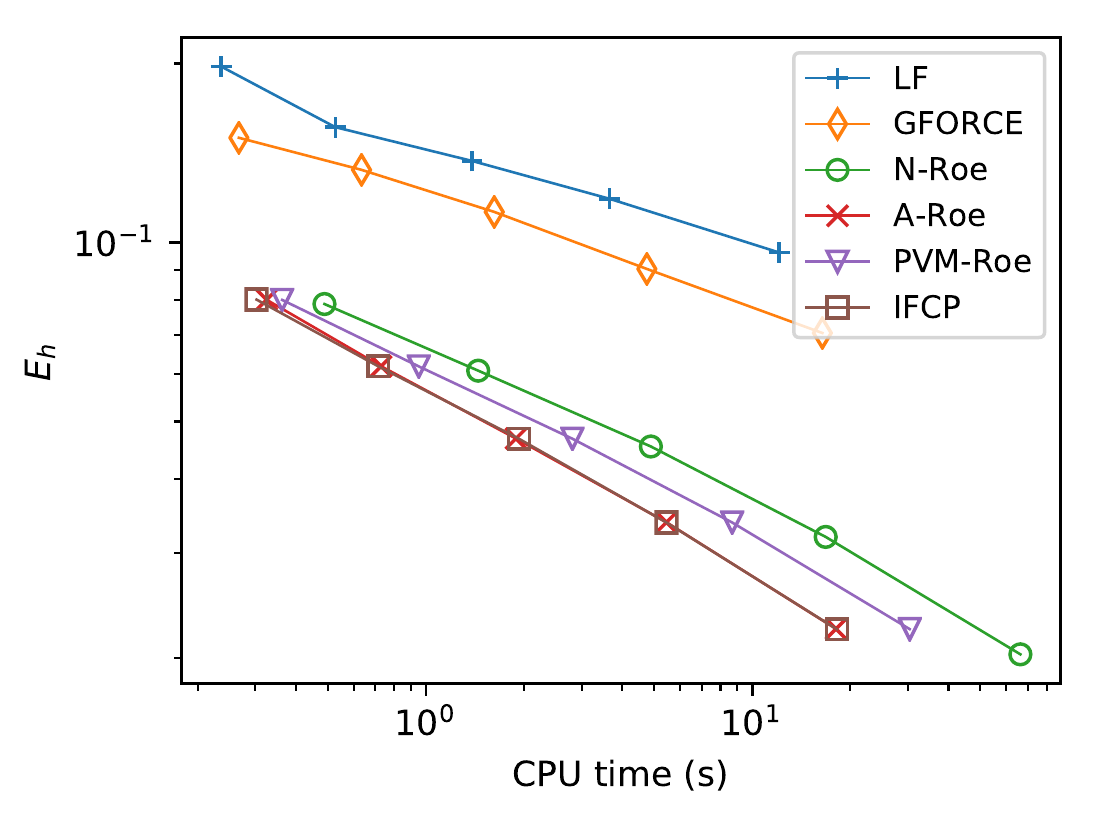}
	\hfill	
	\includegraphics[width=6.5cm]{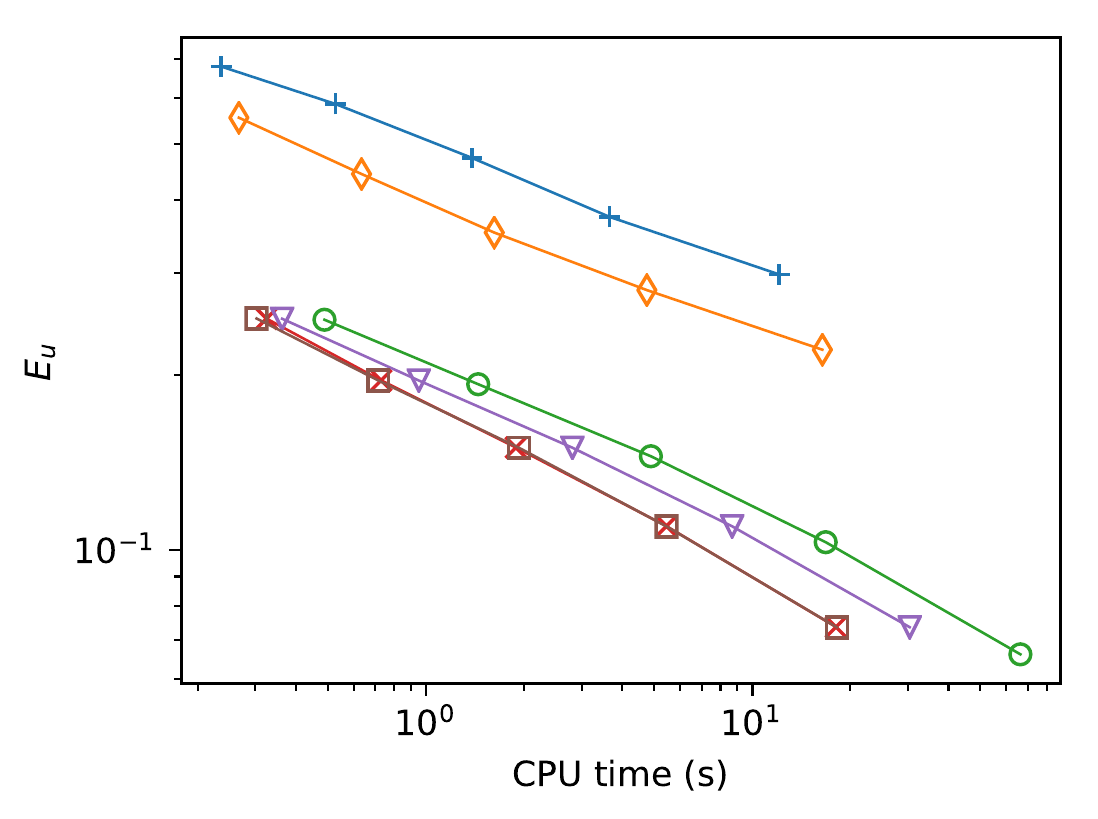}
	\caption{Test I: CPU time vs Error $E_{\Phi}$ for Lax-Friedrichs, GFORCE, N-Roe, PVM-Roe, IFCP, and A-Roe scheme, compared to the reference solution (log-log scale)}
	\label{fig:Err_Test01}
\end{figure}

To further evaluate the efficiency of each scheme, a CPU time vs. normalized root square error $E_{\Phi}$ is presented in Fig. \ref{fig:Err_Test01}:
\begin{equation}
E_{\Phi} = \frac{\sqrt{\sum_{n=1}^{M} \left[ \Phi(x_n, t_{end}) - \Phi^{ref}(x_n, t_{end}) \right]^2}}{\sqrt{\sum_{n=1}^{M} \Phi^{ref}(x_n, t_{end})^2}},
\label{eq:Err}
\end{equation}
where $M$ is number of spatial points, and $\Phi = h, u$, where $h=h_1, h_2$ are computed layer depths, $u=u_1, u_2$ are computed layer velocities, and $\Phi^{ref}$ are the corresponding reference values. The results show that A-Roe method is superior to the LF, GFORCE, N-Roe, and PVM-Roe schemes when efficiency is considered, and almost identical to the IFCP scheme; it has the same accuracy as the N-Roe method, with CPU times much closer to the GFORCE and IFCP schemes. We should note that a square of the pseudo-Jacobian matrix is computed here for the IFPC scheme, which can be avoided to save the computation time (see \cite{fernandez2011intermediate}).

\subsection{Test II: A Riemann problem with flat bottom}

The second case of a two-layer flow through a rectangular channel with flat bottom topography is considered. This test was introduced by \cite{castro2001q} to demonstrate that the uncoupled layer-by-layer approach is unsuitable for time-dependent two-layer exchange flows. It can also be used to evaluate the accuracy of different numerical schemes in computing non-regular time-dependent solutions over a flat bottom (\textit{e.g.}, \citep{castro2010some}). 

The initial free-surface is horizontal and the interface is characterized by two steep fronts. The spatial domain is set to [0, 100], and the initial condition is given by:
\begin{equation}
	h_1(x,0) = 
	\begin{cases}
		0.5 \textrm{ m}, \quad \textrm{if } x < 50 \textrm{ m} \\
		0.55 \textrm{ m}, \quad \textrm{otherwise}
	\end{cases}
	\quad
	h_2(x,0) = 
	\begin{cases}
		0.5 \textrm{ m}, \quad \textrm{if } x < 50 \textrm{ m} \\
		0.45 \textrm{ m}, \quad \textrm{otherwise}
	\end{cases}
\end{equation}
\begin{equation}
	u_1(x,0) = u_2(x,0) = 2.5 \textrm{ m s}^{-1}
\end{equation}

Non-reflective conditions are imposed at the boundaries, and the relative density ratio is set to $r=0.98$. As in the previous example, the solutions are obtained using the Lax-Friedrichs, GFORCE, N-Roe, PVM-Roe, IFCP, and A-Roe numerical schemes. Several grid densities are considered, namely $\Delta x$ = 1, 1/2, 1/4, 1/8, 1/16, and 1/32 m. A fixed time step $\Delta t = 0.1 \Delta x$ s m$^{-1}$ was used in this test, which gives $CFL \approx 0.6$, depending on the scheme and a maximum eigenvalue. 
The reference solution is computed using the N-Roe scheme and a dense grid of 6400 points.

\begin{figure}[htbp]
	\center
	\includegraphics[width=6.5cm]{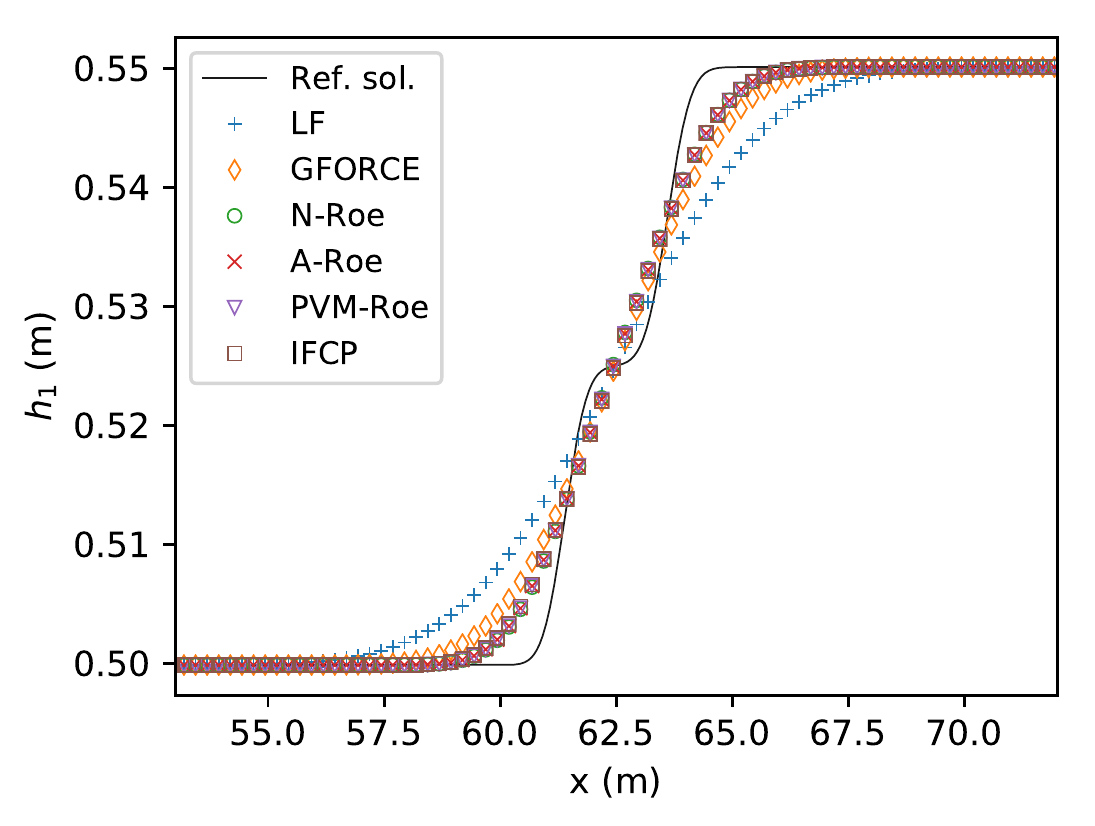}
	\hfill	
	\includegraphics[width=6.5cm]{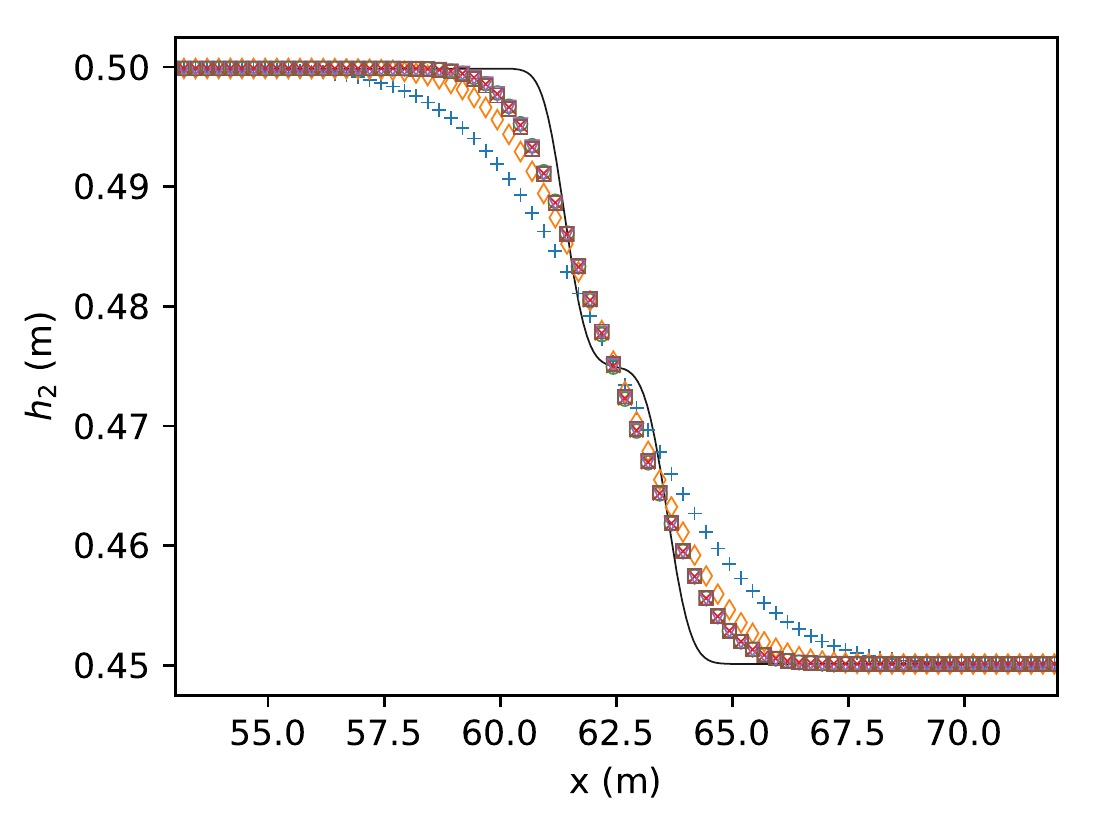}
	\vfill
	\includegraphics[width=6.5cm]{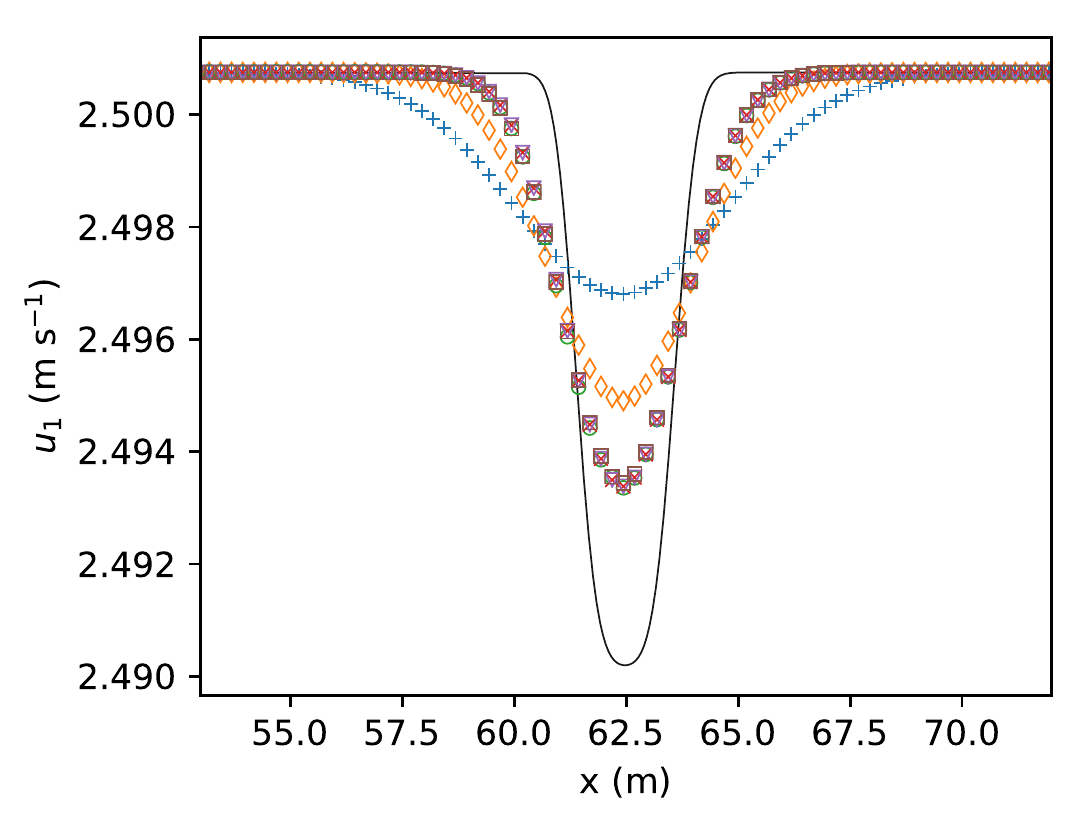}
	\hfill	
	\includegraphics[width=6.5cm]{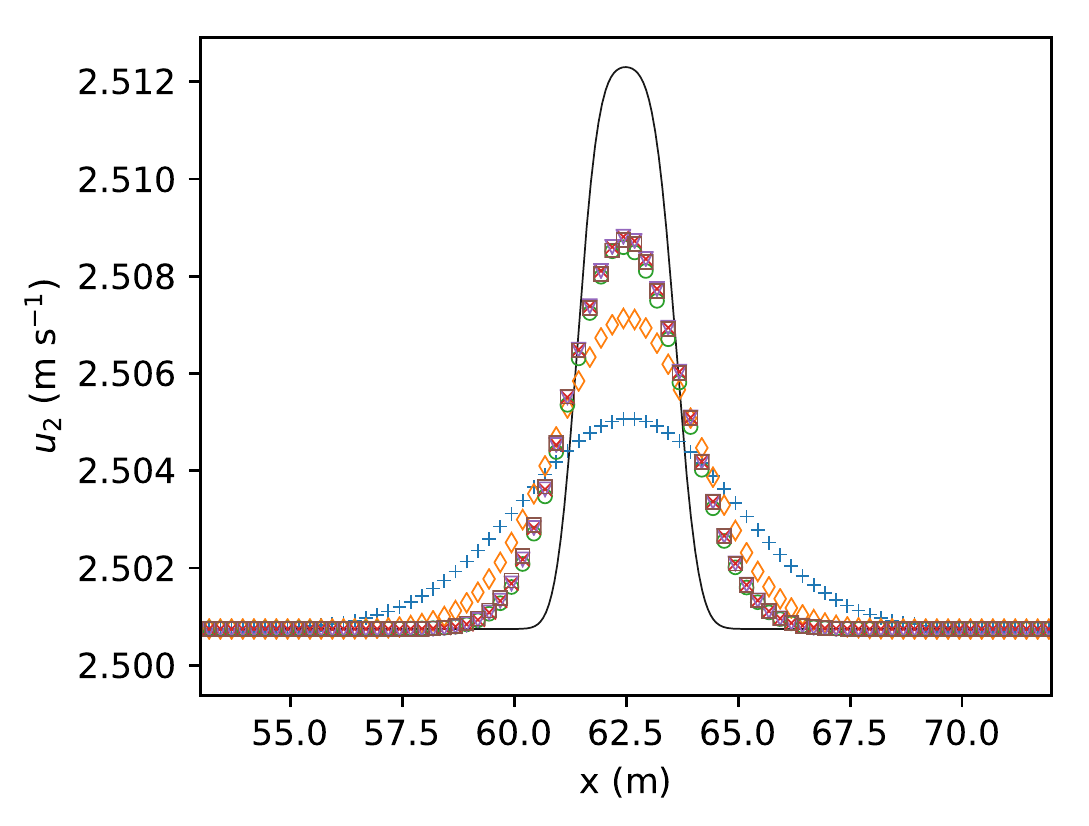}
	\caption{Test II: Results obtained by the LF, GFORCE, N-Roe, PVM-Roe, IFCP, and A-Roe scheme, compared to the reference solution, at $t=5$ s with $\Delta x = 1/8$ m}
	\label{fig:Results_Test02}
\end{figure}

Figure \ref{fig:Results_Test02} compares Lax-Friedrichs, GFORCE, N-Roe, PVM-Roe, IFCP, and A-Roe numerical schemes at $t=5$ s with $\Delta x = 1/8$ m against a reference solution. As expected, the results show that the A-Roe scheme, similarly as the N-Roe, PVM-Roe, and IFCP schemes, provide more accurate values in comparison to GFORCE, and especially to LF scheme, for the same grid density. In comparison to the previous example, GFORCE scheme here behaves better due to smaller differences between the external and internal eigenvalues. Same as in the previous example, N-Roe and A-Roe scheme give almost identical results. The results are in agreement with \citep{castro2010some}, where the same accuracy was found for these numerical schemes.

Table \ref{tab:CPU_Test02} shows CPU times in (s) for different grid densities. Again, the LF and GFORCE schemes have similar CPU times, which are several times lower than the N-Roe and PVM-Roe schemes, while the A-Roe scheme shows CPU times that are much closer to the GFORCE, almost identical to the IFCP scheme and significantly faster than the N-Roe (up to 3.8 times) and the PVM-Roe schemes (up to 83\%).

% Table generated by Excel2LaTeX from sheet 'List1'
\begin{table}[htbp]
	\small
	\centering
	\caption{Test II: CPU times in (s) for different grid sizes obtained by LF, GFORCE, N-Roe, PVM-Roe, IFCP, and A-Roe schemes (best of 5 runs)}
	\begin{tabular}{rcccccc}
		\hline
		No. of points & LF    & GFORCE & N-Roe   & PVM-Roe & IFCP & A-Roe \\
		\hline
	    100   & 0.05  & 0.06  & 0.12  & 0.08  & 0.06  & 0.06 \\
		200   & 0.11  & 0.13  & 0.36  & 0.21  & 0.15  & 0.16 \\
		400   & 0.29  & 0.36  & 1.33  & 0.63  & 0.41  & 0.42 \\
		800   & 0.88  & 1.16  & 4.73  & 2.18  & 1.30  & 1.30 \\
		1600  & 2.98  & 4.08  & 18.39 & 7.97  & 4.51  & 4.53 \\
		3200  & 11.14 & 15.56 & 72.40 & 30.73 & 16.95 & 16.78 \\
		\hline
	\end{tabular}%
	\label{tab:CPU_Test02}%
\end{table}%

\begin{figure}[htbp]
	\center
	\includegraphics[width=6.5cm]{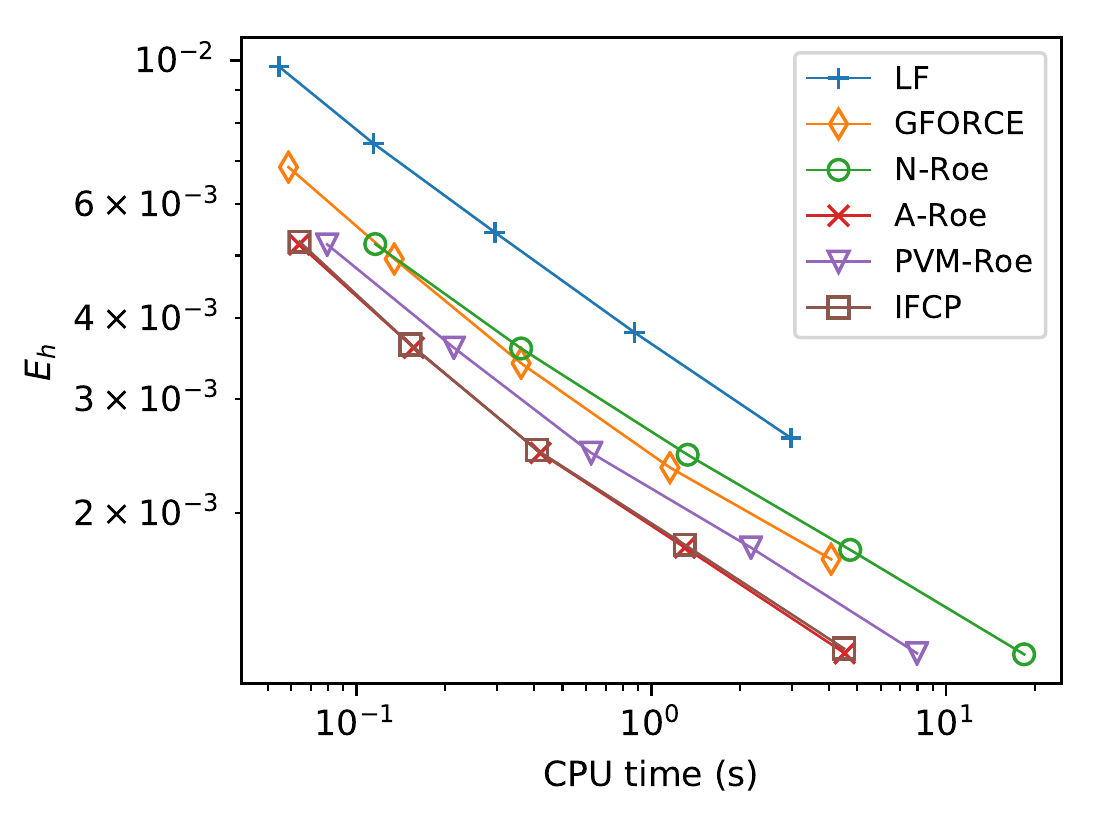}
	\hfill	
	\includegraphics[width=6.5cm]{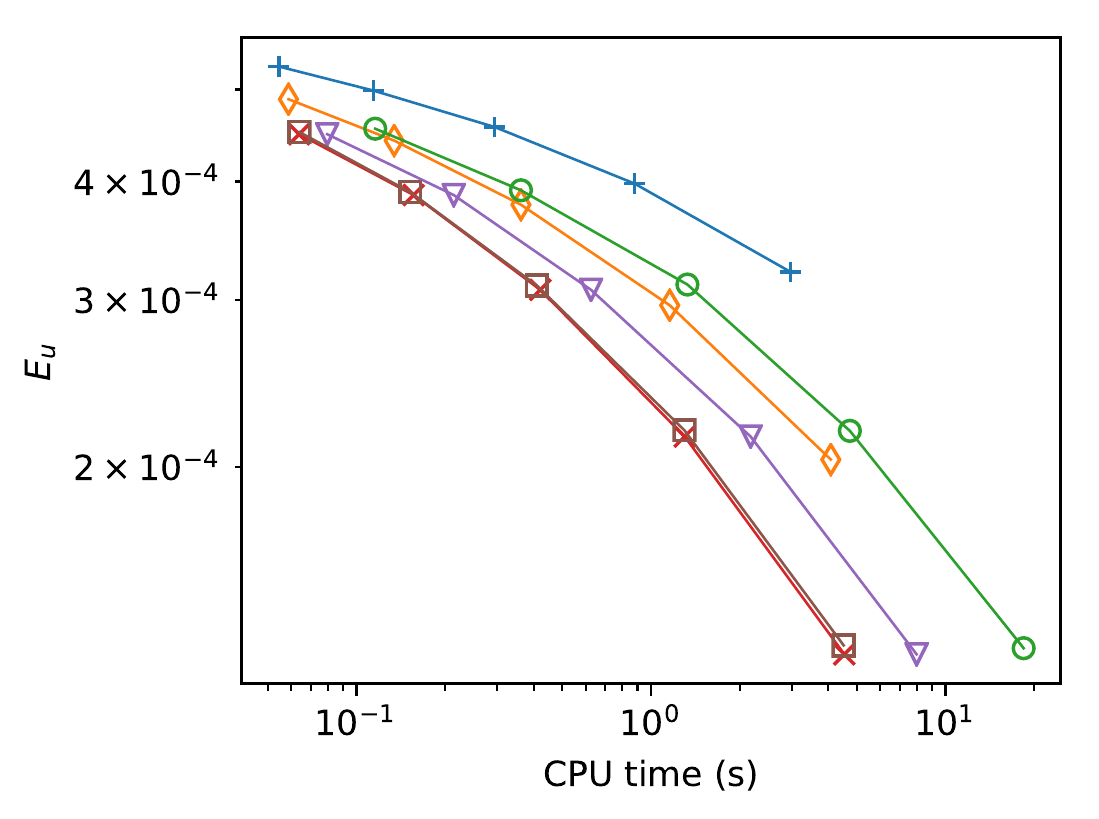}
	\caption{Test II: CPU time vs Error $E_{\Phi}$ for LF, GFORCE, N-Roe, PVM-Roe, IFCP, and A-Roe scheme, compared to the reference solution (log-log scale)}
	\label{fig:Err_Test02}
\end{figure}

To further evaluate the efficiency of each scheme a CPU time vs. normalized root square error Eq. (\ref{eq:Err}) is given in Fig. \ref{fig:Err_Test02}. The results show that A-Roe is better than the LF, GFORCE, N-Roe, and PVM-Roe schemes, and almost identical to the IFCP scheme, when efficiency is considered; it has the same accuracy as N-Roe method, with CPU times closer to the GFORCE and IFCP scheme. 

\subsection{Test III: Wet-dry front over a smooth bottom topography}

A case of a two-layer flow through a rectangular channel with non-flat bottom topography is considered next. This test was introduced by \cite{fernandez2011intermediate} to verify the well-balanced properties of numerical schemes when a non-flat bed and wet-dry fronts appear.

The spatial domain is set to [0, 10], and the bed elevation is defined by the following function:
\begin{equation}
b(x) = 
\begin{cases}
0.0 \textrm{ m}, \quad \textrm{if } x < 5 \textrm{ m} \\
\frac{x-5}{10} \textrm{ m}, \quad \textrm{otherwise}
\end{cases}
\end{equation}
whereas, the initial condition is defined by:
\begin{equation}
h_2(x,0) = 
\begin{cases}
0.6 \textrm{ m}, \quad \textrm{if } 4.5 < x < 5 \textrm{ m} \\
0.0 \textrm{ m}, \quad \textrm{if } x > 7 \textrm{ m} \\
	\max
	\begin{cases}
	0.2 \textrm{ m} - b(x) \\
	0.0 \textrm{ m}
	\end{cases}
\quad \textrm{otherwise}
\end{cases},
\end{equation}
\begin{equation}
h_1(x,0) = 1.0 \textrm{ m} - h_2(x,0) - b(x)
\end{equation}
\begin{equation}
u_1(x,0) = 0.0 \textrm{ m s}^{-1}
\quad
u_2(x,0) = 0.0 \textrm{ m s}^{-1}
\end{equation}

Non-reflective conditions are imposed at the boundaries, the relative density ratio is set to $r=0.99$, spatial grid size is set to $\Delta x$ = 1/20 m, and $CFL = 0.8$. Only the N-Roe and A-Roe schemes are compared, both with an implemented numerical technique for dealing with wet-dry fronts \citep{castro2005numerical}. The wet-dry parameter is set to $\varepsilon = 10^{-3}$ m. The reference solution is computed using the N-Roe scheme and a dense grid of 3200 points. The CPU time of the complete simulation has been found to be 25.4 s for the N-Roe scheme, and 10.7 s for the proposed A-Roe scheme.

\begin{figure}[htbp]
	\center
	\includegraphics[width=6.5cm]{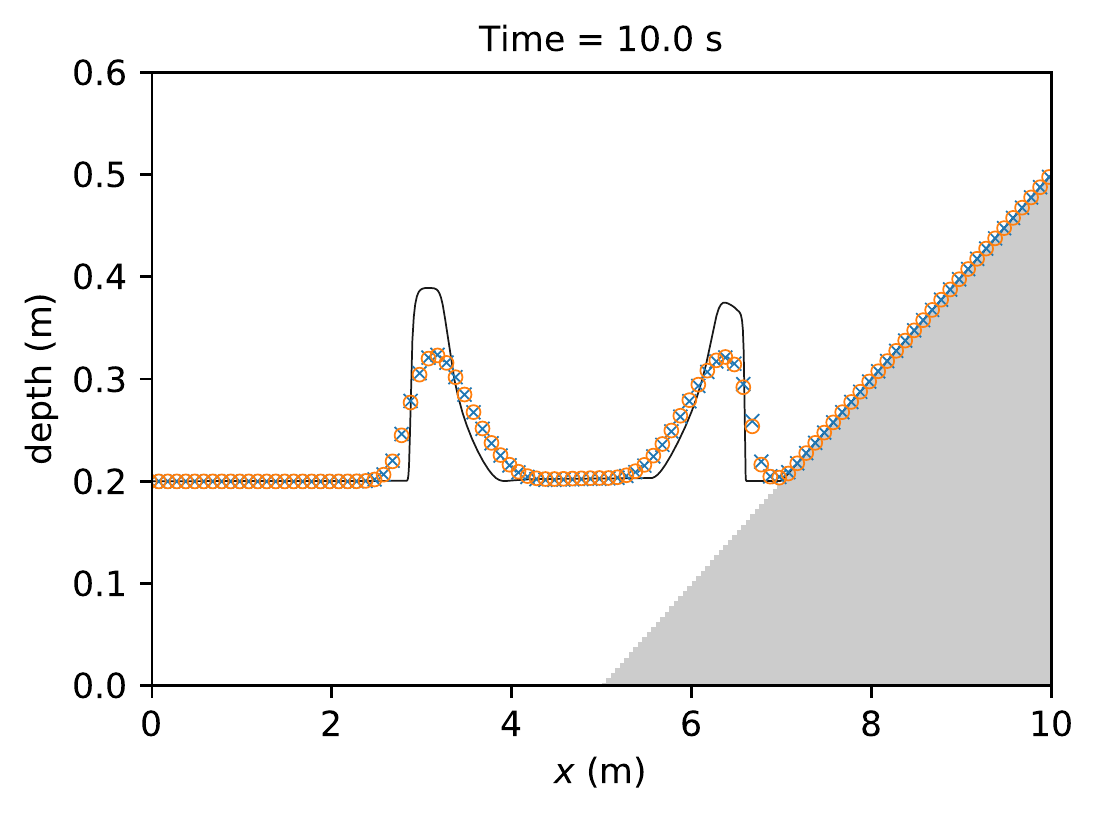}
	\hfill	
	\includegraphics[width=6.5cm]{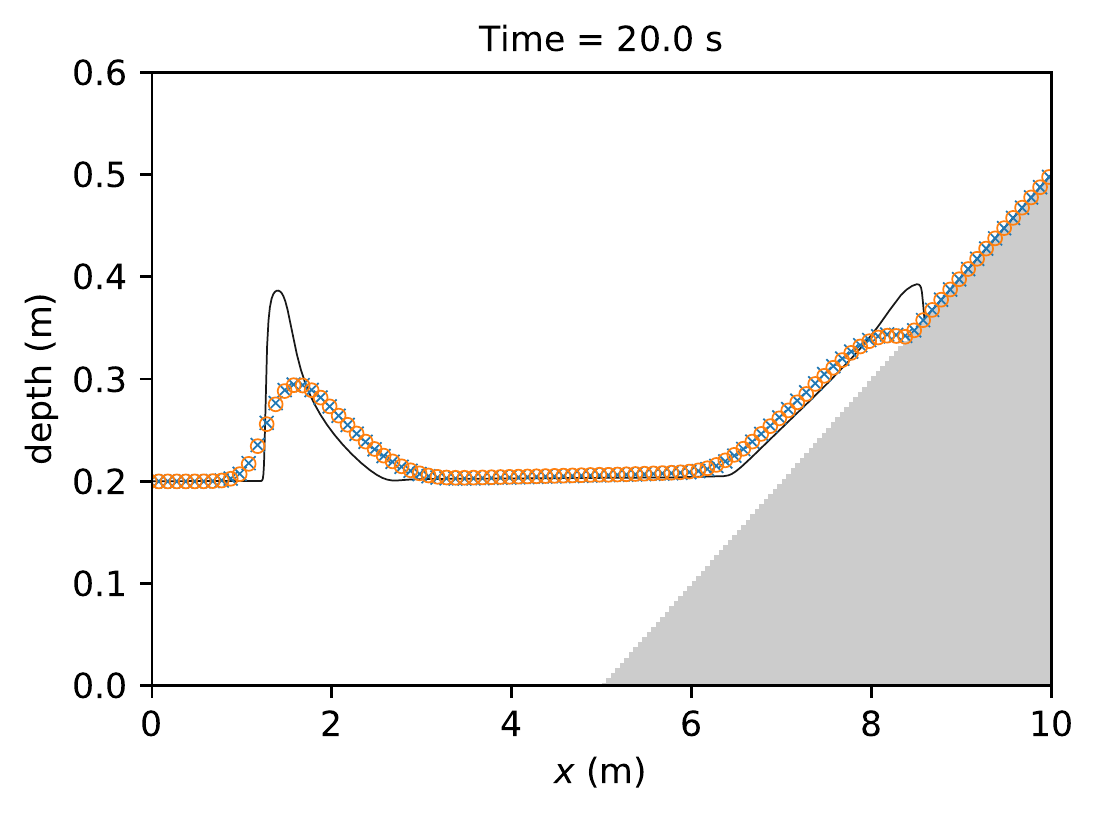}
	\vfill
	\includegraphics[width=6.5cm]{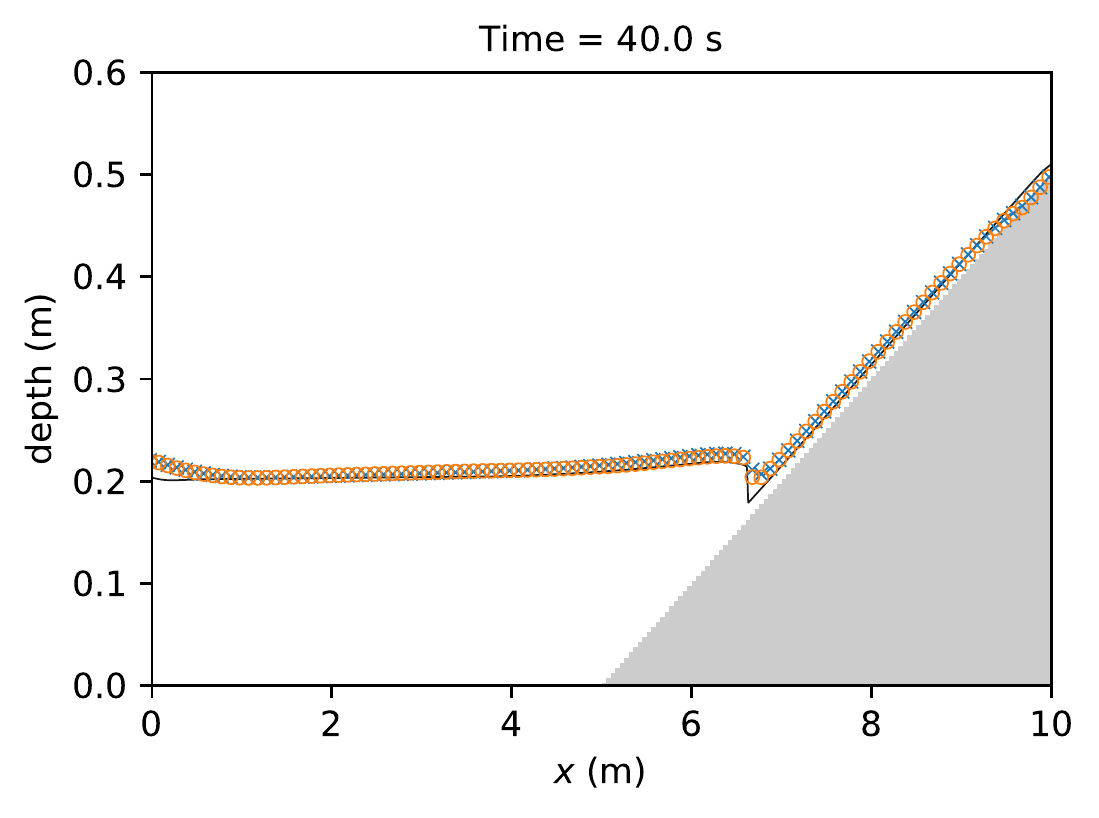}
	\hfill	
	\includegraphics[width=6.5cm]{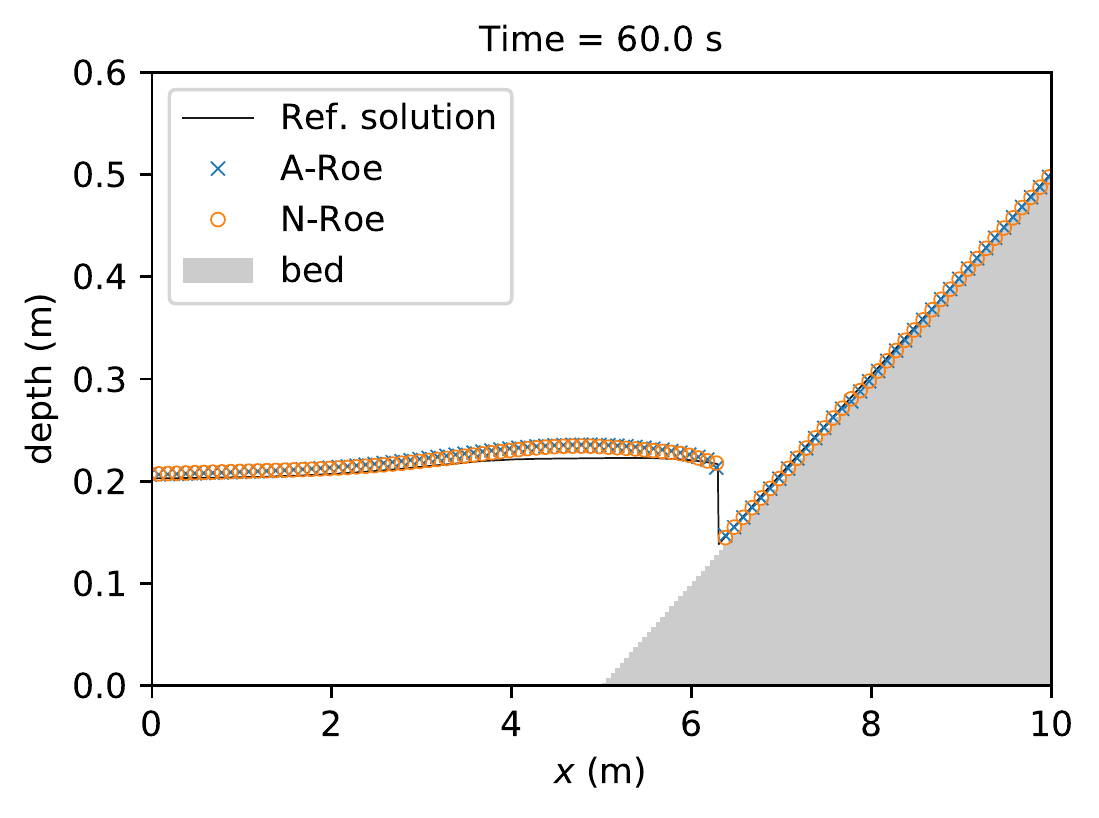}
	\caption{Test III: Results of the interface obtained by N-Roe and A-Roe scheme, compared to the reference solution, at $t=10, 20, 40$ and $60$ s with $\Delta x = 1/20$ m}
	\label{fig:Results_Test04}
\end{figure}

Figure \ref{fig:Results_Test04} shows the evolution of the interface obtained by the N-Roe and A-Roe scheme, compared to the reference solution. Both the N-Roe and A-Roe scheme produce almost identical results at every time step, and both schemes successfully deal with wet-dry fronts. These results are in agreement with those presented by \cite{fernandez2011intermediate}.

\subsection{Test IV: A Riemann problem with a bottom step}

Another case of a two-layer flow through a rectangular channel with non-flat bottom topography is considered. This test was introduced by \cite{fernandez2011intermediate} to examine how numerical schemes deal with bottom steps in the very extreme circumstances for which the SWE theory may cease to be applicable.

The spatial domain is set to [0, 10], and the bed elevation is defined by a single bottom step:
\begin{equation}
b(x) = 
\begin{cases}
0.5 \textrm{ m}, \quad \textrm{if } x < 5 \textrm{ m} \\
0.0 \textrm{ m}, \quad \textrm{otherwise}
\end{cases}
\label{eq:test5_bed}
\end{equation}
whereas the initial condition is defined by:
\begin{equation}
h_2(x,0) = 
\begin{cases}
0.2 \textrm{ m}, \quad \textrm{if } x < 5 \textrm{ m} \\
0.1 \textrm{ m}, \quad \textrm{otherwise}
\end{cases}
\quad
h_1(x,0) = 1.5 \textrm{ m} - h_2(x,0) - b(x)
\end{equation}
\begin{equation}
u_1(x,0) = 0.0 \textrm{ m s}^{-1}
\quad
u_2(x,0) = 0.1 \textrm{ m s}^{-1}
\end{equation}

Non-reflective conditions are imposed at the boundaries, the relative density ratio is set to $r=0.98$, spatial grid size is set to $\Delta x$ = 1/20 m, and $CFL = 0.7$. Again, only the N-Roe and A-Roe schemes are compared here, both with an implemented numerical technique for wet-dry fronts \citep{castro2005numerical} to deal with an emerging bottom step. The reference solution is computed using the N-Roe scheme and a dense grid of 3200 points. The CPU time of the complete simulation has been found to be 0.89 s for the N-Roe scheme, and 0.33 s for the proposed A-Roe scheme. 

\begin{figure}[htbp]
	\center
	\includegraphics[width=6.5cm]{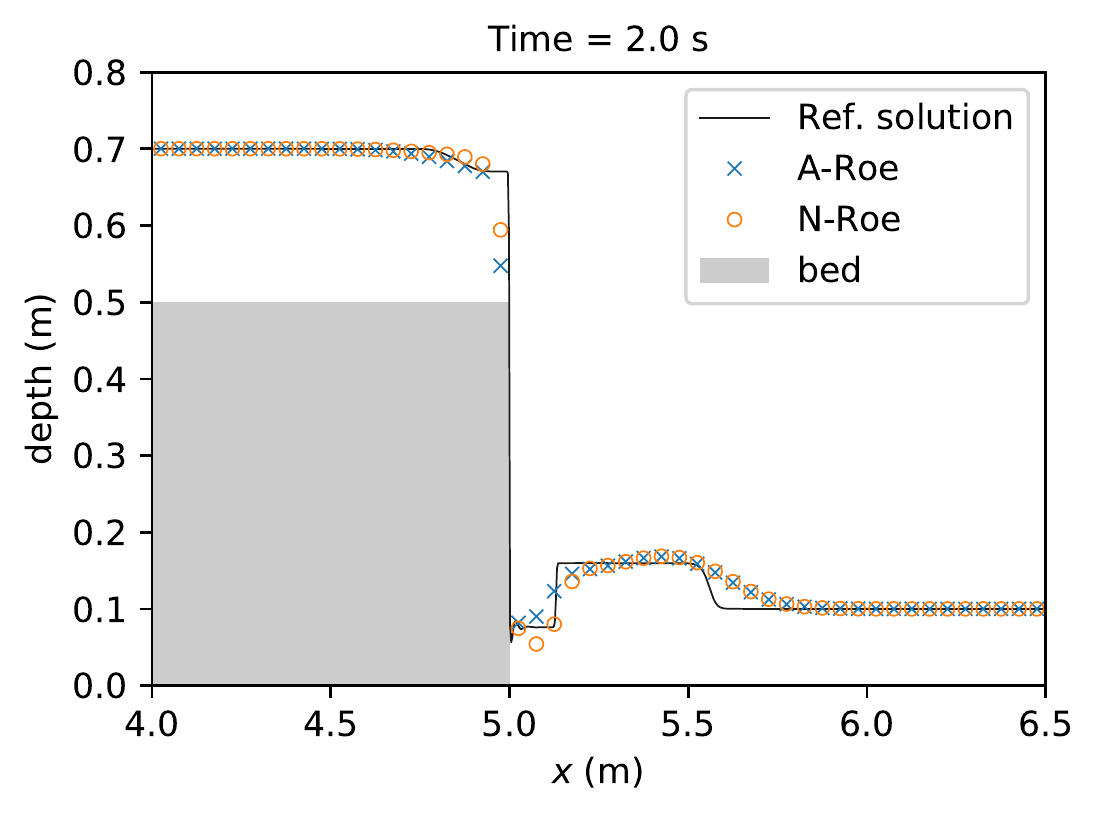}
	\hfill	
	\includegraphics[width=6.5cm]{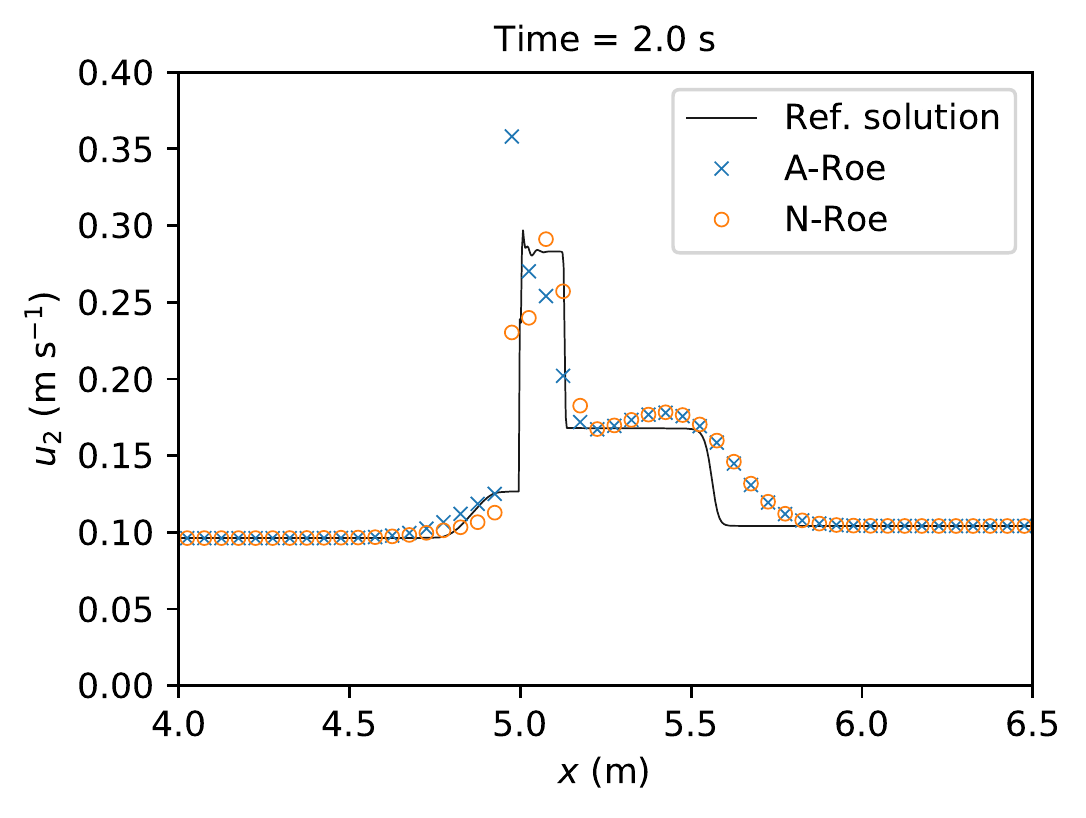}
	\caption{Test IV: Detail of the interface and bottom layer velocity obtained by N-Roe and A-Roe scheme, compared to the reference solution, at $t=2$ s with $\Delta x = 1/20$ m}
	\label{fig:Results_Test03}
\end{figure}

Figure \ref{fig:Results_Test03} shows the interface depth and bottom layer velocity at $t=2.0$ s. The A-Roe and N-Roe scheme produce very similar results, without any appearance of negative depths. 
The position and propagation velocity of the downstream wave, are in agreement with values obtained by \cite{fernandez2011intermediate}. The only difference between the results may be seen immediately downstream from the bed step, which is presumably the result of a different correction technique used to achieve a well-balanced solution and deal with resonant problems in this particular test. Namely, the technique proposed by \cite{castro2010some} sets the velocities at the interface to zero, whereas the hydrostatic reconstruction (HR) used by \cite{fernandez2011intermediate} redefines the geometry source term at the discontinuous interface differently from the technique proposed by \cite{castro2005numerical} applied here.
Nevertheless, the modification of the A-Roe by the HR technique is straightforward, directly following the HR extension of the Roe scheme (see \cite{castro2007well}), because these two schemes differ only by the method in which the eigenstructure is computed and the correction algorithm for the hyperbolicity loss. However, a detailed performance analysis of the A-Roe scheme extended by HR is outside the scope of this manuscript.

\subsection{Test V: Exchange flow with non-hyperbolic initial conditions and $r=0.99$}

The final three numerical tests demonstrate the performance of the proposed hyperbolicity correction. 
The solutions are obtained by the N-Roe method using the real Jordan decomposition, as well as using the A-Roe method with three different implementations of the hyperbolicity correction: (\textit{i}) approximate correction proposed by \cite{castro2011numerical} (A-RoeC), (\textit{ii}) iterative correction based on the full discriminant of the characteristic polynomial prosed by \cite{sarno2017some} (A-RoeS), and (\textit{iii}) iterative correction based on the discriminant of the resolvent cubic equation presented in Section 2.4, which makes an integral part of the analytical solutions for the eigenvalues proposed here (A-Roe).

A two-layer exchange flow through a rectangular channel with flat bottom topography is again considered. This particular test has been proposed by \cite{castro2011numerical} to demonstrate how un-physical oscillations may occur and eventually blow-up the computation when hyperbolicity loss is not treated adequately.
The initial free-surface is horizontal and the interface is characterized by two steep fronts. The spatial domain is set to [-1, 1], and the initial condition is given by:
\begin{equation}
h_1(x,0) = 
\begin{cases}
0.4 \textrm{ m}, \quad \textrm{if } \lvert x \rvert < 0.5 \textrm{ m} \\
0.5 \textrm{ m}, \quad \textrm{otherwise}
\end{cases}
\quad
h_2(x,0) = 1.0 \textrm{ m} - h_1(x,0)
\end{equation}
\begin{equation}
u_1(x,0) = 0.2 \textrm{ m s}^{-1}, \quad u_2(x,0) = -0.3 \textrm{ m s}^{-1}
\end{equation}

Non-reflective conditions are imposed at the boundaries, and the relative density ratio is set to $r=0.99$.
All of the computations are performed using a small grid size $\Delta x = 1/200$ m and a fixed time step $\Delta t = 0.001$ s, which gives $CFL \approx 0.7$. 

\begin{figure}[htbp]
	\center
	\includegraphics[width=6.5cm]{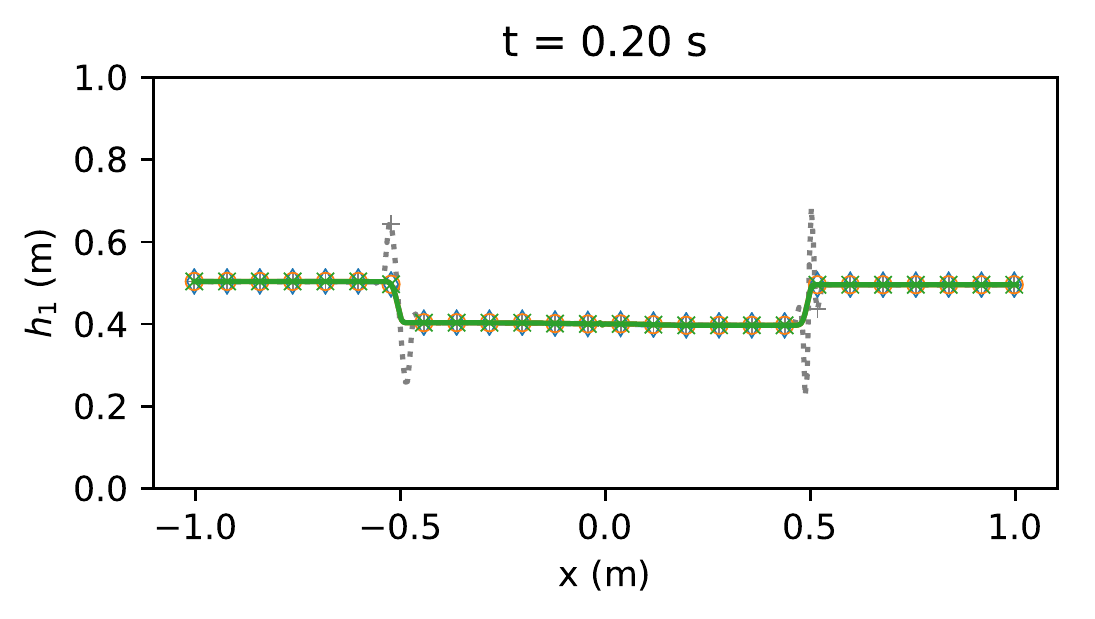}
	\hfill	
	\includegraphics[width=6.5cm]{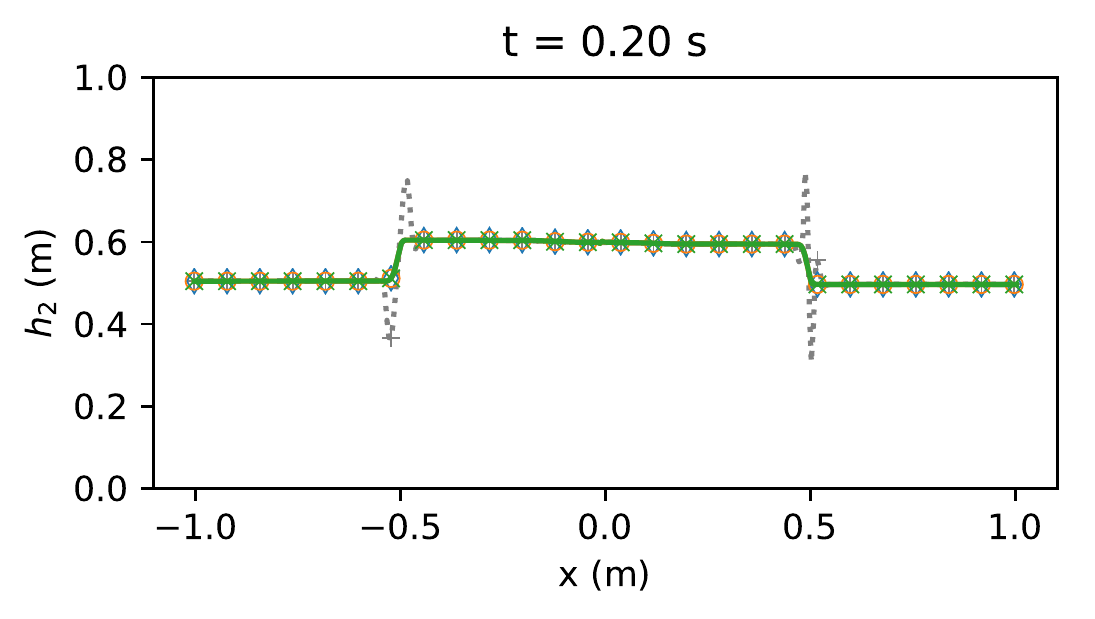}
	\vfill
	\includegraphics[width=6.5cm]{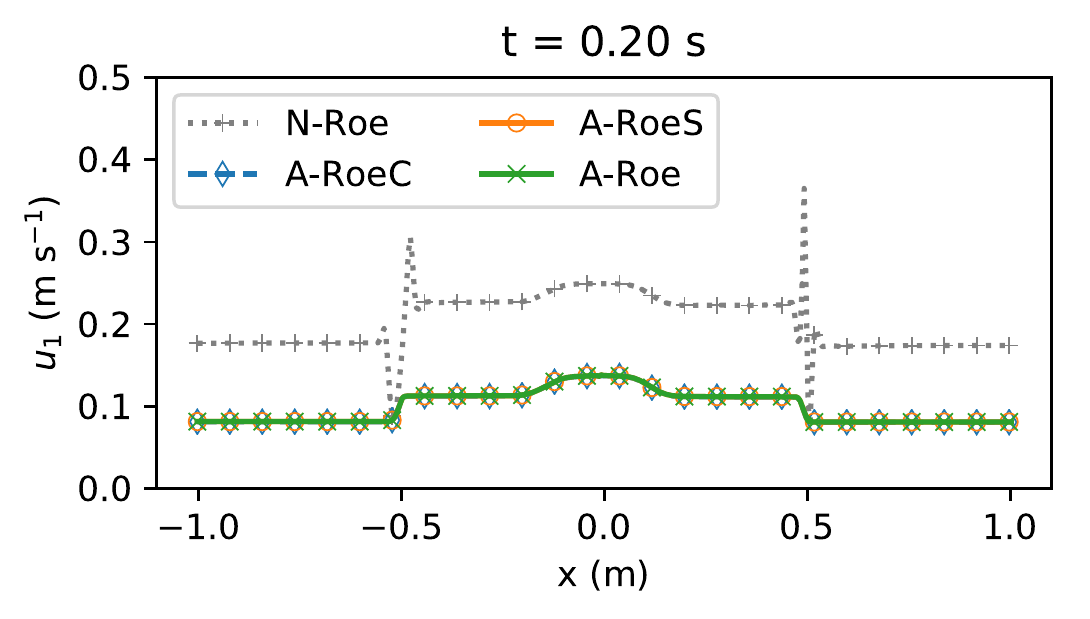}
	\hfill	
	\includegraphics[width=6.5cm]{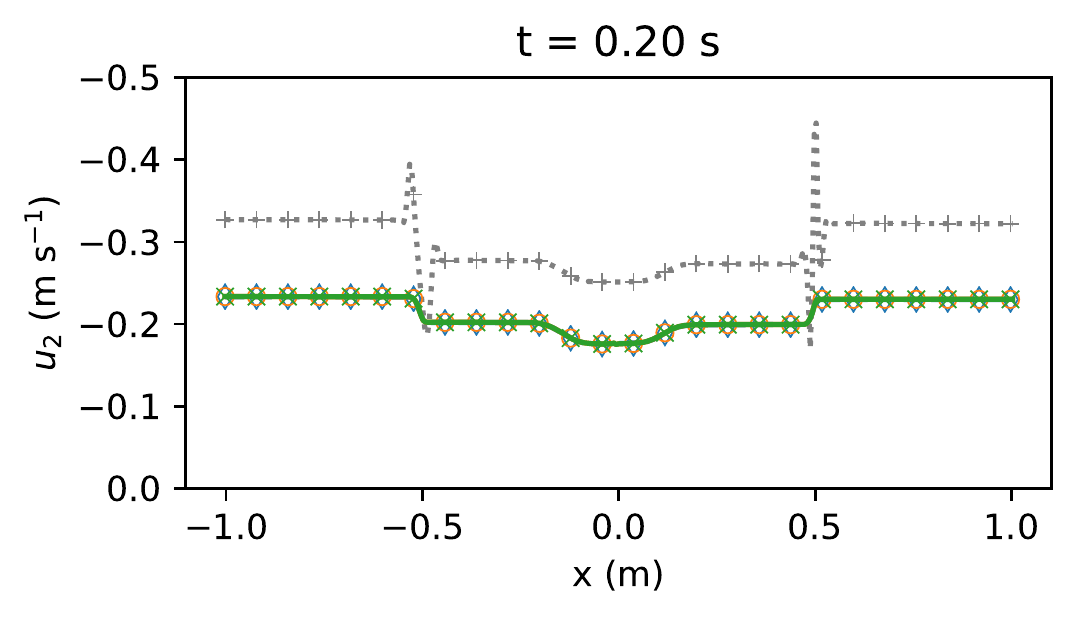}
	\vfill
	\includegraphics[width=6.5cm]{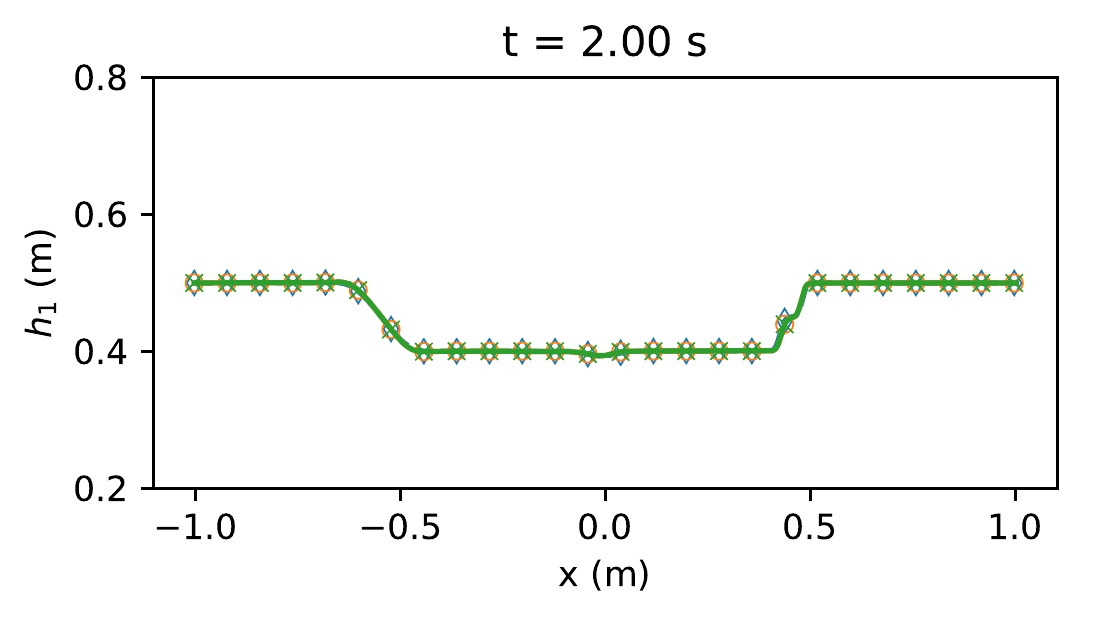}
	\hfill	
	\includegraphics[width=6.5cm]{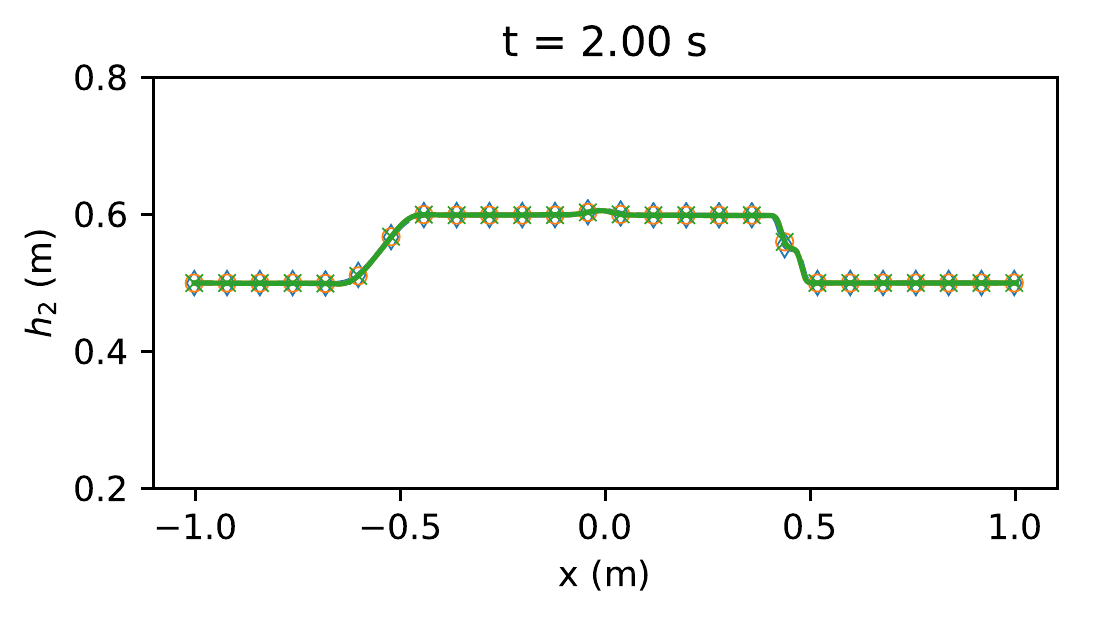}
	\vfill
	\includegraphics[width=6.5cm]{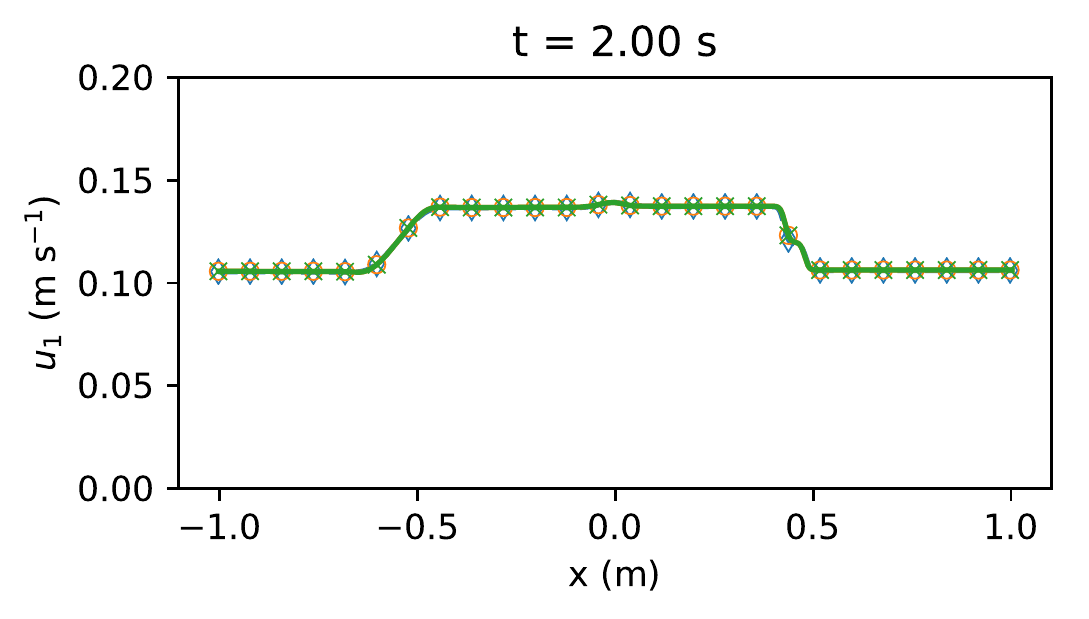}
	\hfill	
	\includegraphics[width=6.5cm]{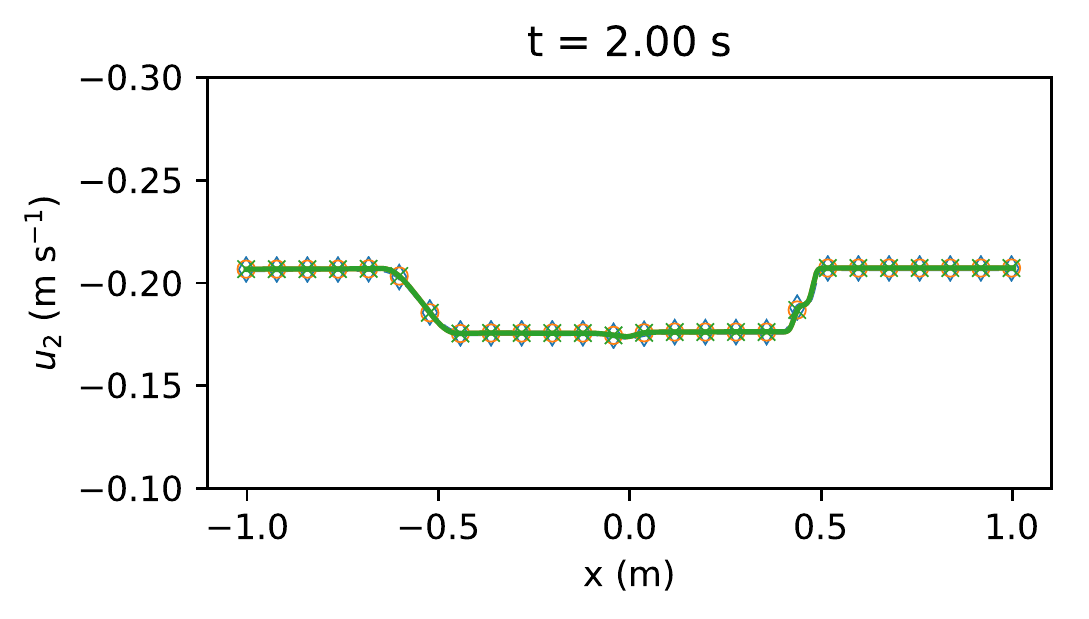}
	\caption{Test V: Upper and lower layer depths and velocities obtained by N-Roe without hyperbolicity correction and A-Roe method with three different implementations of the hyperbolicity correction, at $t=0.2$ and $t=2.0$ s and $\Delta x = 1/200$ m}
	\label{fig:Results_Test05}
\end{figure}

Figure \ref{fig:Results_Test05} shows the upper and lower layer depths and velocities at $t=0.2$ s and at $t=2.0$ s. The results computed by the A-Roe method with the proposed integrated hyperbolicity correction (A-Roe) are practically the same as the results obtained using the iterative correction proposed by \cite{sarno2017some} (A-RoeS) and the results using the approximate correction proposed by \cite{castro2011numerical} (A-RoeC). Note that the N-Roe method without hyperbolicity correction does not change the initial velocities in the layers, but as a consequence, strong oscillations appear at the interface discontinuities. Shortly after $t=0.2$ s the computation by the N-Roe method blows-up. On the other hand, when either of three hyperbolicity corrections is applied, the velocities are reduced shortly after the start of the simulation, but the computation remains stable until a steady-state is reached at $t=2.0$ s. These findings are in agreement with the results obtained by \cite{castro2011numerical}, who found similar differences between the Roe scheme that is based only on the real Jordan decomposition, and the Roe scheme that additionally uses an approximate hyperbolicity correction, as well as \cite{sarno2017some}, who repeated this numerical test and showed that their iterative hyperbolicity algorithm behaves very similarly. 

\begin{figure}[htbp]
	\center
	\includegraphics[width=8cm]{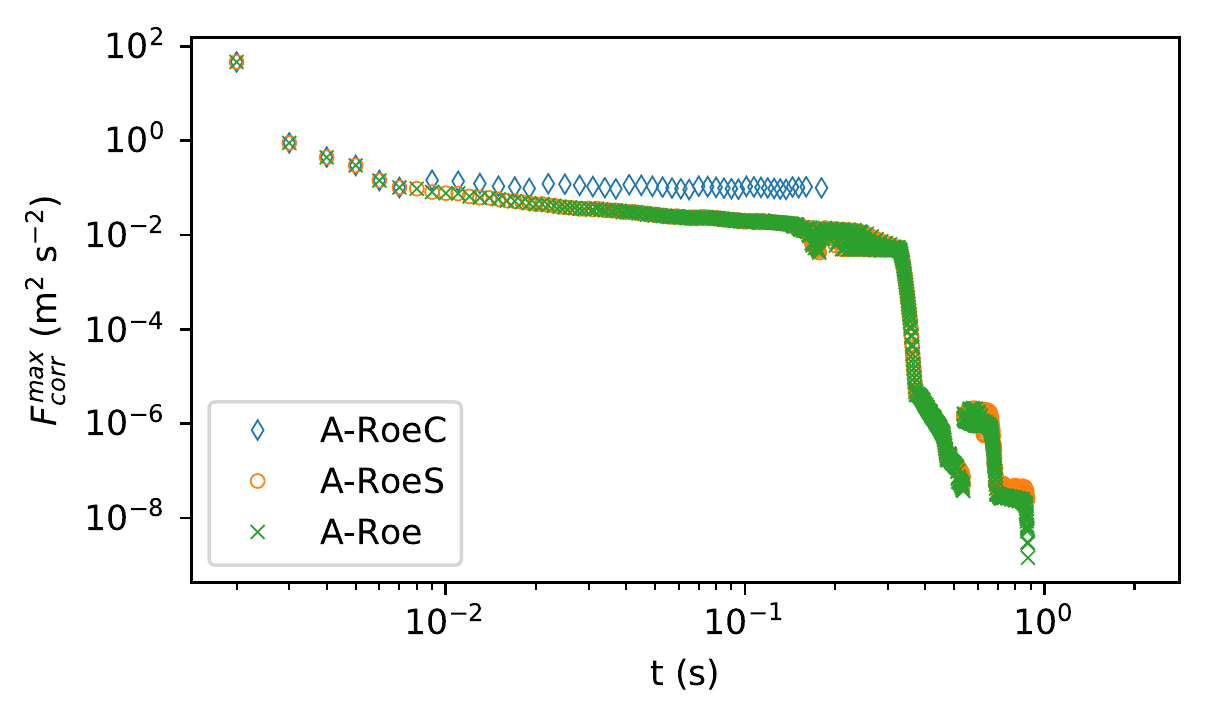}
	\caption{Test V: Comparison of maximum values of $F_{corr}^{max}$ at each time step, computed by different implementations of the hyperbolicity correction}
	\label{fig:F_Test05}
\end{figure}

To examine the behaviour of the proposed iterative correction algorithm in more detail, Fig.~\ref{fig:F_Test05} shows the computed maximum friction $F_{corr}^{max}$ which is added to the system to prevent the hyperbolicity loss and the appearance of complex eigenvalues. Since the initial conditions are in a non-hyperbolic state, a relatively high $F_{corr}^{max}$ is added in the first time step; namely 46.3 m$^2$ s$^{-2}$ by both the A-Roe and the A-RoeS method, and 46.5 m$^2$ s$^{-2}$ by the approximate A-RoeC method. Just after a few time steps $F_{corr}^{max}$ reduces to under $10^{-1}$ m$^2$ s$^{-2}$. These results confirm that the proposed A-Roe algorithm is almost identical to the iterative solution given by \cite{sarno2017some} and very close to the approximate solution given by \cite{castro2011numerical} when $r\approx 1$.

The CPU time of the complete simulation has been found to be 5.14 s for A-RoeC, 8.01 s for the iterative A-RoeS, and 6.4 s for the proposed A-Roe scheme. Clearly, the proposed algorithm A-Roe is noticeably faster than A-RoeS, because the discriminant of the resolvent cubic equation is solved at each time step instead of the full discriminant of a quartic equation.

\subsection{Test VI: Exchange flow with non-hyperbolic initial conditions and $r=0.3$}

Another case of a two-layer exchange flow through a rectangular channel with flat bottom topography is now considered as proposed by \cite{sarno2017some} to illustrate the advantages of the iterative hyperbolicity correction based on the discriminant of the characteristic polynomial over an approximate treatment proposed by \cite{castro2011numerical}. The main idea here is to show that the approximate eigenvalues can produce not only less accurate results, but they can even completely change the two-layer flow structure.
The spatial domain is set to [-1, 1], and the initial condition is given by:
\begin{equation}
h_1(x,0) = 
\begin{cases}
0.4 \textrm{ m}, \quad \textrm{if } \lvert x \rvert < 0.5 \textrm{ m} \\
0.5 \textrm{ m}, \quad \textrm{otherwise}
\end{cases}
\quad
h_2(x,0) = 1.0 \textrm{ m} - h_1(x,0)
\end{equation}
\begin{equation}
u_1(x,0) = 1.0 \textrm{ m s}^{-1}, \quad u_2(x,0) = -3.0 \textrm{ m s}^{-1}
\end{equation}

Non-reflective conditions are imposed at the boundaries, but in contrast to the previous example, the relative density ratio is set to be as low as $r=0.3$. The same numerical schemes are used as in the previous example, with the same grid size and time step.

\begin{figure}[htbp]
	\center
	\includegraphics[width=6.5cm]{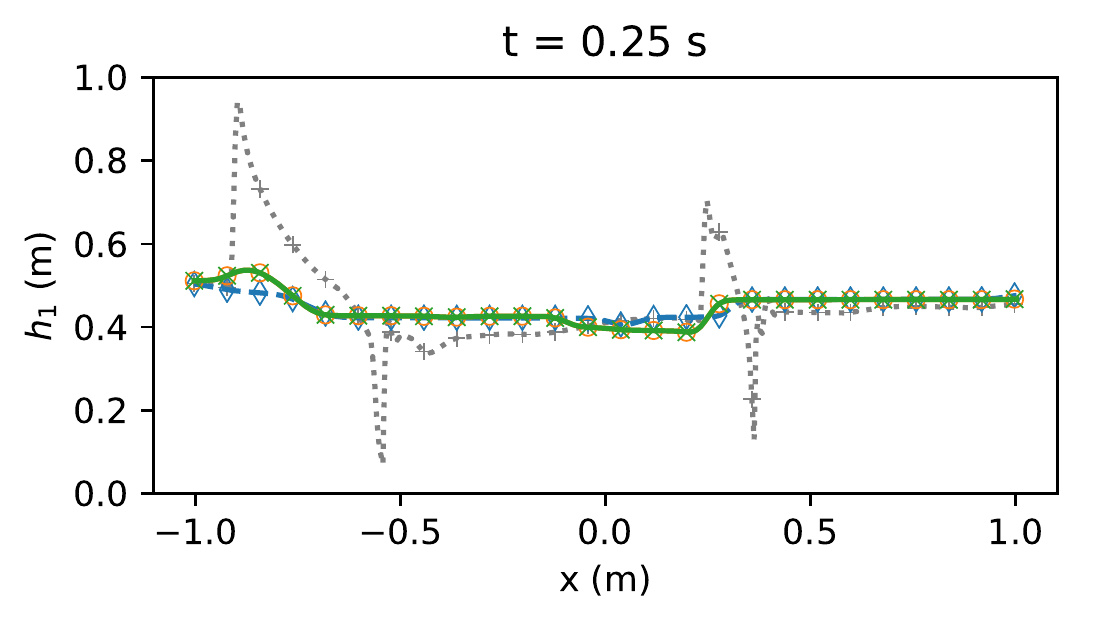}
	\hfill	
	\includegraphics[width=6.5cm]{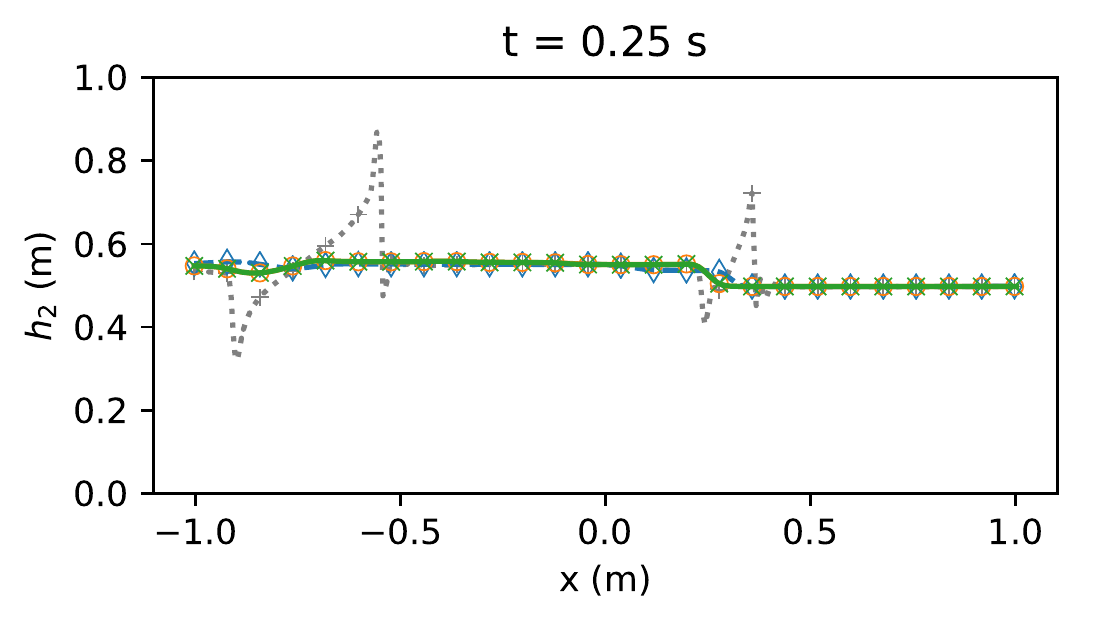}
	\vfill
	\includegraphics[width=6.5cm]{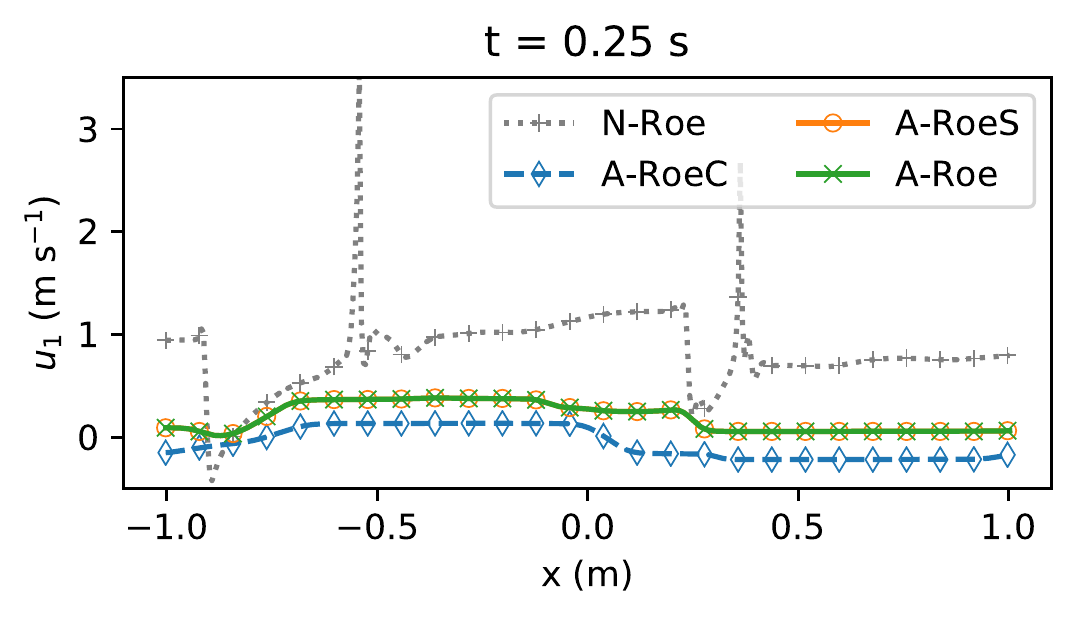}
	\hfill	
	\includegraphics[width=6.5cm]{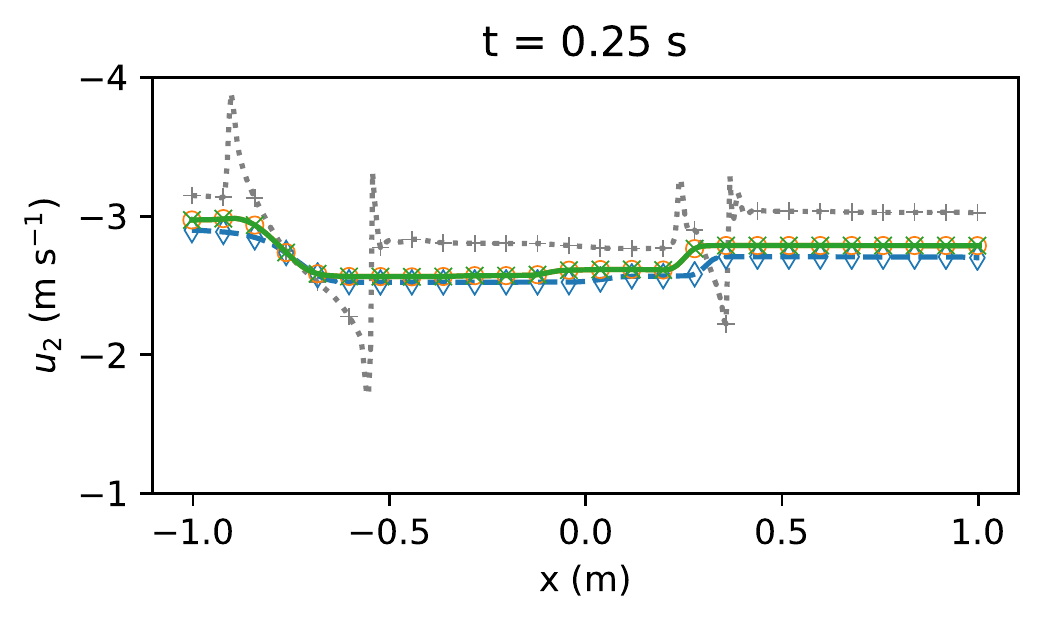}
	\vfill
	\includegraphics[width=6.5cm]{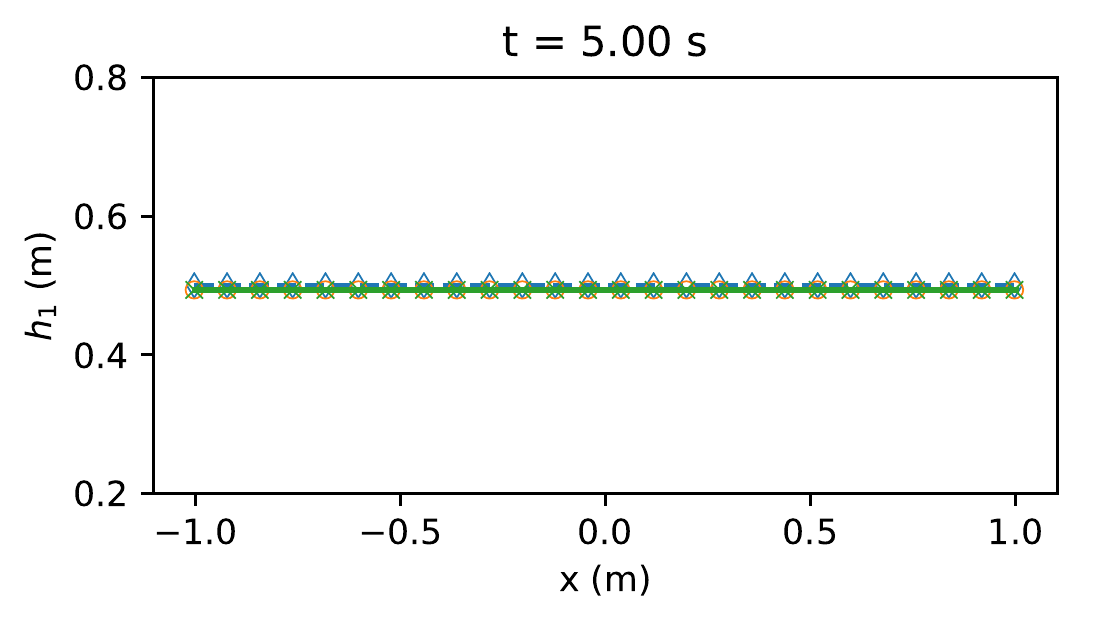}
	\hfill	
	\includegraphics[width=6.5cm]{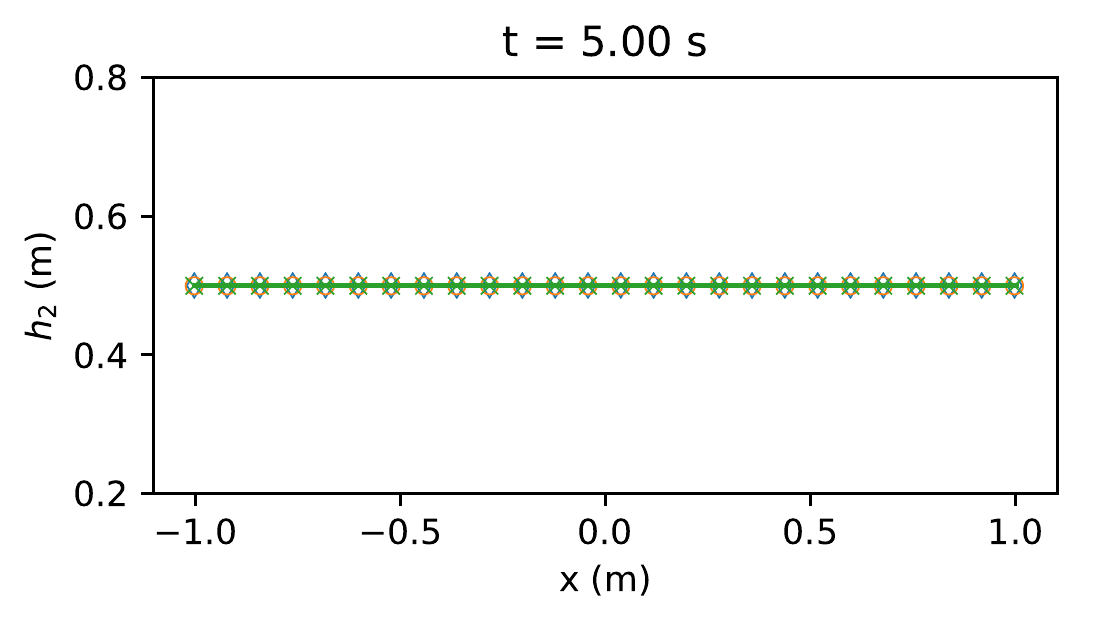}
	\vfill
	\includegraphics[width=6.5cm]{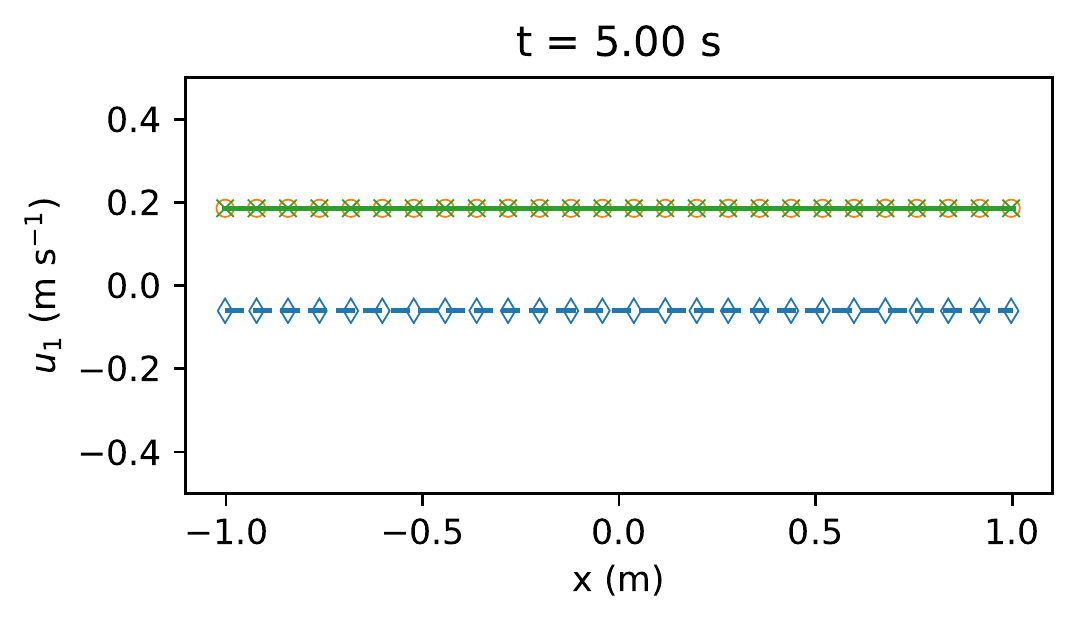}
	\hfill	
	\includegraphics[width=6.5cm]{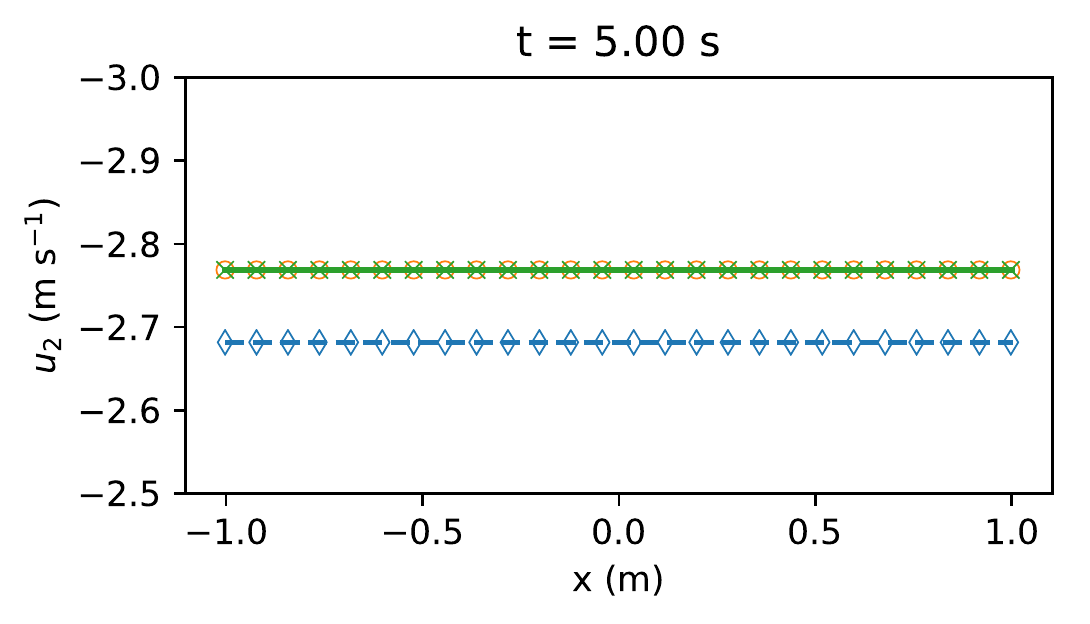}
	\caption{Test VI: Upper and lower layer depths and velocities obtained by N-Roe without hyperbolicity correction and A-Roe method with three different implementations of the hyperbolicity correction, at $t=0.25$ s and $t=5.0$ s for $\Delta x = 1/200$ m}
	\label{fig:Results_Test06}
\end{figure}

Figure \ref{fig:Results_Test06} shows the upper and lower layer depths and velocities at $t=0.25$ s and at $t=5.0$ s. First of all, strong oscillations are noticeable for the N-Roe method without the hyperbolicity correction, which blows-up after $t=0.25$. The results computed by the A-Roe scheme are practically the same as the results computed by the A-RoeS scheme. However, differences are noticeable between the two iterative schemes and the approximate A-RoeC scheme, because of a different way in which the additional friction is computed. As a consequence, the corrected velocities are significantly lower; at the end of the simulation, $u_1 = -0.06$ m s$^{-1}$ is computed by the approximate A-RoeC scheme, in comparison to $u_1 = +0.19$ m s$^{-1}$ computed by the other two iterative schemes (A-Roe and A-RoeS). Not only are velocities lower, but the A-RoeC scheme changes the flow structure, which becomes unidirectional. On the other hand, A-Roe and A-RoeS compute the optimal friction and preserve the correct flow directions. The same behaviour of approximate and iterative hyperbolicity correction was found by \cite{sarno2017some}.

\begin{figure}[htbp]
	\center
	\includegraphics[width=8cm]{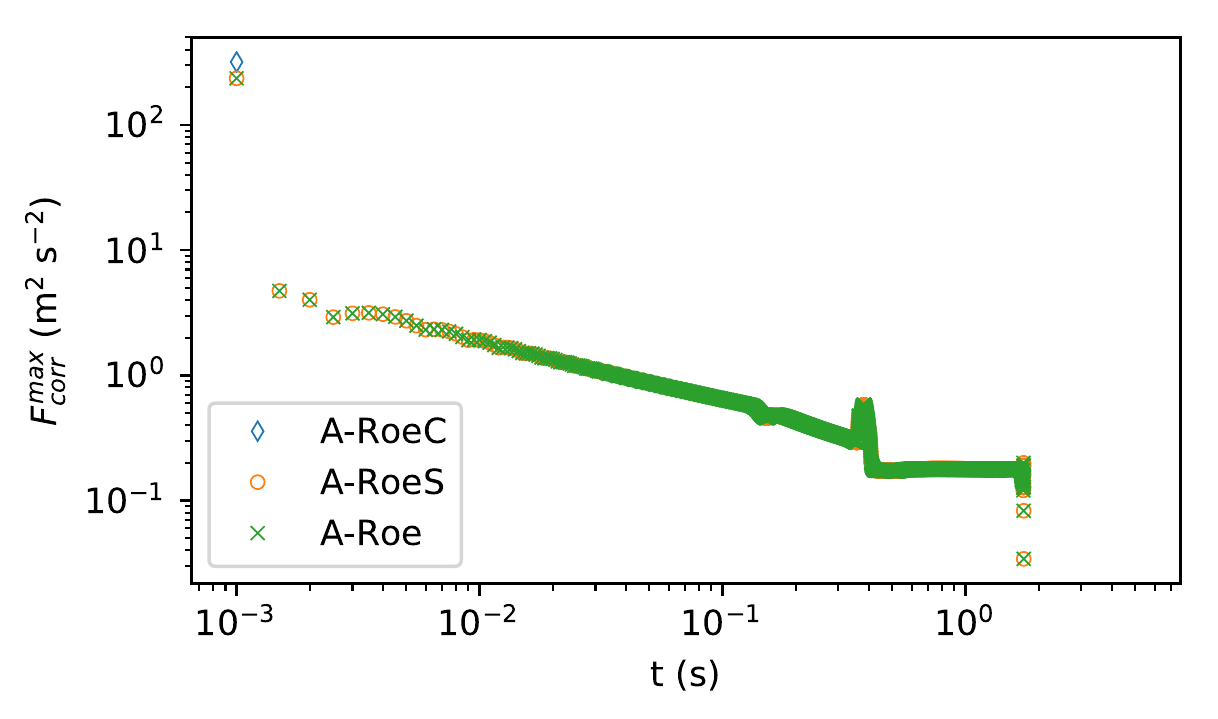}
	\caption{Test VI: Comparison of maximum values of $F_{corr}^{max}$ at each time step, computed by different implementations of the hyperbolicity correction}
	\label{fig:F_Test06}
\end{figure}

To assess the behaviour of three correction algorithms in more detail, the temporal changes of $F_{corr}^{max}$ are shown in Fig~\ref{fig:F_Test06}. As in the previous example, a relatively high $F_{corr}^{max}$ is added in the first time step, namely 235.6 m$^2$ s$^{-2}$ by the A-Roe and A-RoeS scheme, and 318.3 m$^2$ s$^{-2}$ by the A-RoeC scheme with an approximate correction. Due to an overestimated $F_{corr}^{max}$ by the A-RoeC scheme, there is no need for further corrections in the subsequent time steps. However, the flow structure is compromised. On the other hand, $F_{corr}^{max}$ computed by the iterative schemes A-Roe and A-RoeS, is reduced to under 10 m$^2$ s$^{-2}$ after the second time step, and under 1 m$^2$ s$^{-2}$ after $t=0.1$ s. The results also confirm that the proposed A-Roe scheme provides almost identical values as the A-RoeS during the entire simulation.

The CPU time of the complete simulation has been found to be 21.3 s for A-RoeC, 46.7 s for the iterative A-RoeS, and 36.2 s for the proposed A-Roe scheme. Again, the results show that the proposed algorithm A-Roe is noticeably faster than A-RoeS.

\subsection{Test VII: Exchange flow over smooth bottom topography and with hyperbolic initial conditions}

A final case of a two-layer exchange flow through a rectangular channel with non-flat smooth bottom topography is now considered to carefully compare the influence of the A-Roe hyperbolicity correction against frictionless solution (N-Roe scheme with a real Jordan decomposition). In this example, a transcritical flow eventually develops with an internal shock.

The spatial domain is set to [0, 10], and the bed elevation is defined by the following function:
\begin{equation}
b(x) = 0.5 \exp \left( - (x-5)^2 \right)
\end{equation}
whereas, the initial condition is defined by:
\begin{equation}
h_2(x,0) = 0.8 \textrm{ m} - b(x),
\quad
h_1(x,0) = 1.2 \textrm{ m} - h_2(x,0) - b(x)
\end{equation}
\begin{equation}
u_1(x,0) = 0.15 \textrm{ m s}^{-1},
\quad
u_2(x,0) = -0.2 \textrm{ m s}^{-1}
\end{equation}
In contrast to previous two examples, here we have hyperbolic initial conditions.
Non-reflective conditions are imposed at the boundaries, and the relative density ratio is set to $r=0.98$.
All of the computations are performed using a small grid size $\Delta x = 1/100$ m and $CFL = 0.7$. 

\begin{figure}[htbp]
	\center
	\includegraphics[width=6.5cm]{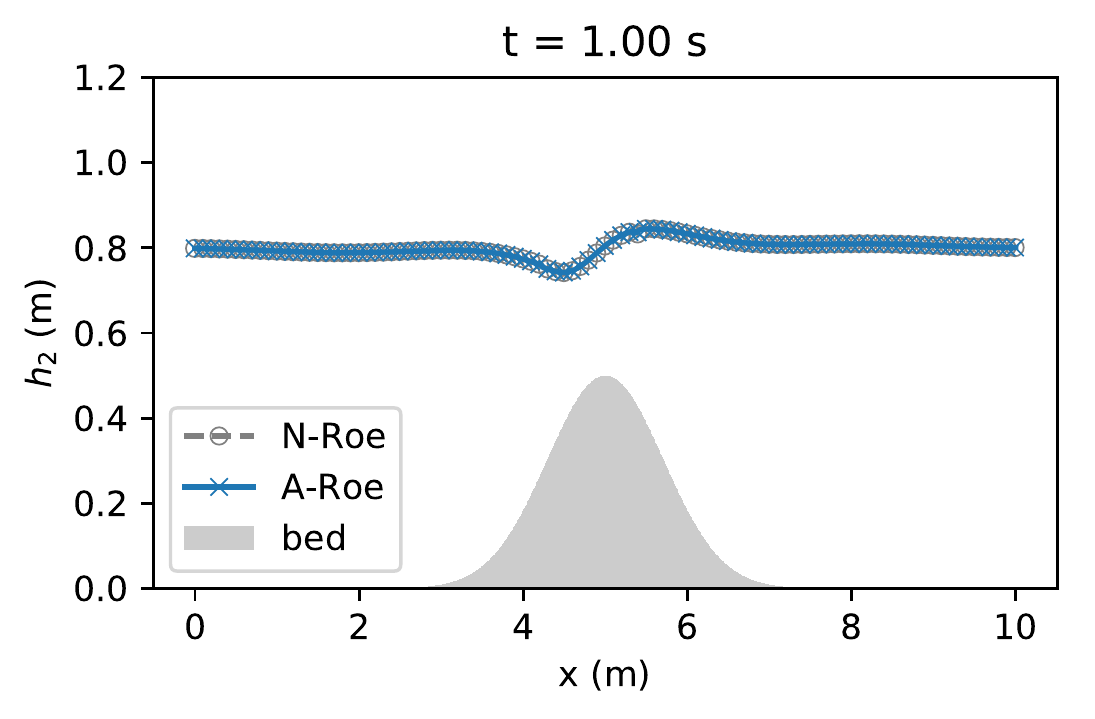}
	\hfill	
	\includegraphics[width=6.5cm]{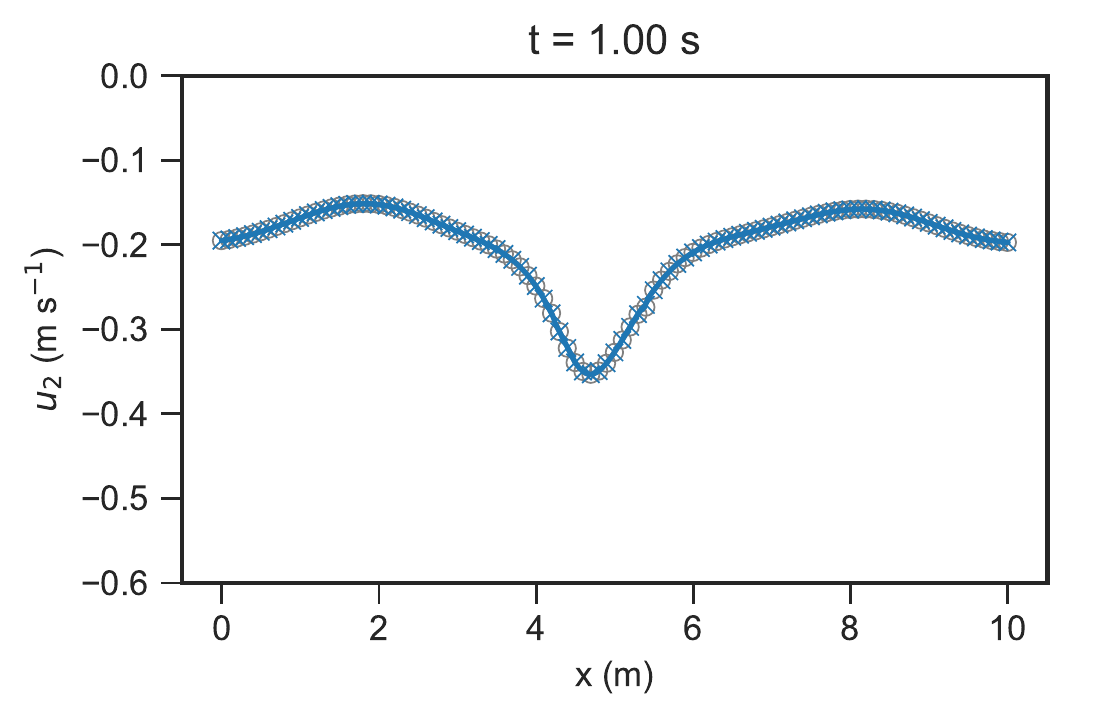}
	\vfill
	\includegraphics[width=6.5cm]{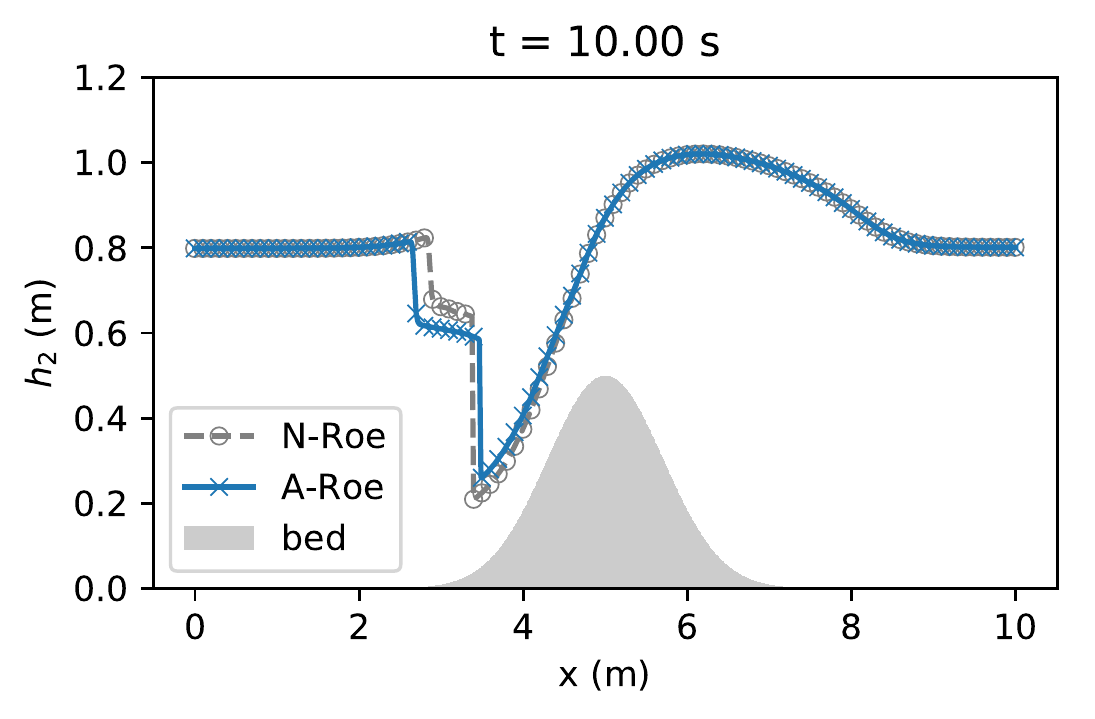}
	\hfill	
	\includegraphics[width=6.5cm]{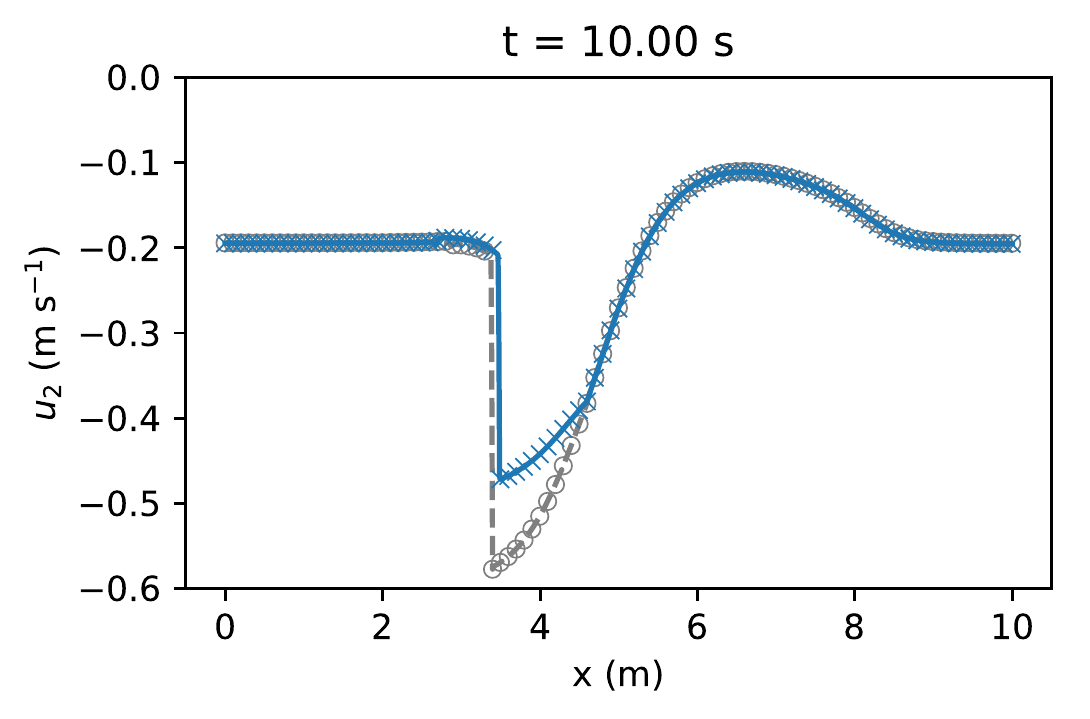}
	\vfill
	\includegraphics[width=6.5cm]{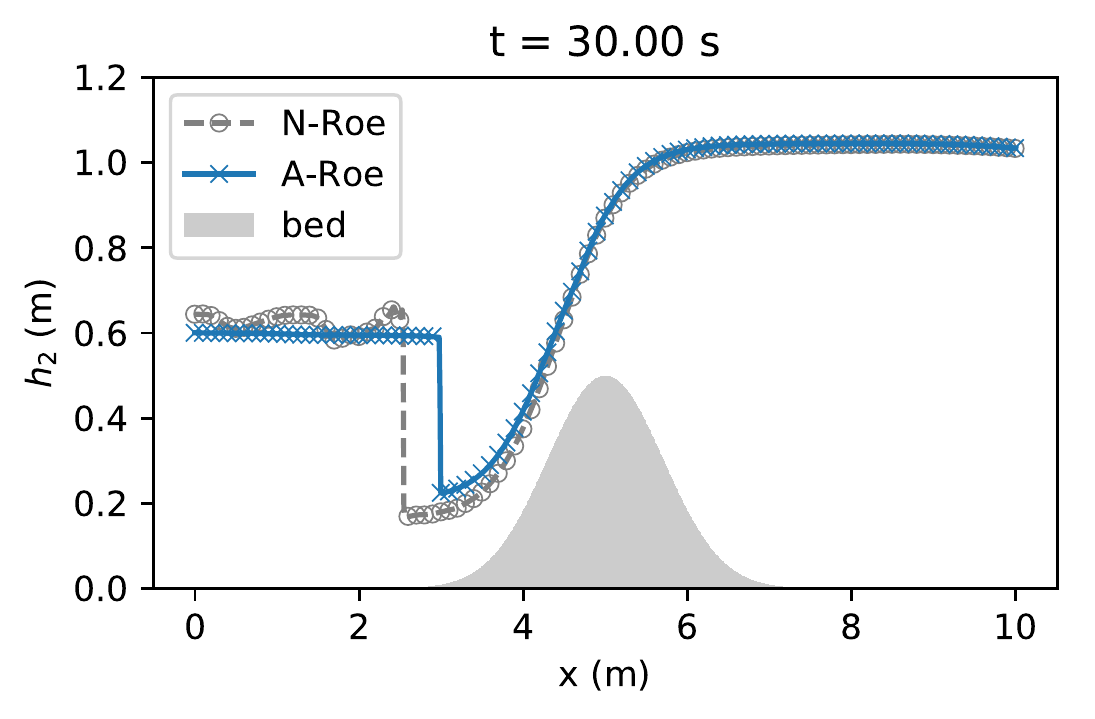}
	\hfill	
	\includegraphics[width=6.5cm]{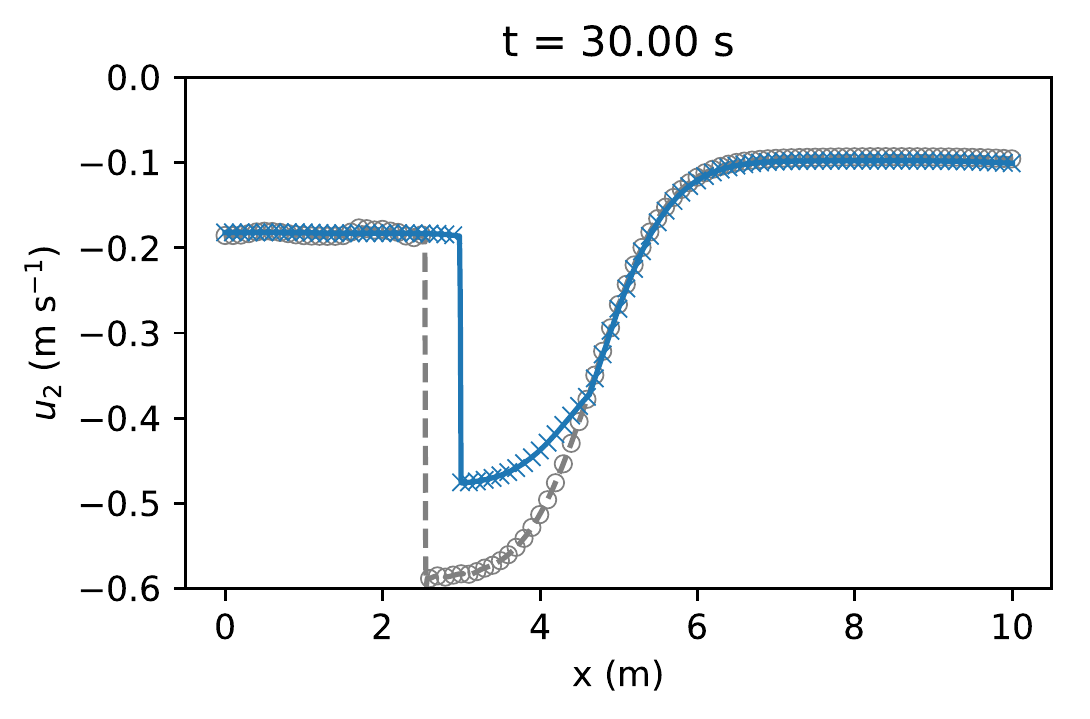}
	\caption{Test VII: Evolution of the interface and lower layer velocities obtained by N-Roe without hyperbolicity correction and A-Roe with hyperbolicity correction, at $t=$ 1, 10 and 30 s, and $\Delta x = 1/100$ m}
	\label{fig:Results_Test07}
\end{figure}

Figure \ref{fig:Results_Test07} shows the evolution of the interface and lower layer velocity at $t=$ 1, 10 and 30 s. The results computed by the A-Roe method with the proposed hyperbolicity correction are almost identical to the N-Roe scheme without the correction up until $t=1.0$ s, when non-hyperbolic conditions develop. Both schemes are able to compute the results until steady-state conditions are reached at $t=30$ s. However, N-Roe scheme develops spurious oscillations which grow in time.
The CPU time of the complete simulation has been found to be 190 s for A-Roe, and 239 s for the N-Roe scheme.

\begin{figure}[htbp]
	\center
	\includegraphics[width=6.5cm]{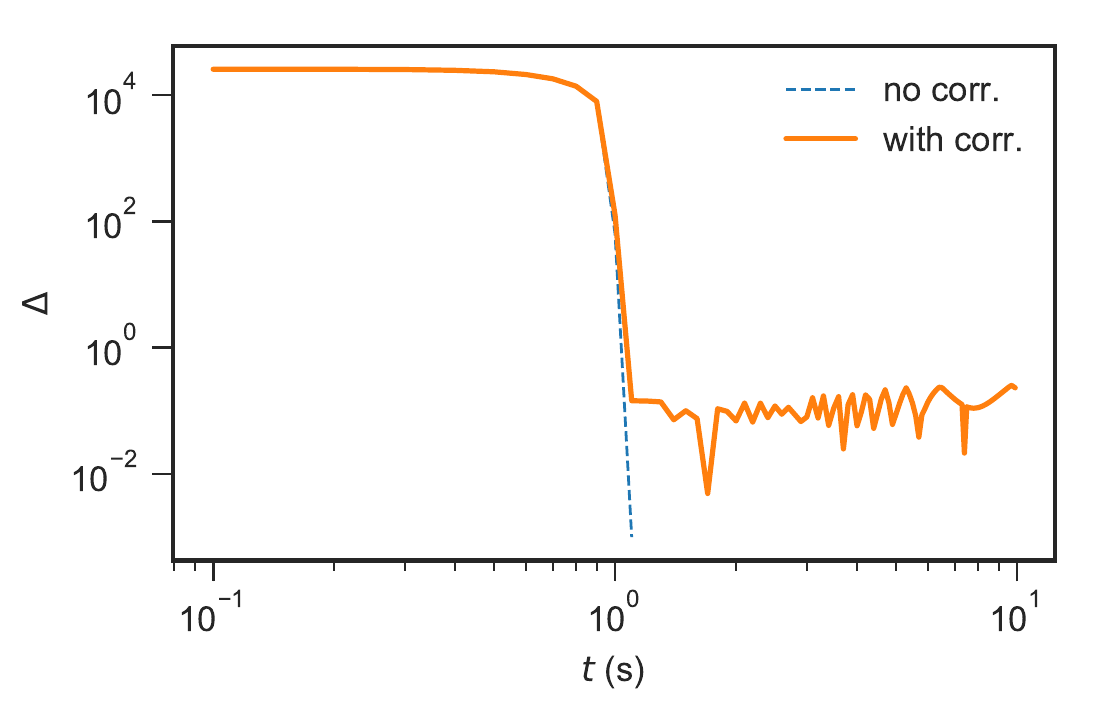}
	\hfill
	\includegraphics[width=6.5cm]{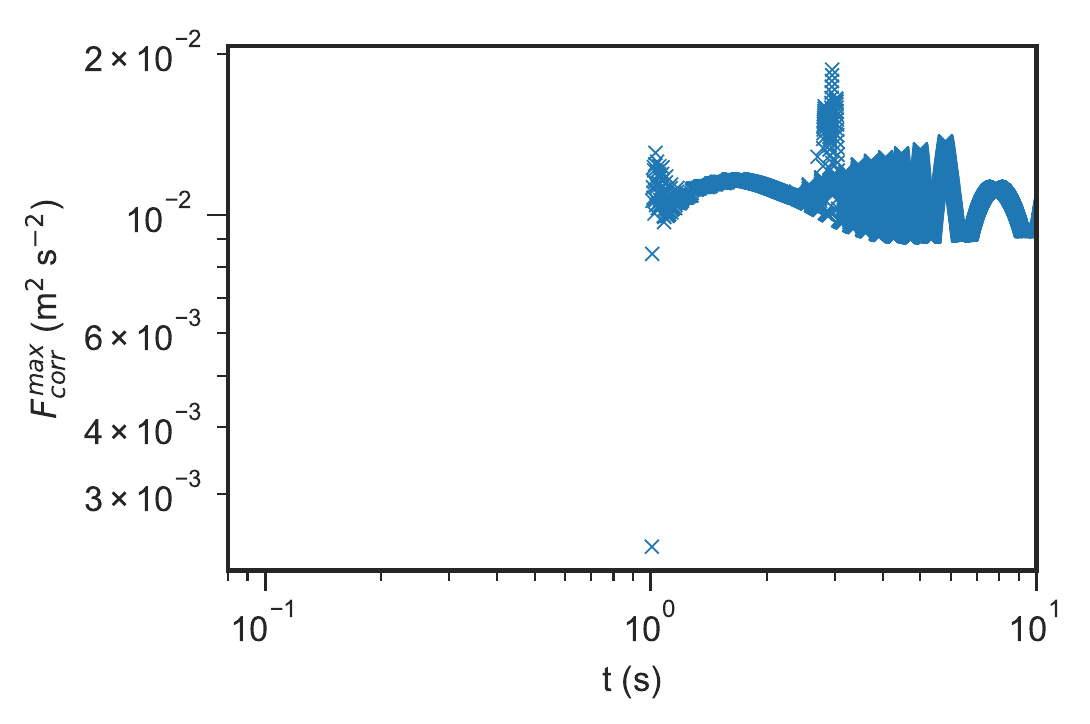}
	\vfill
	\includegraphics[width=6.5cm]{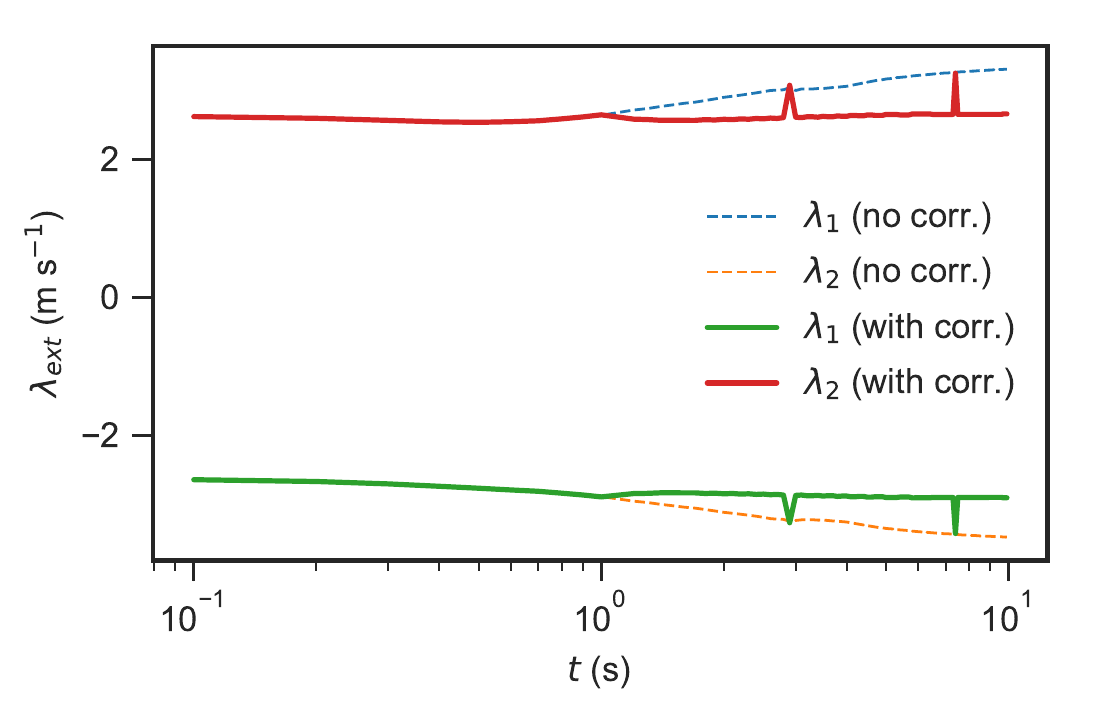}
	\hfill
	\includegraphics[width=6.5cm]{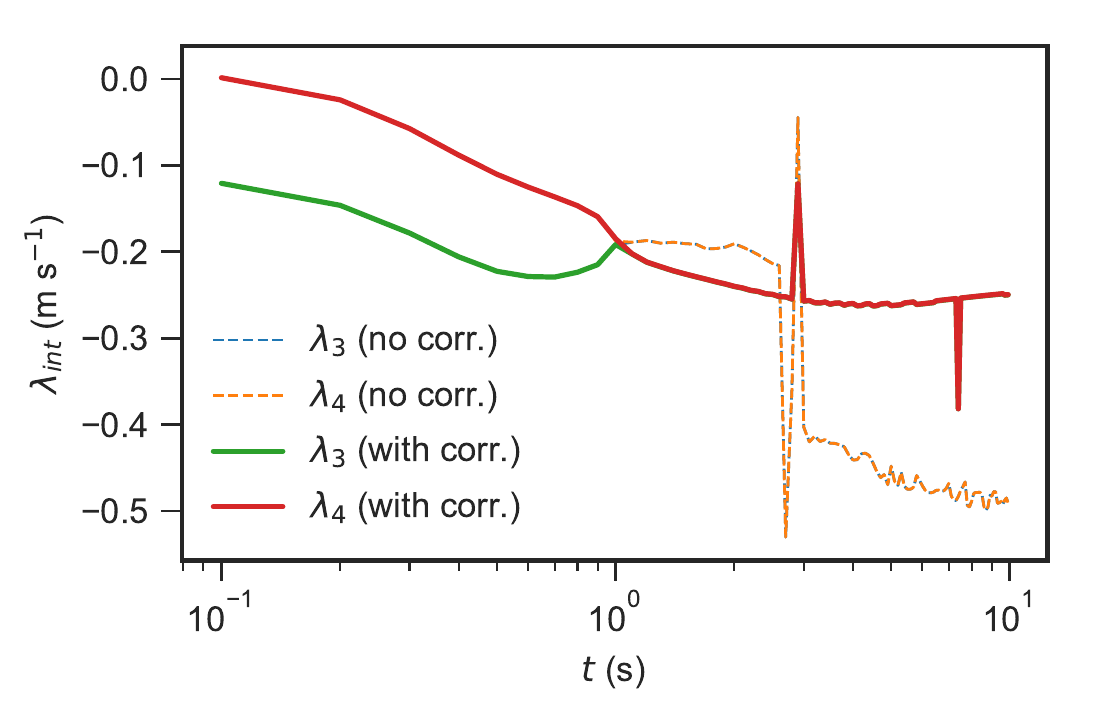}
	\caption{Test VII: Evolution of the discriminant, $F_{corr}^{max}$, and the external and internal eigenvalues}
	\label{fig:F_Test07}
\end{figure}

To examine the behaviour of the proposed iterative correction algorithm in more detail, Fig.~\ref{fig:F_Test07} shows the evolution of the discriminant $\Delta$, computed correction $F_{corr}^{max}$, as well as external and internal eigenvalues (waves), when The N-Roe scheme without correction and the A-Roe scheme with hyperbolicity correction are applied. 

Since the initial conditions are in a hyperbolic state, at the beginning of the simulation, $\Delta$ is positive and of the same order for both schemes (Fig.~\ref{fig:F_Test07}). At $t=1.0$ s the velocity difference increases and a loss of hyperbolicity occurs. From this point forward, the numerical scheme without correction produces negative $\Delta$, although real eigenvalues are recovered through real Jordan decomposition. On the other hand, A-Roe with hyperbolicity correction maintains a positive discriminant by applying extra friction of the order $\sim 10^{-2}$ m$^2$ s$^{-1}$.

Similarly, the external and internal eigenvalues are identical up to $t=1.0$ s for both schemes (Fig.~\ref{fig:F_Test07}). After this point, when a loss of hyperbolicity occurs, the A-Roe scheme produces different results from the N-Roe scheme. Without correction, the external eigenvalues grow, and the internal eigenvalues collapse to a single value which increases over time. On the other hand, hyperbolicity correction implemented in the A-Roe scheme maintains constant external wave velocities, and, although it appears that double internal eigenvalues are also present here, the correction algorithm actually preserves some small difference between them (due to the fact that $\Delta$ is always larger than zero).

\section{Conclusion}

In this study, a new implementation of the Roe scheme for solving two-layer shallow-water equations has been introduced. The proposed method is based on an analytical formulation for the eigenstructure of the quasi-Jacobian matrix. This analytical expression is derived from the explicit Ferrari's solution to the characteristic polynomial, which is a significantly faster alternative to numerical eigensolvers. The analysis of the accuracy and computational speed of the closed-form quartic solver, presented in this paper, suggests that it can be considered as reliable as numerical eigenstructure solvers and up to 20 times faster. 

The efficiency of the proposed A-Roe scheme was also examined in terms of its accuracy and computational speed and compared to the Roe scheme in which the viscosity matrix is computed numerically (N-Roe), two incomplete Riemann solvers (Lax-Friedrich and GFORCE), as well as two PVM schemes (PVM-Roe and IFCP). For a fixed computational grid (both in space and time), the proposed A-Roe scheme is up to 4 times faster than the N-Roe scheme, while maintaining the same accuracy of the solution. The A-Roe scheme is also faster than the PVM-Roe scheme (up to 83\%). In comparison to the LF and GFORCE, the A-Roe scheme is somewhat computationally slower (30-60\%), but significantly more accurate. When compared to the IFCP scheme, the A-Roe is slightly more accurate with very similar computational speeds.

In addition to its computational speed, a significant advantage of the A-Roe method is an integrated correction algorithm for keeping the solutions of two-layer shallow-water equations inside the hyperbolic domain. It ensures that only real eigenvalues are considered in the process of the Roe linearisation. The iterative algorithm uses the Illinois solver and is based on the numerical treatment for the loss of hyperbolicity proposed by \cite{sarno2017some}, which in contrast to \cite{castro2012hyperbolicity} is applicable for any density ratio. The only difference is that the hyperbolicity loss prediction and correction are based on the sign of the discriminant of a resolvent cubic equation and that both actions are implemented at the intermediate step when the eigenstructure is calculated. 
Numerical tests of exchange flow show that the proposed algorithm is as accurate as the iterative approach by \cite{sarno2017some} regardless of the density ratio, but requires 25-30\% less computational time. The approximate algorithm by \cite{castro2012hyperbolicity} is 25-60\% faster than the proposed one; however, in the case of small density ratios it may fail to preserve the exchange flow structure and produce unphysical results. 

To conclude, the A-Roe scheme proves to be an efficient alternative to a numerical implementation of the Roe scheme tested here for two-layer shallow-water flows; it is as accurate but computationally much faster. The proposed scheme gives more precise results for all values of $r$ and therefore it has a wider range of possible applications in comparison to approximate expressions.
The efficiency of the proposed scheme should not depend on a specific problem and it should increase with the number of cells.
Although the A-Roe method has been tested here only for two-layer shallow-water flows, it can easily be applied to some other non-conservative hyperbolic systems defined by four coupled partial differential equations, such as two-phase fluids.
Furthermore, the extension to two-dimensional problems or higher-order schemes is straightforward following the same approach as for any Roe scheme.

\appendix

\section{Analytic solution to the eigenstructure}

\subsection{Solution to a quartic equation}
\label{sec:A1}

Let us consider a general normalized 4th order polynomial equation (quartic)
\begin{equation}
x^4 + ax^3 + bx^2 + cx + d = 0.
\label{eq:quartic}
\end{equation}
To find the analytical solution to roots of Eq.~(\ref{eq:quartic}), first the cubic term $x^3$ is eliminated and the general polynomial is converted into a so-called \textit{depressed quartic} by a change of variables. Following Ferrari's method \citep{abramowitz1972}, a substitution $x = y - a/4$ is introduced, which gives a depressed polynomial
\begin{equation}
y^4 + py^2 + qy + r = 0,
\label{eq:depressed}
\end{equation}
where
\begin{equation}
p = b - 6\left(\frac{a}{4}\right)^2,
\end{equation}
\begin{equation}
q = c - 2b\left(\frac{a}{4}\right) + 8\left(\frac{a}{4}\right)^3,
\end{equation}
\begin{equation}
r = d - c \left(\frac{a}{4}\right) + b\left(\frac{a}{4}\right)^2 -3\left(\frac{a}{4}\right)^4. 
\end{equation}

The depressed polynomial can be rewritten as
\begin{equation}
\left(y^2 + \frac{p}{2}\right)^2 = -qy + \frac{p^2}{4} - r.
\label{eq:derpessed1}
\end{equation}
Next, expression $2zy^2 + zp + z^2$ is added to both sides of Eq.~(\ref{eq:derpessed1}), which after some regrouping gives
\begin{equation}
\left( y^2 + \frac{p}{2} + z \right)^2 = 2zy^2 -qy + z^2 + zp + \frac{p^2}{4} - r.
\label{eq:depressed2}
\end{equation}
When $z$ is chosen to be any non-zero root $z_0$ of the so-called \textit{resolvent cubic} equation
\begin{equation}
8z^3 + 8pz^2 + (2p^2 - 8r)z - q^2 = 0,
\label{eq:cubic1}
\end{equation}
the right-hand side of Eq.~(\ref{eq:depressed2}) can be written as a perfect square; therefore, Eq.~(\ref{eq:depressed2}) becomes
\begin{equation}
\left( y^2 + \frac{p}{2} + z_0 \right)^2 = \left(y\sqrt{2z_0} - \frac{q}{2\sqrt{2z_0}}\right)^2.
\label{eq:depressed3}
\end{equation}
And finally, Eq.~(\ref{eq:depressed3}) can be written as a factorized quadratic equation
\begin{equation}
\left( y^2 + \sqrt{2z_0}y + \frac{p}{2} + z_0 - \frac{q}{2\sqrt{2z_0}}\right)\left( y^2 - \sqrt{2z_0}y + \frac{p}{2} + z_0 + \frac{q}{2\sqrt{2z_0}}\right) = 0,
\end{equation}
which is easily solved by a quadratic formula.

Therefore, the solutions to the roots of the general quartic Eq.~(\ref{eq:quartic}) are given by
\begin{equation}
x_{1,2} = -\frac{a}{4} - \frac{1}{2}\sqrt{2z_0} \pm \frac{1}{2}\sqrt{- \left( 2p + 2z_0 - \frac{2q}{\sqrt{2z_0}} \right)},
\label{eq:solution1a}
\end{equation}
\begin{equation}
x_{3,4} = -\frac{a}{4} + \frac{1}{2}\sqrt{2z_0} \pm \frac{1}{2}\sqrt{- \left( 2p + 2z_0 + \frac{2q}{\sqrt{2z_0}} \right)}.
\label{eq:solution2b}
\end{equation}

For a general normalized 3rd order polynomial equation (cubic)
\begin{equation}
x^3 + \alpha x^2 + \beta x + \gamma = 0,
\label{eq:cubic}
\end{equation}
a real solution is given by Cardano's formula \citep{abramowitz1972}
\begin{equation}
x_0 = s_1 + s_2 - \frac{\alpha}{3},
\label{eq:x0}
\end{equation}
with
\begin{equation}
s_1 = \sqrt[3]{R + \sqrt{R^2 + Q^3 }},
\label{eq:S}
\end{equation}
\begin{equation}
s_2 = \sqrt[3]{R - \sqrt{R^2 + Q^3}},
\end{equation}
where
\begin{equation}
Q = \frac{3 \beta - \alpha^2}{9},
\end{equation}
\begin{equation}
R = \frac{9\alpha \beta - 27 \gamma - 2 \alpha^3}{54}.
\end{equation}
Note that Eq.~(\ref{eq:x0}) may be also written as either $x_0 = s_1 - \frac{Q}{s_1} -\frac{\alpha}{3}$ or  $x_0 = s_2 - \frac{Q}{s_2} -\frac{\alpha}{3}$, which is computationally more convenient since only $s_1$ or $s_2$ needs to be computed. Furthermore, if $Q=0$ then we have to choose $s_1$ if $R>0$ and $s_2$ if $R<0$ to get non-zero value. 
Therefore, the solution to the resolvent cubic Eq.~(\ref{eq:cubic1}) is given as
\begin{equation}
z_0 = s - \frac{Q}{s} -\frac{p}{3} ,
\end{equation}
where
\begin{equation}
s = \sqrt[3]{R + \textrm{sign}(R) \sqrt{R^2 + Q^3}},
\end{equation}
\begin{equation}
Q = \frac{3 (p^2/4 - r) - p^2}{9} = \frac{-b^2 - 12d + 3ac}{36},
\end{equation}
\begin{equation}
R = \frac{9 p (p^2/4 - r) + 27 q^2/8 - 2p^3}{54} = \frac{27a^2d - 9abc + 2b^3 - 72bd + 27c^2}{432}.
\end{equation}

To eliminate redundant divisions and optimize computation of Eq.~(\ref{eq:solution1a}) and (\ref{eq:solution2b}), the root of the resolvent cubic equation is expressed via
\begin{equation}
2z_0 =  \frac{1}{3} \left( S + \frac{\Delta_0}{S}  - 2p \right),
\label{eq:cubic_Solution}
\end{equation}
where
\begin{equation}
S = 6s =  \sqrt[3]{\frac{\Delta_1 + \textrm{sign}(\Delta_1) \sqrt{\Delta_1^2 - 4\Delta_0^3 }}{2}},
\label{eq:S_sol}
\end{equation}
\begin{equation}
\Delta_0 = - 36 Q = b^2 + 12d - 3ac, 
\end{equation}
\begin{equation}
\Delta_1 = 432 R = 27a^2d - 9abc + 2b^3 - 72bd + 27c^2.
\end{equation}

Note that $\Delta_1^2 - 4 \Delta_0^3 = -\frac{27}{64} \mathcal{D}_{cubic} = -27 \mathcal{D}_{quartic}$, which is a much simpler expression for the discriminant of the resolvent cubic equation $\mathcal{D}_{cubic}$ and especially the discriminant of the quartic equation $\mathcal{D}_{quartic}$ given by Eq.~(\ref{eq:discriminant}). Therefore, if $\Delta_1^2 - 4 \Delta_0^3 < 0$, three resolvent cubic roots are all real and the quartic roots are either all complex or all real. 
In this case, Eq.~(\ref{eq:cubic_Solution}) can be solved trigonometrically \citep{lambert1906}, which is computationally faster than computing the cube root required in Eq.~(\ref{eq:S_sol}):
\begin{equation}
z_0 =  \frac{1}{3} \left( \sqrt{\Delta_0} \cos \frac{\phi}{3}  - p \right),
\label{eq:cubic_trig}
\end{equation}
where
\begin{equation}
\phi = \arccos \left( \frac{\Delta_1}{2 \sqrt{\Delta_0^3}}\right).
\end{equation}

To summarize, the real solution to the quartic equation can be simplified as follows:
\begin{equation}
x_{1,2} = \frac{ - \frac{a}{2} \pm \sqrt{Z} - \sqrt{- A - Z \mp \frac{B}{\sqrt{Z}} } }{2} ,
\label{eq:A_solution1}
\end{equation}
\begin{equation}
x_{3,4} = \frac{- \frac{a}{2} \pm \sqrt{Z} + \sqrt{- A - Z \mp \frac{B}{\sqrt{Z}} }}{2} ,
\label{eq:A_solution2}
\end{equation}
where
\begin{equation}
Z =  2z_0 = \frac{1}{3} \left( 2 \sqrt{\Delta_0} \cos \frac{\phi}{3}  - A \right),
\label{eq:A_Zcoeff}
\end{equation}
with
\begin{equation}
A = 2p = 2b - \frac{3a^2}{4},
\label{eq:A_Acoeff}
\end{equation}
\begin{equation}
B = 2q = 2c - ab + \frac{a^3}{4} .
\label{eq:A_Bcoeff}
\end{equation}

\subsection{Explicit solution to the inverse of the eigenvector matrix}
\label{sec:A2}

Inverse of matrix $\bb{K}$, whose columns are eigenvectors, is derived from
\begin{equation}
	\bb{K}^{-1} = \frac{1}{\textrm{det}(\bb{K})}\textrm{adj}(\bb{K}),
\end{equation}
which, after some regrouping and simplifications, gives
\begin{equation}
\bb{K}^{-1} =
	\begin{bmatrix}
	\bb{k}_1 & \bb{k}_2 & \bb{k}_3 & \bb{k}_4,
	\end{bmatrix}^{T}
\end{equation}
with
\begin{equation}
\bb{k}_k =
\begin{Bmatrix}
\dfrac{(c_1^2 - u_1^2)\delta_k + \xi_{k}}{\zeta_{k}} & 
-\dfrac{c_1^2 - u_1^2 - 2u_1\delta_k + \kappa_{k}}{\zeta_{k}}	& 
\dfrac{c_1^2 \delta_k}{\zeta_{k}} 			& 
-\dfrac{c_1^2 }{\zeta_{k}} ,
\end{Bmatrix}
\label{eq:Kinv}
\end{equation}
$k=1,..,4$, where
%
%\begin{equation}
%\begin{aligned}
%\delta_{1} = \lambda_2 + \lambda_3 + \lambda_4 - 2u_1\\
%\delta_{2} = \lambda_1 + \lambda_3 + \lambda_4 - 2u_1 \\
%\delta_{3} = \lambda_1 + \lambda_2 + \lambda_4 - 2u_1 \\
%\delta_{4} = \lambda_1 + \lambda_2 + \lambda_3 - 2u_1
%\end{aligned}
%\end{equation}
%
\begin{equation}
\delta_k = \sum_{j=1,j\neq k}^{4}\lambda_j - 2u_1,
\end{equation}
%
%\begin{equation}
%\begin{aligned}
%\xi_{1} = \lambda_2 \lambda_3 \lambda_4 \\
%\xi_{2} = \lambda_1 \lambda_3 \lambda_4 \\
%\xi_{3} = \lambda_1 \lambda_2 \lambda_4 \\
%\xi_{4} = \lambda_1 \lambda_2 \lambda_3
%\end{aligned}
%\end{equation}
%
\begin{equation}
\xi_k = \prod_{j=1,j\neq k}^{4}\lambda_j ,
\end{equation}
%
%\begin{equation}
%\begin{aligned}
%\kappa_{1} = \lambda_2 \lambda_3  + \lambda_2 \lambda_4  + \lambda_3 \lambda_4 \\
%\kappa_{2} = \lambda_1 \lambda_3  + \lambda_1 \lambda_4  + \lambda_3 \lambda_4 \\
%\kappa_{3} = \lambda_1 \lambda_2  + \lambda_1 \lambda_4  + \lambda_2 \lambda_4 \\
%\kappa_{4} = \lambda_1 \lambda_2  + \lambda_1 \lambda_3  + \lambda_2 \lambda_3
%\end{aligned}
%\end{equation}
%
\begin{equation}
\kappa_k = \sum_{j=1,j\neq k}^{4} \prod_{i=1,i\neq j,k}^{4} \lambda_i,
\end{equation}
%
%\begin{equation}
%\begin{aligned}
%\zeta_{1} = (\lambda_2 - \lambda_1)(\lambda_3 - \lambda_1)(\lambda_4 - \lambda_1) \\
%\zeta_{2} = (\lambda_1 - \lambda_2)(\lambda_3 - \lambda_2)(\lambda_4 - \lambda_2) \\
%\zeta_{3} = (\lambda_1 - \lambda_3)(\lambda_2 - \lambda_3)(\lambda_4 - \lambda_3) \\
%\zeta_{4} = (\lambda_1 - \lambda_4)(\lambda_2 - \lambda_4)(\lambda_3 - \lambda_4) 
%\end{aligned}
%\end{equation}
%
\begin{equation}
\zeta_k = \prod_{j=1,j\neq k}^{4}(\lambda_j -\lambda_k).
\end{equation}

\section*{Acknowledgements}

This work has been fully supported by the University of Rijeka under the project number 17.06.2.1.02 (River-Sea Interaction in the Context of Climate Change).

%\section*{References}
\small
\bibliographystyle{elsarticle-harv} 
\bibliography{Qiqqa2BibTexExport}

\begin{thebibliography}{48}
\expandafter\ifx\csname natexlab\endcsname\relax\def\natexlab#1{#1}\fi
\expandafter\ifx\csname url\endcsname\relax
  \def\url#1{\texttt{#1}}\fi
\expandafter\ifx\csname urlprefix\endcsname\relax\def\urlprefix{URL }\fi

\bibitem[{Abgrall and Karni(2009)}]{abgrall2009two}
Abgrall, R., Karni, S., 2009. Two-layer shallow water system: a relaxation
  approach. SIAM Journal on Scientific Computing 31~(3), 1603--1627.
\newline\urlprefix\url{https://doi.org/10.1137/06067167X}

\bibitem[{Abramowitz and Stegun(1965)}]{abramowitz1972}
Abramowitz, M., Stegun, I.~A., 1965. Handbook of mathematical functions with
  formulas, graphs, and mathematical tables. Dover Publications, New York.

\bibitem[{Adduce et~al.(2011)Adduce, Sciortino, and
  Proietti}]{adduce2011gravity}
Adduce, C., Sciortino, G., Proietti, S., 2011. {Gravity currents produced by
  lock exchanges: Experiments and simulations with a two-layer shallow-water
  model with entrainment}. Journal of Hydraulic Engineering 138~(2), 111--121.
\newline\urlprefix\url{https://doi.org/10.1061/(ASCE)HY.1943-7900.0000484}

\bibitem[{Anderson et~al.(1999)Anderson, Bai, Bischof, Blackford, Demmel,
  Dongarra, Du~Croz, Greenbaum, Hammarling, McKenney, and Sorensen}]{lapack}
Anderson, E., Bai, Z., Bischof, C., Blackford, S., Demmel, J., Dongarra, J.,
  Du~Croz, J., Greenbaum, A., Hammarling, S., McKenney, A., Sorensen, D., 1999.
  {LAPACK} Users' Guide, 3rd Edition. Society for Industrial and Applied
  Mathematics, Philadelphia, PA.

\bibitem[{Bermudez and Vazquez(1994)}]{bermudez1994upwind}
Bermudez, A., Vazquez, M.~E., 1994. Upwind methods for hyperbolic conservation
  laws with source terms. Computers \& Fluids 23~(8), 1049--1071.
\newline\urlprefix\url{https://doi.org/10.1016/0045-7930(94)90004-3}

\bibitem[{Bouchut and Zeitlin(2010)}]{bouchut2010robust}
Bouchut, F., Zeitlin, V., 2010. A robust well-balanced scheme for multi-layer
  shallow water equations. Discrete and Continuous Dynamical Systems-Series B
  13~(4), 739--758.
\newline\urlprefix\url{http://dx.doi.org/10.3934/dcdsb.2010.13.739}

\bibitem[{Canestrelli et~al.(2012)Canestrelli, Fagherazzi, and
  Lanzoni}]{canestrelli2012mass}
Canestrelli, A., Fagherazzi, S., Lanzoni, S., 2012. A mass-conservative
  centered finite volume model for solving two-dimensional two-layer shallow
  water equations for fluid mud propagation over varying topography and dry
  areas. Advances in Water Resources 40, 54--70.
\newline\urlprefix\url{https://doi.org/10.1016/j.advwatres.2012.01.009}

\bibitem[{Canestrelli and Toro(2012)}]{canestrelli2012restoration}
Canestrelli, A., Toro, E.~F., 2012. {Restoration of the contact surface in
  FORCE-type centred schemes II: Non-conservative one-and two-layer
  two-dimensional shallow water equations}. Advances in Water Resources 47,
  76--87.
\newline\urlprefix\url{https://doi.org/10.1016/j.advwatres.2012.03.018}

\bibitem[{Carraro et~al.(2018)Carraro, Valiani, and
  Caleffi}]{carraro2018efficient}
Carraro, F., Valiani, A., Caleffi, V., 2018. {Efficient analytical
  implementation of the DOT Riemann solver for the de Saint Venant-Exner
  morphodynamic model}. Advances in Water Resources 113, 189--201.
\newline\urlprefix\url{https://doi.org/10.1016/j.advwatres.2018.01.011}

\bibitem[{Castro and Fern{\'a}ndez-Nieto(2012)}]{castro2012class}
Castro, M.~J., Fern{\'a}ndez-Nieto, E.~D., 2012. {A class of computationally
  fast first order finite volume solvers: PVM methods}. SIAM Journal on
  Scientific Computing 34~(4), A2173--A2196.
\newline\urlprefix\url{https://dx.doi.org/10.1137/100795280}

\bibitem[{Castro et~al.(2009)Castro, Fern{\'a}ndez-Nieto, Ferreiro, Par{\'e}s,
  et~al.}]{castro2009two}
Castro, M.~J., Fern{\'a}ndez-Nieto, E.~D., Ferreiro, A., Par{\'e}s, C., et~al.,
  2009. {Two-dimensional sediment transport models in shallow water equations.
  A second order finite volume approach on unstructured meshes}. Computer
  Methods in Applied Mechanics and Engineering 198~(33-36), 2520--2538.
\newline\urlprefix\url{https://doi.org/10.1016/j.cma.2009.03.001}

\bibitem[{Castro et~al.(2011)Castro, Fern{\'a}ndez-Nieto, Gonz{\'a}lez-Vida,
  and Par{\'e}s-Madronal}]{castro2011numerical}
Castro, M.~J., Fern{\'a}ndez-Nieto, E.~D., Gonz{\'a}lez-Vida, J.~M.,
  Par{\'e}s-Madronal, C., 2011. Numerical treatment of the loss of
  hyperbolicity of the two-layer shallow-water system. Journal of Scientific
  Computing 48~(1-3), 16--40.
\newline\urlprefix\url{https://doi.org/10.1007/s10915-010-9427-5}

\bibitem[{Castro et~al.(2005)Castro, Ferreiro, Garc{\'\i}a-Rodr{\'\i}guez,
  Gonz{\'a}lez-Vida, Mac{\'\i}as, Par{\'e}s, and
  V{\'a}zquez-Cend{\'o}n}]{castro2005numerical}
Castro, M.~J., Ferreiro, A.~F., Garc{\'\i}a-Rodr{\'\i}guez, J.~A.,
  Gonz{\'a}lez-Vida, J.~M., Mac{\'\i}as, J., Par{\'e}s, C.,
  V{\'a}zquez-Cend{\'o}n, M.~E., 2005. The numerical treatment of wet/dry
  fronts in shallow flows: application to one-layer and two-layer systems.
  Mathematical and Computer Modelling 42~(3), 419--439.
\newline\urlprefix\url{https://doi.org/10.1016/j.mcm.2004.01.016}

\bibitem[{Castro et~al.(2012)Castro, Frings, Noelle, Par\'es, and
  Puppo}]{castro2012hyperbolicity}
Castro, M.~J., Frings, J.~T., Noelle, S., Par\'es, C., Puppo, G., 2012. On the
  hyperbolicity of two-and three-layer shallow water equations. Hyperbolic
  Problems. Theory, Numerics and Applications 1, 337--345.
\newline\urlprefix\url{http://dx.doi.org/10.1142/9789814417099_0030}

\bibitem[{Castro et~al.(2004)Castro, Garc{\i}a-Rodr{\i}guez, Gonz{\'a}lez-Vida,
  Mac{\i}as, Par{\'e}s, and V{\'a}zquez-Cend{\'o}n}]{castro2004numerical}
Castro, M.~J., Garc{\i}a-Rodr{\i}guez, J.~A., Gonz{\'a}lez-Vida, J.~M.,
  Mac{\i}as, J., Par{\'e}s, C., V{\'a}zquez-Cend{\'o}n, M.~E., 2004. Numerical
  simulation of two-layer shallow water flows through channels with irregular
  geometry. Journal of Computational Physics 195~(1), 202--235.
\newline\urlprefix\url{https://doi.org/10.1016/j.jcp.2003.08.035}

\bibitem[{Castro et~al.(2001)Castro, Mac{\'\i}as, and Par{\'e}s}]{castro2001q}
Castro, M.~J., Mac{\'\i}as, J., Par{\'e}s, C., 2001. {A Q-scheme for a class of
  systems of coupled conservation laws with source term. Application to a
  two-layer 1-D shallow water system}. ESAIM: Mathematical Modelling and
  Numerical Analysis 35~(1), 107--127.
\newline\urlprefix\url{https://doi.org/10.1051/m2an:2001108}

\bibitem[{Castro et~al.(2010)Castro, Pardo, Par{\'e}s, and
  Toro}]{castro2010some}
Castro, M.~J., Pardo, A., Par{\'e}s, C., Toro, E., 2010. On some fast
  well-balanced first order solvers for nonconservative systems. Mathematics of
  Computation 79~(271), 1427--1472.
\newline\urlprefix\url{http://doi.org/10.1090/S0025-5718-09-02317-5}

\bibitem[{Castro et~al.(2007)Castro, Pardo~Milan{\'e}s, and
  Par{\'e}s}]{castro2007well}
Castro, M.~J., Pardo~Milan{\'e}s, A., Par{\'e}s, C., 2007. Well-balanced
  numerical schemes based on a generalized hydrostatic reconstruction
  technique. Mathematical Models and Methods in Applied Sciences 17~(12),
  2055--2113.
\newline\urlprefix\url{https://doi.org/10.1142/S021820250700256X}

\bibitem[{Chakir et~al.(2009)Chakir, Ouazar, and Taik}]{chakir2009roe}
Chakir, M., Ouazar, D., Taik, A., 2009. {Roe scheme for two-layer shallow water
  equations: Application to the Strait of Gibraltar}. Mathematical Modelling of
  Natural Phenomena 4~(5), 114--127.
\newline\urlprefix\url{https://doi.org/10.1051/mmnp/20094508}

\bibitem[{Dowell and Jarratt(1971)}]{dowell1971modified}
Dowell, M., Jarratt, P., 1971. A modified regula falsi method for computing the
  root of an equation. BIT Numerical Mathematics 11~(2), 168--174.

\bibitem[{Doyle et~al.(2011)Doyle, Hogg, and Mader}]{doyle2011two}
Doyle, E.~E., Hogg, A.~J., Mader, H.~M., 2011. A two-layer approach to
  modelling the transformation of dilute pyroclastic currents into dense
  pyroclastic flows. Proceedings of the Royal Society of London A:
  Mathematical, Physical and Engineering Sciences 467~(2129), 1348--1371.
\newline\urlprefix\url{http://dx.doi.org/10.1098/rspa.2010.0402}

\bibitem[{Fern{\'a}ndez-Nieto et~al.(2008)Fern{\'a}ndez-Nieto, Bouchut, Bresch,
  Castro, and Mangeney}]{fernandez2008new}
Fern{\'a}ndez-Nieto, E.~D., Bouchut, F., Bresch, D., Castro, M.~J., Mangeney,
  A., 2008. {A new Savage--Hutter type model for submarine avalanches and
  generated tsunami}. Journal of Computational Physics 227~(16), 7720--7754.
\newline\urlprefix\url{https://doi.org/10.1016/j.jcp.2008.04.039}

\bibitem[{Fern{\'a}ndez-Nieto et~al.(2011)Fern{\'a}ndez-Nieto, Castro, and
  Par{\'e}s}]{fernandez2011intermediate}
Fern{\'a}ndez-Nieto, E.~D., Castro, M.~J., Par{\'e}s, C., 2011. {On an
  intermediate field capturing Riemann solver based on a parabolic viscosity
  matrix for the two-layer shallow water system}. Journal of Scientific
  Computing 48~(1-3), 117--140.
\newline\urlprefix\url{https://doi.org/10.1007/s10915-011-9465-7}

\bibitem[{Fjordholm(2012)}]{fjordholm2012energy}
Fjordholm, U.~S., 2012. Energy conservative and stable schemes for the
  two-layer shallow water equations. Hyperbolic Problems: Theory, Numerics and
  Applications 17, 414.
\newline\urlprefix\url{https://doi.org/10.1142/9789814417099_0039}

\bibitem[{Flocke(2015)}]{flocke2015}
Flocke, N., 2015. Algorithm 954: An accurate and efficient cubic and quartic
  equation solver for physical applications. ACM Transactions on Mathematical
  Software (TOMS) 41~(4), 30.
\newline\urlprefix\url{https://doi.org/10.1145/2699468}

\bibitem[{Kesserwani et~al.(2008)Kesserwani, Ghostine, Vazquez, Ghenaim, and
  Mos{\'e}}]{kesserwani2008riemann}
Kesserwani, G., Ghostine, R., Vazquez, J., Ghenaim, A., Mos{\'e}, R., 2008.
  {Riemann solvers with Runge--Kutta discontinuous Galerkin schemes for the 1D
  shallow water equations}. Journal of Hydraulic Engineering 134~(2), 243--255.
\newline\urlprefix\url{https://doi.org/10.1061/(ASCE)0733-9429(2008)134:2(243)}

\bibitem[{Kim and LeVeque(2008)}]{kim2008two}
Kim, J., LeVeque, R.~J., 2008. Two-layer shallow water system and its
  applications. In: Proceedings of the Twelth International Conference on
  Hyperbolic Problems, Maryland. pp. 1--8.

\bibitem[{Krvavica et~al.(2018)Krvavica, Ko{\v{z}}ar, and
  O{\v{z}}ani{\'c}}]{krvavica2018relevance}
Krvavica, N., Ko{\v{z}}ar, I., O{\v{z}}ani{\'c}, N., 2018. The relevance of
  turbulent mixing in estuarine numerical models for two-layer shallow water
  flow. Coupled Systems Mechanics 7~(1), 95--109.
\newline\urlprefix\url{https://doi.org/10.12989/csm.2018.7.1.095}

\bibitem[{Krvavica et~al.(2017{\natexlab{a}})Krvavica, Ko{\v{z}}ar,
  Trava{\v{s}}, and O{\v{z}}ani{\'c}}]{krvavica2017numerical}
Krvavica, N., Ko{\v{z}}ar, I., Trava{\v{s}}, V., O{\v{z}}ani{\'c}, N.,
  2017{\natexlab{a}}. {Numerical modelling of two-layer shallow water flow in
  microtidal salt-wedge estuaries: Finite volume solver and field validation}.
  Journal of Hydrology and Hydromechanics 65~(1), 49--59.
\newline\urlprefix\url{https://doi.org/10.1515/johh-2016-0039}

\bibitem[{Krvavica et~al.(2017{\natexlab{b}})Krvavica, Trava{\v{s}}, and
  O{\v{z}}ani{\'c}}]{krvavica2016salt}
Krvavica, N., Trava{\v{s}}, V., O{\v{z}}ani{\'c}, N., 2017{\natexlab{b}}.
  {Salt-Wedge Response to Variable River Flow and Sea-Level Rise in the
  Microtidal Rje{\v{c}}ina River Estuary, Croatia}. Journal of Coastal Research
  33~(4), 802--814.
\newline\urlprefix\url{https://doi.org/10.2112/JCOASTRES-D-16-00053.1}

\bibitem[{Kurganov and Petrova(2009)}]{kurganov2009central}
Kurganov, A., Petrova, G., 2009. Central-upwind schemes for two-layer shallow
  water equations. SIAM Journal on Scientific Computing 31~(3), 1742--1773.
\newline\urlprefix\url{https://doi.org/10.1137/080719091}

\bibitem[{La~Rocca et~al.(2012)La~Rocca, Adduce, Sciortino, Pinzon, and
  Boniforti}]{la2012two}
La~Rocca, M., Adduce, C., Sciortino, G., Pinzon, A.~B., Boniforti, M.~A., 2012.
  {A two-layer, shallow-water model for 3D gravity currents}. Journal of
  Hydraulic Research 50~(2), 208--217.
\newline\urlprefix\url{https://doi.org/10.1080/00221686.2012.667680}

\bibitem[{Lambert(1906)}]{lambert1906}
Lambert, W.~D., 1906. A generalized trigonometric solution of the cubic
  equation. The American Mathematical Monthly 13~(4), 73--76.

\bibitem[{Liu et~al.(2015)Liu, Yoshikawa, Miyazu, and
  Watanabe}]{liu2015influence}
Liu, H., Yoshikawa, N., Miyazu, S., Watanabe, K., 2015. Influence of saltwater
  wedges on irrigation water near a river estuary. Paddy and Water Environment
  13~(2), 179--189.
\newline\urlprefix\url{https://doi.org/10.1007/s10333-014-0419-1}

\bibitem[{Ljubenkov(2015)}]{ljubenkov2015hydrodynamic}
Ljubenkov, I., 2015. {Hydrodynamic modeling of stratified estuary: case study
  of the Jadro River (Croatia)}. Journal of Hydrology and Hydromechanics
  63~(1), 29--37.
\newline\urlprefix\url{https://doi.org/10.1515/johh-2015-0001}

\bibitem[{Luca et~al.(2009)Luca, Hutter, Kuo, and Tai}]{luca2009two}
Luca, I., Hutter, K., Kuo, C., Tai, Y., 2009. Two-layer models for shallow
  avalanche flows over arbitrary variable topography. International Journal of
  Advances in Engineering Sciences and Applied Mathematics 1~(2), 99--121.
\newline\urlprefix\url{https://doi.org/10.1007/s12572-010-0006-7}

\bibitem[{Majd and Sanders(2014)}]{majd2014lhllc}
Majd, M.~S., Sanders, B.~F., 2014. {The LHLLC scheme for two-layer and
  two-phase transcritical flows over a mobile bed with avalanching, wetting and
  drying}. Advances in water resources 67, 16--31.
\newline\urlprefix\url{https://doi.org/10.1016/j.advwatres.2014.02.002}

\bibitem[{Murillo and Garc{\'\i}a-Navarro(2010)}]{murillo2010exner}
Murillo, J., Garc{\'\i}a-Navarro, P., 2010. {An Exner-based coupled model for
  two-dimensional transient flow over erodible bed}. Journal of Computational
  Physics 229~(23), 8704--8732.
\newline\urlprefix\url{https://doi.org/10.1016/j.jcp.2010.08.006}

\bibitem[{Par{\'e}s(2006)}]{pares2006numerical}
Par{\'e}s, C., 2006. Numerical methods for nonconservative hyperbolic systems:
  a theoretical framework. SIAM Journal on Numerical Analysis 44~(1), 300--321.
\newline\urlprefix\url{https://doi.org/10.1137/050628052}

\bibitem[{Par\'es and Castro(2004)}]{pares2004well}
Par\'es, C., Castro, M.~J., 2004. On the well-balance property of roe's method
  for nonconservative hyperbolic systems. applications to shallow-water
  systems. ESAIM: Mathematical Modelling and Numerical Analysis 38~(5),
  821--852.
\newline\urlprefix\url{https://doi.org/10.1051/m2an:2004041}

\bibitem[{Pelanti et~al.(2008)Pelanti, Bouchut, and Mangeney}]{pelanti2008roe}
Pelanti, M., Bouchut, F., Mangeney, A., 2008. {A Roe-type scheme for two-phase
  shallow granular flows over variable topography}. ESAIM: Mathematical
  Modelling and Numerical Analysis 42~(5), 851--885.
\newline\urlprefix\url{https://doi.org/10.1051/m2an:2008029}

\bibitem[{Rosatti et~al.(2008)Rosatti, Murillo, and
  Fraccarollo}]{rosatti2008generalized}
Rosatti, G., Murillo, J., Fraccarollo, L., 2008. {Generalized Roe schemes for
  1D two-phase, free-surface flows over a mobile bed}. Journal of Computational
  Physics 227~(24), 10058--10077.
\newline\urlprefix\url{https://doi.org/10.1016/j.jcp.2008.08.007}

\bibitem[{Sarno et~al.(2017)Sarno, Carravetta, Martino, Papa, and
  Tai}]{sarno2017some}
Sarno, L., Carravetta, A., Martino, R., Papa, M., Tai, Y.-C., 2017. Some
  considerations on numerical schemes for treating hyperbolicity issues in
  two-layer models. Advances in Water Resources 100, 183--198.
\newline\urlprefix\url{https://doi.org/10.1016/j.advwatres.2016.12.014}

\bibitem[{Schijf and Sch{\"o}nfled(1953)}]{schijf1953theoretical}
Schijf, J., Sch{\"o}nfled, J., 1953. Theoretical considerations on the motion
  of salt and fresh water. In: Proceedings Minnesota International Hydraulic
  Convention. IAHR.

\bibitem[{Strobach(2010)}]{strobach2010}
Strobach, P., 2010. The fast quartic solver. Journal of computational and
  applied mathematics 234~(10), 3007--3024.
\newline\urlprefix\url{https://doi.org/10.1016/j.cam.2010.04.015}

\bibitem[{Strobach(2015)}]{strobach2015}
Strobach, P., 2015. {The Low-Rank LDLT Quartic Solver}. AST-Consulting
  Technical Report, DOI 10~(2.1), 3955--7440.

\bibitem[{Toro(2013)}]{toro2013riemann}
Toro, E.~F., 2013. Riemann solvers and numerical methods for fluid dynamics: a
  practical introduction. Springer Science \& Business Media.

\bibitem[{Wikipedia(2018)}]{wikipedia}
Wikipedia, 2018. Quartic function --- wikipedia{,} the free encyclopedia.
  [Online; accessed 27-January-2018].
\newline\urlprefix\url{https://en.wikipedia.org/w/index.php?title=Quartic_function&oldid=821571464}

\end{thebibliography}

\end{document}